\documentclass[10pt,twoside]{arxiv_article}

\usepackage[table]{xcolor}
\usepackage[latin9]{inputenc}
\usepackage{amssymb}
\usepackage{amsfonts}
\usepackage{amsmath}
\usepackage{amsthm}
\usepackage{commath}
\usepackage{fancybox}
\usepackage{fancyhdr}
\usepackage{graphicx}
\usepackage{subcaption}
\usepackage[hyphens]{url}
\usepackage{arydshln}
\usepackage{multirow}
\usepackage{pdflscape}
\usepackage{tikz}
\usepackage{comment}
\usepackage{nicefrac}
\usepackage{subcaption}
\usepackage{dsfont}
\usepackage{algorithmic}
\usepackage{algorithm}
\usepackage{smartdiagram}
\usepackage{lipsum}
\usepackage[super]{nth}
\usepackage{natbib}
\usepackage[colorlinks = true, linkcolor = blue, urlcolor = blue,
            citecolor = blue, anchorcolor = blue]{hyperref}
\usepackage{ragged2e}
\usepackage{dcolumn,booktabs}
\newcolumntype{d}[1]{D{.}{.}{#1}}

\newlength{\figurewidth}
\newlength{\figureheight}
\def\figureskip{\vskip 10pt plus 2pt minus 2pt\relax}

\newtheorem{remark}{Remark}

\def\limfunc#1{\mathop{\rm #1}}
\def\func#1{\mathop{\rm #1}}

\newcommand{\TsIII}{\hspace{3pt}}
\newcommand{\TsV}{\hspace{5pt}}
\newcommand{\TsVIII}{\hspace{8pt}}

\DeclareMathAlphabet\mathbfcal{OMS}{cmsy}{b}{n}
\newcommand*\LogL{\ensuremath{\boldsymbol\ell}}

\DeclareMathAlphabet{\mathpzc}{OT1}{pzc}{m}{it}

\newcommand{\WACI}{\mathrm{WACI}}
\newcommand{\intensity}{\mathcal{CI}}

\usetikzlibrary{3d, decorations.text, shapes.arrows, positioning, fit, backgrounds}
\usetikzlibrary{shapes, calc, arrows, decorations.markings}

\def\VC{\mathrm{VC}}
\def\PP{\mathrm{PP}}
\def\NA{\mathrm{NA}}
\def\BGS{\mathrm{BGS}}
\def\BMG{\mathrm{BMG}}
\def\mkt{\mathrm{mkt}}
\def\bmg{\mathrm{bmg}}
\def\smb{\mathrm{smb}}
\def\hml{\mathrm{hml}}
\def\wml{\mathrm{wml}}
\def\AS{\mathrm{AS}}

\begin{document}

\title{\textbf{\color{amundi_blue}Measuring and Managing Carbon Risk in Investment Portfolios}%
\footnote{We are grateful to Martin Nerlinger from the University of Augsburg, who
provided us with the time series of the BMG risk factor (see
\url{https://carima-project.de/en/downloads} for more details about this carbon
risk factor). We would also like to thank Melchior Dechelette and Bruno
Taillardat for their helpful comments.}}
\author{
{\color{amundi_dark_blue} Th\'eo Roncalli} \\
Quantitative Research \\
Amundi Asset Management, Paris \\
\texttt{theo.roncalli@amundi.com} \and
{\color{amundi_dark_blue} Th\'eo Le Guenedal} \\
Quantitative Research \\
Amundi Asset Management, Paris \\
\texttt{theo.leguenedal@amundi.com} \and
{\color{amundi_dark_blue} Fr\'ed\'eric Lepetit} \\
Quantitative Research \\
Amundi Asset Management, Paris \\
\texttt{frederic.lepetit@amundi.com} \and
{\color{amundi_dark_blue} Thierry Roncalli} \\
Quantitative Research \\
Amundi Asset Management, Paris \\
\texttt{thierry.roncalli@amundi.com} \and
{\color{amundi_dark_blue} Takaya Sekine} \\
Quantitative Research \\
Amundi Asset Management, Paris \\
\texttt{takaya.sekine@amundi.com}}

\date{\color{amundi_dark_blue}August 2020}

\maketitle

\begin{abstract}
This article studies the impact of carbon risk on stock pricing. To address
this, we consider the seminal approach of \citet{Gorgen-2019}, who proposed
estimating the carbon financial risk of equities by their carbon beta. To
achieve this, the primary task is to develop a brown-minus-green (or
BMG) risk factor, similar to \citet{Fama-1992}. Secondly, we must estimate
the carbon beta using a multi-factor model. While \citet{Gorgen-2019}
considered that the carbon beta is constant, we propose a time-varying
estimation model to assess the dynamics of the carbon risk. Moreover, we test
several specifications of the BMG factor to understand which climate
change-related dimensions are priced in by the stock market. In the second
part of the article, we focus on the carbon risk management of investment
portfolios. First, we analyze how carbon risk impacts the construction of a
minimum variance portfolio. As the goal of this portfolio is to reduce
unrewarded financial risks of an investment, incorporating the carbon risk
into this approach fulfils this objective. Second, we propose a new framework
for building enhanced index portfolios with a lower exposure to carbon risk
than capitalization-weighted stock indices. Finally, we explore how carbon
sensitivities can improve the robustness of factor investing portfolios.
\end{abstract}

\medskip


\clearpage

\section{Introduction}

The general approach to managing the carbon risk of an investment portfolio is
to reduce or control the portfolio's carbon footprint, for instance by considering
CO$_2$ emissions. This approach supposes that the carbon risk will
materialize and having a portfolio with a lower exposure to CO$_2$ emissions
will help to avoid some future losses. The main assumption of this approach is
then to postulate that firms currently with high carbon footprints will be
penalized in the future in comparison with firms currently with low carbon
footprints.\smallskip

In this article, we use an alternative approach. We define
carbon risk from a financial point of view, and we consider that the
carbon risk of equities corresponds to the market risk
priced in by the stock market. This carbon financial risk can be
decomposed into a common (or systematic) risk factor and a specific
(or idiosyncratic) risk factor. Since identifying the
specific risk is impossible, we focus on the common risk factor that
drives the carbon risk. The objective is then to build a
market-based carbon risk measure to manage this market
risk in investment portfolios. This is exactly the framework
proposed by \citet{Gorgen-2019} in their seminal paper.\smallskip

\citet{Gorgen-2019} proposed extending the Fama-French-Carhart model by
including a brown-minus-green (or BMG) risk factor. Using the sorted portfolios
technique popularized by \citet{Fama-1992}, they build a factor-mimicking
portfolio based on a scoring model and more than fifty carbon risk variables.
They then defined the carbon financial risk of a stock using its price
sensitivity to the BMG factor or its carbon beta. In this paper, we explore the
original approach of \citet{Gorgen-2019} and estimate a time-varying model in
order to analyze the dynamics of the carbon risk. Moreover, we make the
distinction between relative and absolute carbon risk. Relative carbon risk may
be viewed as an extension or forward-looking measure of the carbon footprint,
where the objective is to be more exposed to green firms than to brown firms.
In this case, this is equivalent to promoting stocks with a negative carbon
beta over stocks with a positive carbon beta. Absolute carbon risk considers
that both large positive and negative carbon beta values incur a financial risk
that must be reduced. This is an agnostic or neutral method, contrary to the
first method which is more related to investors' moral values. In this paper,
an important issue concerns the climate change-related dimensions that are
priced in by the financial market. According to \citet{Delmas-2013}, the
concept of environmental performance encompasses several dimensions but there
is no consensus on universally accepted environmental performance indicators.
To address this issue, we compare pricing models with different criteria:
current carbon footprint, carbon management, climate change score and
environmental risk.\smallskip

The carbon beta of a stock can be interpreted as its carbon-related
systematic risk. Therefore, it contains financial information that
is extremely useful from a trading point of view. In particular, it
can be used to improve the construction of a minimum variance
portfolio, the main goal of which is to avoid unrewarded risks. It can also
be used in the investment scope of enhanced indexing or factor
investing. For these different illustrations, we develop an
analytical framework to better understand the impact of carbon
betas.\smallskip

This paper is organized as follows. Section Two presents the seminal approach
of \citet{Gorgen-2019}, and reviews the pricing impact of the carbon risk
factor. Besides the static analysis, we also consider a dynamic approach where
the carbon beta is estimated using the Kalman filter. Then we test the
contribution of the different climate-change related dimensions. Section Three
is dedicated to investment portfolio management considering the information
deduced from the carbon beta values. First, we focus on the minimum variance
portfolio, before extending the analytical results to enhanced index portfolios
and explaining how carbon betas can also be used in a factor investing
framework. Finally, Section Four offers some concluding remarks.

\section{Measuring carbon risk}

To manage a portfolio's carbon risk, it is important to measure
carbon risk at the company level. There are different ways to measure this
risk, including the fundamental and market approaches. In this paper, we will favor the second
approach because it provides a better assessment of the impact of
climate-related transition risks on each company's stock price. Moreover,
the market-based approach allows us to mitigate the issue of a lack of climate
change-relevant information. In what follows, we present this latest approach
by using the mimicking portfolio for carbon risk developed by
\citet{Gorgen-2019}. We also discuss the different climate change-relevant
dimensions to determine which dimensions are priced in by
the market.\smallskip

\subsection{The Carima approach}
\label{section:carima}

The goal of the carbon risk management (Carima) project, developed by
\citet{Gorgen-2019}, is to develop \textquoteleft\textsl{a quantitative tool in
order to assess the opportunities of profits and the risks of losses that occur
from the transition process}\textquoteright. The
Carima approach combines a market-based approach and a fundamental approach.
Indeed, the carbon risk of a firm or a portfolio is measured by considering the
dynamics of stock prices which are partly determined by climate policies and
transition processes towards a green economy. Nevertheless, a prior fundamental
approach is important to quantify carbon risk. In a practical manner, the
fundamental approach consists in defining a carbon risk score for each stock of
a universe using a set of objective measures, whereas the market approach
consists in building a brown minus green or BMG carbon risk factor, and
computing the risk sensitivity of stock prices with respect to this BMG factor.
Therefore, the carbon factor is derived from climate change-relevant
information from numerous firms.

\subsubsection{Construction of the BMG factor}

The development of the BMG factor is based on a large amount of climate-relevant
information provided by different databases. In the
following, we report the methodology used by the Carima project to construct
the BMG factor and thereby obtain a deeper understanding of the
results\footnote{A more exhaustive presentation is available in the Carima
manual, which can be downloaded at the following address:
\url{https://carima-project.de/downloads}.}. Two steps are required to
develop this new common risk factor: (1) the development of a scoring system to
determine if a firm is green, neutral or brown and (2) the construction
of a mimicking factor portfolio for carbon risk which has a long exposure to
brown firms and a short exposure to green firms.\smallskip

The first step consists in defining a brown green score (BGS) with a
fundamental approach to assess the carbon risk of different firms. This scoring
system uses four ESG databases over the period from 2010 to 2016: Thomson
Reuters ESG, MSCI ESG Ratings, Sustainalytics ESG ratings and
the Carbon Disclosure Project (CDP) climate change questionnaire. Overall, $55$
carbon risk proxy variables are retained\footnote{The governance and social
variables of a traditional ESG analysis or even certain environmental variables
such as waste recycling, water consumption or toxic emissions have been
deleted.}. Then, \citet{Gorgen-2019} classified the variables into three
different dimensions that may affect the stock value of a firm in the event of
unexpected shifts towards a low carbon economy:
\begin{enumerate}
\item Value chain (impact of a climate policy or a cap and trade system on
    the different activities of a firm: inbound logistics and supplier
    chain, manufacturing production, sales, etc.);

\item Public perception (external environmental image of a firm: ratings,
    controversies, disclosure of environmental information, etc.);

\item Adaptability (capacity of the firm to shift towards a low carbon
    strategy without substantial efforts and losses).
\end{enumerate}
The value chain dimension mainly deals with current emissions while the
adaptability dimension reflects potential future emissions determined
in particular by emission reduction targets and environmental R\&D spending.
The Carima project considers that the higher the variable, the browner the
firm. Hence each variable (except the dummies) is transformed into a dummy
derived with respect to the median, meaning that $1$ corresponds to a brown
value and $0$ corresponds to a green value. Then, three scores are created
and correspond to the average of all variables contained in each dimension:
the \textit{value chain} $\VC$, the \textit{public perception} $\PP$ and the
\textit{non-adaptability} $\NA$. It follows that each score has a range
between $0$ and $1$. \citet{Gorgen-2019} proposed defining the brown green
score (BGS) by the following equation:
\begin{equation}
\BGS_{i}\left( t\right) =\frac{2}{3}\left( 0.7\cdot \VC_{i}\left( t\right)
+0.3\cdot \PP_{i}\left( t\right) \right) +\frac{\NA_{i}\left( t\right) }{3}%
\left( 0.7\cdot \VC_{i}\left( t\right) +0.3\cdot \PP_{i}\left( t\right)
\right)   \label{eq:BGS}
\end{equation}
The higher the BGS value, the browner the firm. The value chain and public
perception axes directly influence stock prices in the case of unexpected
changes in the transition process. However, \citet{Gorgen-2019} considered that
the impact of the value chain score is more important than the impact of the
public perception score. The adaptability axis influences the equity value in a
different way. Indeed, it mitigates the upward or downward impacts of the two
other axes. The less adaptable a firm is, the greater the impact of an
unexpected acceleration in the transition process. In total, almost $1\,650$
firms are retained thanks to sufficient data covered.\smallskip

The second step consists in constructing a BMG carbon risk factor. Here the
Carima project considers an average BGS for each stock that corresponds to the
mean value of the BGS over the period in question, from 2010 to 2016. The
construction of the BMG factor follows the methodology of \citet{Fama-1992,
Fama-1993}, which consists in splitting the stocks into six portfolios:
\begin{equation*}
\begin{tabular}{c|ccc}
        & Green & Neutral & Brown\\
\hline
Small   & SG    & SN      & SB \\
Big     & BG    & BN      & BB \\
\end{tabular}
\end{equation*}
where the classification is based on the terciles of the aggregating BGS and
the median market capitalization. Then, the return of the BMG factor
is defined as follows:
\begin{equation}
R_{\bmg}\left(t\right) = \frac{1}{2} \left(R_{\mathrm{SB}}\left(t\right) +
R_{\mathrm{BB}}\left(t\right)\right) - \frac{1}{2} \left(R_{\mathrm{SG}}\left(t\right)+
R_{\mathrm{BG}}\left(t\right)\right)
\end{equation}
where the returns of each portfolio is value-weighted by market
capitalization. The BMG factor can then be integrated as a new common risk
factor into a multi-factor model. Some statistical details are reported in
Table \ref{tab:carima1} on page \pageref{tab:carima1}, whereas the historical
cumulative performance of the BMG factor is showed in Figure
\ref{fig:carima1} on page \pageref{fig:carima1}. According to the factor
developed for the Carima project, brown firms slightly outperform
green firms from 2010 to the end of 2012. During the next three years, the
cumulative return fell by almost $35\%$ because of the unexpected path in the
transition process towards a low carbon economy. From 2016 to the end of the
study period, brown firms created a slight excess performance. Overall,
the best-in-class green stocks outperform the worst-in-class green stocks over
the study period with an annual return of $2.52\%$.

\subsubsection{Advantages and limits}

Many advantages can be attributed to the BMG factor. Some biases in the
construction of ESG databases are offset since the BGS scores are derived
from several databases. Moreover, the tests performed by \citet{Gorgen-2019}
showed that there are no significant country-specific or sector-specific
effects\footnote{In the following, we will find some sector-specific effects
for short periods.}. Even though the BMG factor has many benefits, it can
be subject to some disadvantages, starting with the treatment of variables.
The transformation of continuous and discrete variables into a dummy variable
with respect to the median value fixes the problem of extreme values, but
does not differentiate between values based on their distance from the median. Besides, the most
important problem is that no rebalancing takes place. Some tests performed
by \citet{Gorgen-2019} showed that less than five percent of firms shifted
between the green, neutral and brown portfolios during the study period but such
a decision presents some consistency problems in the long-run. For instance,
the results obtained by the average BGS score for the 2010-2016 period have
been generalized for the following two years.\smallskip

Another limit involves the size-specific effects in the BMG factor. Table
\ref{tab:carima2} reports the correlation matrix of common risk factors during
the sample period. While the value (HML) and momentum (WML) factors are not
significantly correlated to the size (SMB) factor, the BMG factor is influenced
by size characteristics. Mitigating this problem can be difficult since the
carbon risk factor has been derived from the methodology of \citet{Fama-1992}.
The most plausible explanation of this correlation is that among the studied
firms, the green firms have the largest market capitalizations as we can see in
Figure \ref{fig:te8} on page \pageref{fig:te8}. In this case, when big firms
outperform small firms, both the SMB and BMG factor returns decrease.
Furthermore, preventing the BMG factor from capturing size-specific effect is
an important but difficult matter to solve.\smallskip

\begin{table}[tbph]
\centering
\caption{Correlation matrix of factor returns (in \%)}
\label{tab:carima2}
\begin{tabular}{c|*{5}{d{3.4}}}
Factor     & \multicolumn{1}{c}{MKT}  &  \multicolumn{1}{c}{SMB} &  \multicolumn{1}{c}{HML}  &  \multicolumn{1}{c}{WML}  & \multicolumn{1}{c}{BMG} \\
\hline
 MKT & 100.00^{***} &              &              &              &              \\
 SMB &   1.41       & 100.00^{***} &              &              &              \\
 HML &  11.51       &  -8.93       & 100.00^{***} &              &              \\
 WML & -14.59       &   3.87       & -41.43^{***} & 100.00^{***} &              \\
 BMG &   5.33       &  20.33^{**}  &  27.41^{***} & -21.28^{**}  & 100.00^{***} \\
\end{tabular}
\begin{flushleft}
{\small \textit{Source}: \citet{Gorgen-2019}.}
\end{flushleft}
\vspace*{-10pt}
\end{table}

Selecting numerous variables allows us to avoid some important dependencies on
a variable and incorporate a lot of climate change-relevant information.
Nevertheless, we have double counting problems. For instance, the carbon
emissions score and the climate change theme score in the MSCI ESG Ratings are
both taken into account when developing the BGS score but the carbon emissions
score is integrated into the climate change theme score. Moreover, some
variables in the public perception dimension are not exclusive to the carbon
risk dimension, such as the ESG score developed by Sustainalytics ESG ratings
or the Industry-adjusted Overall score developed by MSCI ESG Ratings.\smallskip

\subsubsection{Static analysis}
\label{section:carima-static}

The first carbon risk objective is to assess the relevance of the
BMG factor during the study period. To do this, we follow the analysis of
\citet{Gorgen-2019}, but our analysis is slightly different because of the
investment universe. Indeed, \citet{Gorgen-2019} considered a universe of
$39\,500$ stocks, whereas we only consider the stocks that were present in the
MSCI World index during the 2010-2018 period. As a result, our
investment universe has less than $2\,000$ stocks, but we think that a
restricted universe makes more sense than a very large universe. Indeed,
the computation of a market beta is already difficult for some small and micro
stocks because of OTC pricing and low trading activity. Therefore,
calculating a carbon beta is even more difficult for such
equities.\smallskip

It may be worthwhile to compare different common factor models to measure the
information gain related to the carbon risk factor. The first studied model
is the CAPM model introduced by \citet{Sharpe-1964} which is defined by:
\begin{equation}
R_{i}\left(t\right) = \alpha_{i} + \beta_{\mkt,i} R_{\mkt}\left(t\right) +
\varepsilon_{i}\left(t\right)
\label{eq:carima_static1}
\end{equation}
where $R_{i}\left(t\right)$ is the return of asset $i$, $\alpha_{i}$ is the
alpha of the asset $i$, $R_{\mkt}\left(t\right)$ is the return of the market
factor, $\beta_{\mkt,i}$ is the systematic risk (or the market beta) of stock
$i$ and $\varepsilon_{i}\left(t\right)$ is the idiosyncratic risk. We may
also consider that the risk is multi-dimensional with the model developed by
\citet{Fama-1992}:
\begin{equation}
R_{i}\left(t\right) = \alpha_{i} + \beta_{\mkt,i} R_{\mkt}\left(t\right) +
\beta_{\smb,i} R_{\smb}\left(t\right) + \beta_{\hml,i} R_{\hml}\left(t\right) +
\varepsilon_{i}\left(t\right)
\label{eq:carima_static2}
\end{equation}
where $R_{\smb}\left(t\right)$ is the return of the size (or small minus big)
factor, $\beta_{\smb,i}$ is the SMB sensitivity (or the size beta) of stock
$i$, $R_{\hml}\left(t\right)$ is the return of the value (or high minus low)
factor and $\beta_{\hml,i}$ is the HML sensitivity (or the value beta) of
stock $i$. Nevertheless, these two models do not include the carbon risk.
Furthermore, we also consider the MKT+BMG model:
\begin{equation}
R_{i}\left(t\right) = \alpha_{i} + \beta_{\mkt,i} R_{\mkt}\left(t\right) +
\beta_{\bmg,i} R_{\bmg}\left(t\right) + \varepsilon_{i}\left(t\right)
\label{eq:carima_static3}
\end{equation}
and the extended Fama-French (FF+BMG) model:
\begin{equation}
R_{i}\left(t\right) = \alpha_{i} + \beta_{\mkt,i} R_{\mkt}\left(t\right) +
\beta_{\smb,i} R_{\smb}\left(t\right) + \beta_{\hml,i} R_{\hml}\left(t\right) +
\beta_{\bmg,i} R_{\bmg}\left(t\right) + \varepsilon_{i}\left(t\right)
\label{eq:carima_static4}
\end{equation}
where $R_{\bmg}\left(t\right)$ is the return of the carbon risk factor and
$\beta_{\bmg,i}$ is the BMG sensitivity of stock $i$. Another well-known
model is the four-factor model (4F) developed by \citet{Carhart-1997}. This model
corresponds to the following equation:
\begin{equation}
R_{i}\left(t\right) = \alpha_{i} + \beta_{\mkt,i} R_{\mkt}\left(t\right) +
\beta_{\smb,i} R_{\smb}\left(t\right) + \beta_{\hml,i} R_{\hml}\left(t\right) +
\beta_{\wml,i} R_{\wml}\left(t\right) + \varepsilon_{i}\left(t\right)
\label{eq:carima_static5}
\end{equation}
where $R_{\wml}\left(t\right)$ is the return of the momentum (or winners minus
losers) factor and $\beta_{\wml,i}$ is the WML sensitivity of stock $i$. Again,
we may include the carbon risk factor to obtain a five-factor (4F+BMG) model:
\begin{eqnarray}
R_i\left(t\right) &=& \alpha_{i} + \beta_{\mkt,i} R_{\mkt}\left(t\right) +
\beta_{\smb,i} R_{\smb}\left(t\right) + \beta_{\hml,i} R_{\hml}\left(t\right) + \notag \\
& & \beta_{\wml,i} R_{\wml}\left(t\right) + \beta_{\bmg,i} R_{\bmg}\left(t\right) +
\varepsilon_{i}\left(t\right)
\label{eq:carima_static6}
\end{eqnarray}
In minimum variance or enhanced index portfolios, we assume that the factor returns are
uncorrelated.\smallskip

\begin{table}[tbph]
\centering
\caption{Comparison of cross-section regressions (in \%)}
\label{tab:carima3}
\begin{tabular}{lc*{3}{d{2.1}}}
\hline
 & Adjusted $\mathfrak{R}^{2}$ & \multicolumn{3}{c}{$F$-test} \\
 & difference                  & \multicolumn{1}{c}{$10\%$} & \multicolumn{1}{c}{$5\%$} & \multicolumn{1}{c}{$1\%$} \\
\hline
CAPM vs FF      & 1.74 & 34.6 & 25.5 & 13.5 \\
CAPM vs MKT+BMG & 1.74 & 21.2 & 15.6 &  9.2 \\ \hdashline
FF vs FF+BMG    & 1.73 & 22.5 & 17.5 &  9.7 \\
FF vs FF+WML    & 0.22 & 6.6  &  3.0 &  0.8 \\ \hdashline
4F vs 4F+BMG    & 1.76 & 23.6 & 18.6 & 10.0 \\
\hline
\end{tabular}
\vspace*{-20pt}
\end{table}

Risk factor model estimates were performed on single stocks
during the 2010-2018 period\footnote{In this article, we only consider the
stocks that were in the MSCI World index for at least three years during the
2010-2018 period. Moreover, we do not consider the returns for the
period during which the stock is outside the index.}. In Table
\ref{tab:carima3}, we have reported a comparison between the common factor
models and their nested models by computing the average difference of the
adjusted $\mathfrak{R}^{2}$ and the proportion of stocks for which the Fisher
test is significant at $10\%$, $5\%$ and $1\%$. According to the first two
tests, we remark that the Fama-French and MKT+BMG models significantly increase
the explanatory power compared to the CAPM model. This increase is
significant for almost $26\%$ and $16\%$ of the stocks respectively for Models
(\ref{eq:carima_static2}) and (\ref{eq:carima_static3}) at the threshold of
$5\%$. Nevertheless, the difference between the two models for the Fisher test
declines when the threshold significance is $1\%$ and the effect on the
explanatory power is at the same level for the SMB and HML factors together and
the BMG factor alone. The following two tests study the relevance of the carbon
factor against the momentum factor when they are added to the Fama and French
model. We remark that the sensitivity of the stock returns to the carbon factor
is higher than the sensitivity of the stocks returns to the momentum factor.
The last test confirms the relevance of the BMG factor when we add the BMG
factor to the Carhart model. Overall, we confirm the original
results\footnote{See Table IA.2 in \citet{Gorgen-2019}.} obtained by
\citet{Gorgen-2019}, meaning that the carbon factor plays a key role in the
variation of stock returns.\smallskip

\begin{figure}[tbph]
\centering
\caption{Box plots of the carbon sensitivities}
\label{fig:carima5a}
\includegraphics[width = \figurewidth, height = \figureheight]{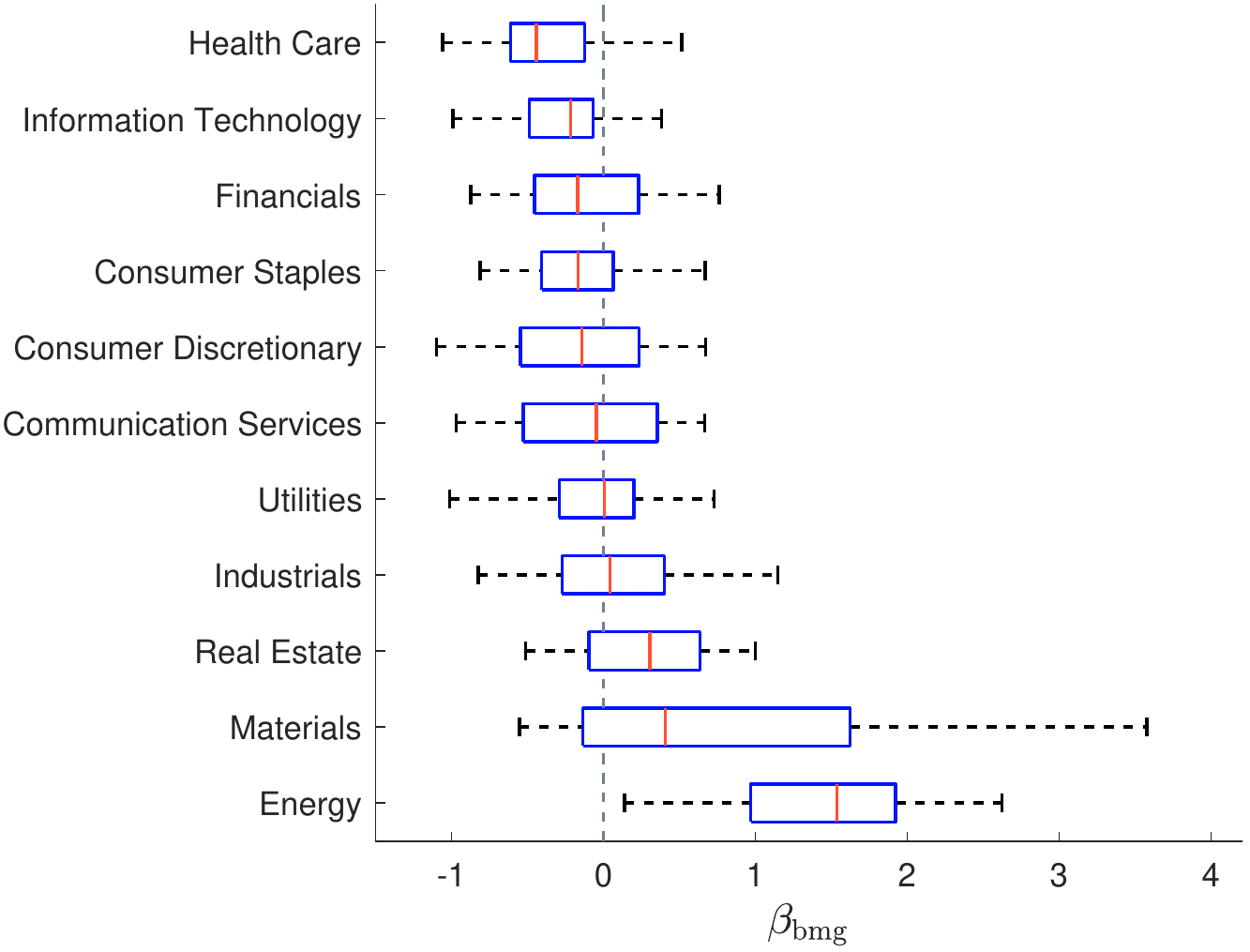}
\end{figure}

Figure \ref{fig:carima5a} reports the sector\footnote{Sector taxonomy is based
on the Global Industry Classification Standard (GICS).} analysis of the carbon
beta $\hat{\beta}_{\bmg,i}$ estimated with Model (\ref{eq:carima_static3}). The
box plots provide the median, the quartiles and the $5\%$ and $95\%$ quantiles
of the carbon beta. The energy, materials, real estate and, to a lesser extent,
industrial sectors are negatively impacted by an unexpected acceleration in the
transition process towards a green economy, certainly because these four
sectors are responsible for a large part of greenhouse gas emissions (GHG).
Indeed, the energy and the materials sectors have a large scope 1 mainly
because of oil and gas drilling and refining for the former and the extraction
and processing of raw materials for the latter\footnote{Moreover, scope 3 of
the basic materials sector is very large.}. Overall, the energy sector is the
most sensitive to an unexpected acceleration in the transition process but the
carbon beta range is widest for the materials sector, which indicates a high
heterogenous risk for this sector. The latter is mostly influenced by the
growth in material demand per capita. In terms of the industrial sector,
construction and transport are responsible for much of global final energy
consumption, which leads to a high carbon risk for this sector. In the real
estate sector, the firms have a large scope 2 and energy efficiency can be
improved in many cases. If we consider a long-run investment, a transition
process that reduces climate change can protect households from physical risks
like climate hazards. Nevertheless, a short-run vision supposes that a climate
policy negatively impacts households that over-consume. Therefore, real estate
investment trusts are highly sensitive to climate-related policies. One
surprising result involves utility firms which do not have a substantial
positive carbon beta whereas their scope 1 is on average the largest of any
sector. This overall neutral carbon sensitivity for utilities is explained by
their carbon emissions management and efforts to reduce carbon exposure.
Indeed, \citet{Leguenedal-2020} have shown that utilities -- power generation
according to the Sectoral Decarbonization Approach (SDA) -- have been
aggressive in their inflexion of carbon intensity trajectories.\smallskip

If we consider the sectors positively impacted by an unexpected shift towards a
green economy, these primarily include health care, information technology and
consumer staples because of their low GHG emissions. Financials are also part
of this group, but the interpretation of the carbon risk differs. Indeed, the
carbon risk of financial institutions is less connected to their GHG emissions
than their investments and financing programs. The greener a financial
institution's investment, the lower its carbon beta. The low value of the
median beta implies that financial firms integrate carbon risk into their
investment strategies or that financials are not significantly disadvantaged by
the relative carbon risk.

\begin{remark}
These results are coherent, but slightly different from those obtained by
\citet{Gorgen-2019}. Certainly, this difference comes from the investment
universe, which is more liquid in our case. For instance, we obtain less high median carbon betas --
except for the energy and materials sectors -- since our universe of stocks
includes only the world's biggest firms.
\end{remark}

We have also reported in Figure \ref{fig:carima5b} on page
\pageref{fig:carima5b} the box plots for four other investment
universes: Eurozone, Europe ex EMU, North America and Japan. The
energy sector remains the most negatively impacted by an unexpected
acceleration in the transition process regardless of the
region under review. The integration of carbon risk in the Eurozone is
substantial, especially in the financial sector where green
investments are widely taken into account. The inclusion of this
carbon risk is also significant in Europe ex EMU and Japan, whereas it is
very mixed in North America. Concerning the real estate sector,
BMG risk is highly integrated in Europe ex EMU\footnote{Almost all
the companies with a negative carbon beta are English, while the one
remaining with a positive carbon beta is Swiss.} and slightly in
North America because of their large exposure to climate
risks\footnote{Real estate in United States is especially
exposed to rising sea levels and hurricanes.}, while the
integration of carbon risk is different in Japan because
real estate investment trusts are more short-sighted despite the
vulnerability of this sector\footnote{Real estate in Japan is
exposed to the physical risk of typhoons.}.\smallskip

\subsubsection{Dynamic analysis}
\label{section:carima-dynamic}

In this section, we suppose that the risks are time-varying. For instance,
carbon risk may evolve with the introduction of a climate-related policy,
a firm's environmental controversy, a change in the firm's environmental strategy,
a greater integration of carbon risk into portfolio strategies, etc. Therefore, we
use the following dynamic common factor model\footnote{The beta estimates are
based on the state space model (SSM) and the Kalman filter algorithm
described in Appendix \ref{appendix:kalman-filter} on page
\pageref{appendix:kalman-filter}.}:
\begin{equation}
R_{i}\left( t\right) =R\left( t\right) ^{\top }\beta _{i}\left( t\right)
+\varepsilon _{i}\left( t\right)   \label{eq:carima_dynamic1}
\end{equation}
where $R\left( t\right) =\left( 1,R_{\mkt}\left( t\right) ,R_{\bmg}\left(
t\right) \right) $ is the vector of factor returns, $\beta _{i}\left(
t\right) $ is the vector of factor betas\footnote{In this model,
we only consider the dynamics of market and carbon risks. We
have also performed the same analysis with the 4F+BMG model, but the
results are noisier.}:%
\begin{equation}
\beta _{i}\left( t\right) =\left(
\begin{array}{c}
\alpha _{i}\left( t\right)  \\
\beta _{\mkt,i}\left( t\right)  \\
\beta _{\bmg,i}\left( t\right)
\end{array}
\right)   \label{eq:carima_dynamic2}
\end{equation}
and $\varepsilon _{i}\left( t\right) $ is a white noise. We assume that the
state vector $\beta _{i}\left( t\right) $ follows a random walk process:
\begin{equation}
\beta_{i}\left( t\right) =\beta_{i}\left( t-1\right) +\eta_{i}\left(
t\right)   \label{eq:carima_dynamic3}
\end{equation}
where $\eta_{i}\left( t\right) \sim \mathcal{N}\left( \mathbf{0}_{3},\Sigma
_{\beta ,i}\right) $ is the white noise vector and $\Sigma_{\beta ,i}$ is
the covariance matrix of the white noise. Several specifications of $\Sigma
_{\beta ,i}$ may be used\footnote{%
For instance, we can assume that $\left( \Sigma_{\beta ,i}\right) _{1,1}$ is
equal to zero, implying that the alpha coefficient $\alpha _{i}\left( t\right)
$ is constant.}, but we assume that $\Sigma _{\beta ,i}$ is a diagonal matrix
in the following. As previously, the time-varying risk factor model is used on
single stocks during the 2010-2018 period\footnote{As previously, we only
consider the stocks that were in the MSCI World index for at least three years
during the 2010-2018 period and we take into account only the returns for the
period during which the stock is in the MSCI World index.}. Below, we provide
the average of two forecast error criteria between the OLS model and the SSM
model:
\begin{equation*}
\begin{tabular}{ccc}
\hline
Model & OLS      & SSM      \\ \hline
MAE   & $4.95\%$ & $4.63\%$ \\
RMSE  & $6.45\%$ & $6.01\%$ \\
\hline
\end{tabular}
\end{equation*}
We notice that the time-varying risk factor model reduces the forecast error.
On average, the monthly return error is equal to $4.95\%$ in the OLS model
while it is equal to $4.63\%$ in the SSM model. Overall, the SSM model reduces
the mean absolute error value of the last observation date by $12.17\%$ with
respect to the OLS model.\smallskip

In Table \ref{tab:carima8}, we have reported the proportion of firms for
which the $t$-student test of the estimation of the covariance matrix $%
\Sigma _{\beta ,i}$ is significant at $10\%$, $5\%$ and $1\%$ confidence
levels. We notice that the coefficients of the covariance matrix are
significant for a substantial number of firms implying that between 10\% and
15\% of stocks present time-varying market and carbon risks.\smallskip

\begin{figure}[tbph]
\centering
\caption{Dynamics of the average relative carbon risk $\beta_{\bmg,\mathcal{R}}\left( t\right)$ by region}
\label{fig:carima10b}
\includegraphics[width = \figurewidth, height = \figureheight]{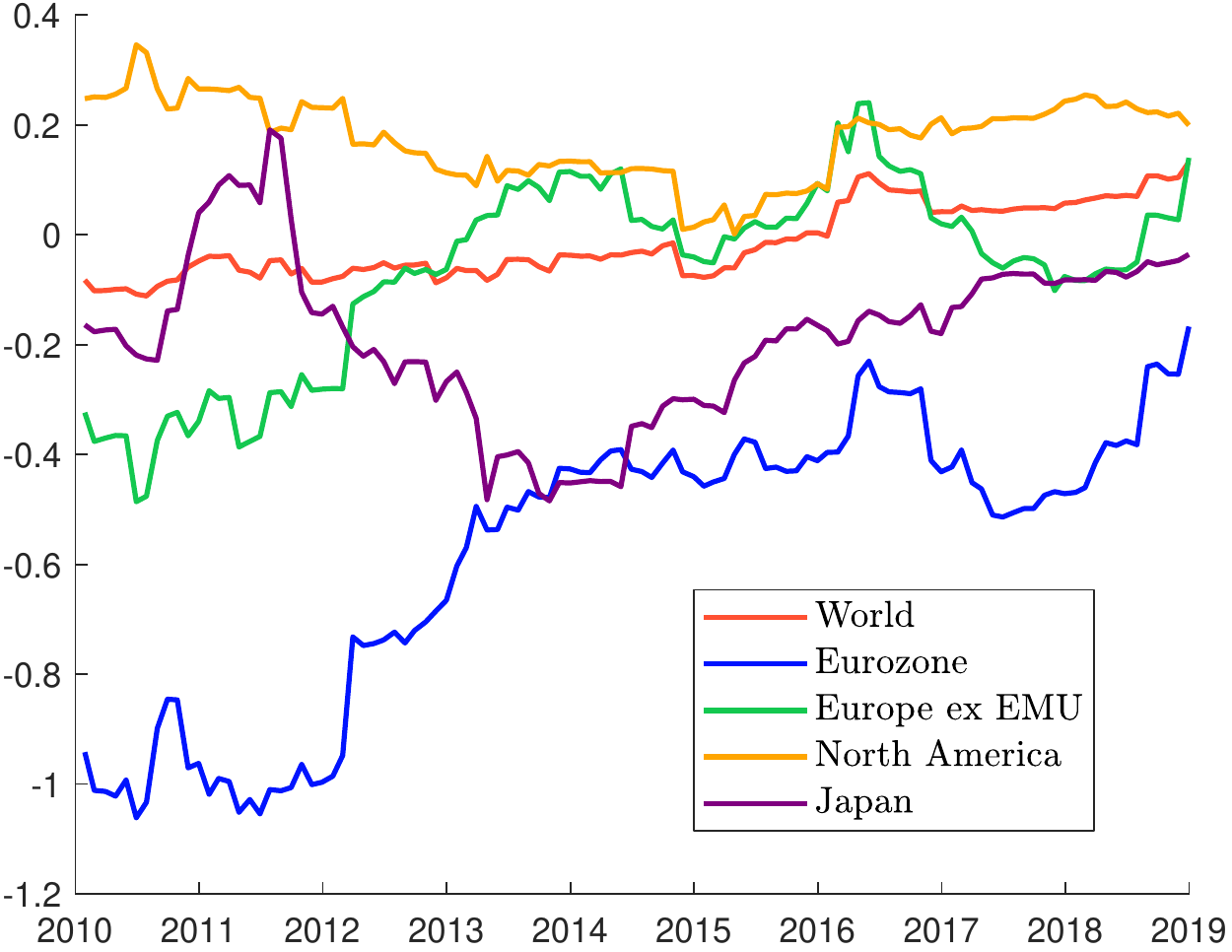}
\end{figure}

In Figure \ref{fig:carima10b}, we have reported the variation of the average
carbon beta by region\footnote{The average carbon beta
$\beta_{\bmg,\mathcal{R}}\left( t\right) $ for the region $\mathcal{R}$
at time $t$ is calculated as follows:
\begin{equation*}
\beta_{\bmg,\mathcal{R}}\left( t\right) =\frac{\sum_{i\in \mathcal{R}%
}\beta _{\bmg,i}\left( t\right) }{\limfunc{card}\mathcal{R}}
\end{equation*}%
}. Whatever the study period, the carbon beta $\beta_{\bmg,\mathcal{R}}\left(
t\right) $ is positive in North America, which implies that American stocks are
negatively influenced by an acceleration in the transition process towards a
green economy. The average carbon beta is always negative in the
Eurozone\footnote{In Japan, it is also negative most of the time.}. Overall,
the Eurozone has always a lower average carbon beta than the world as a whole,
whereas the opposite is true for North America. Nevertheless, the sensitivity
of European equity returns to carbon risk dramatically increases and the BMG
betas are getting closer for North America and the Eurozone. In Europe ex EMU,
the BMG beta is higher than in the Eurozone but their trends are very similar.
Regarding the Japanese firms, the trend has tracked the world as a whole since
2013 but with a lower carbon beta. We notice that the carbon risk is not driven
by climate agreements in the short run. For instance, the 2030 climate and
energy framework, which includes EU-wide targets and policy objectives for the
period from 2021 to 2030 does not influence the average European carbon beta in
2014 certainly because of the lack of binding commitments. Another example is
the 2015 Paris Climate Agreement, which does not include fiscal pressure
mechanisms. Because of the differences between expectations and constraints,
the Paris Climate Agreement has not been followed by a significant increase in
the carbon beta and has been concomitant with the outperformance of brown
stocks one quarter later\footnote{We recall that the performance of the brown
minus green portfolio is given in Figure \ref{fig:carima1} on page
\pageref{fig:carima1}.}. In February 2016, the global increase of the carbon
beta is related to a sector-specific effect. Indeed, the materials sector has
largely outperformed because of a substantial increase in gold, silver and zinc
prices whereas the market index has decreased. Furthermore, some firms in the
materials sector were not driven by the market for a short period, implying a
sharp increase in the carbon beta. This explains that the carbon beta returns
to its long-term trend some months later.\smallskip

\begin{table}[t]
\centering
\caption{Significance test frequency for the white noise covariance matrix (in \%)}
\label{tab:carima8}
\begin{tabular}{cccc}
\hline
Factor         &       $10\%$ &        $5\%$ &  $1\%$ \\ \hline
$\alpha$       & ${\TsV}7.97$ & ${\TsV}4.10$ & $0.84$ \\
$\beta_{\mkt}$ &      $15.95$ &      $10.22$ & $3.93$ \\
$\beta_{\bmg}$ &      $10.00$ & ${\TsV}5.90$ & $1.85$ \\
\hline
\end{tabular}
\vspace*{-10pt}
\end{table}

\begin{figure}[tbph]
\centering
\caption{Dynamics of the average absolute carbon risk
$\left\vert \beta\right\vert _{\bmg,\mathcal{R}}\left( t\right) $ by region}
\label{fig:carima10c}
\includegraphics[width = \figurewidth, height = \figureheight]{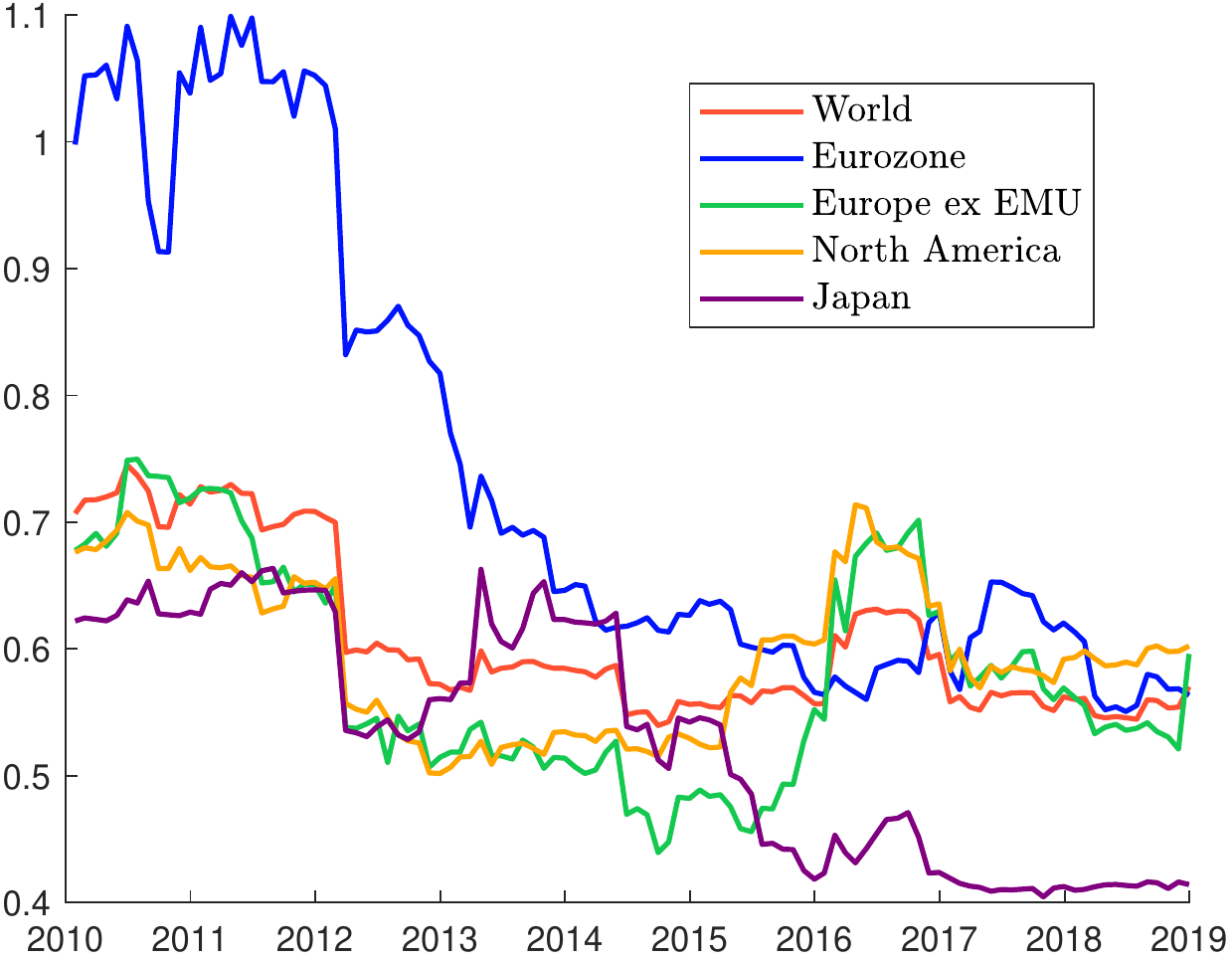}
\end{figure}

\begin{remark}
In terms of the results obtained, one issue concerns the increase in the average
carbon beta in Europe, and we wonder if this is due to a geographical area
effect or a carbon sensitivity specific effect -- the firms with the most
negative carbon betas may be less influenced by the BMG factor over time. To
answer this question, we use the method of sorting portfolios, which has been
popularized by \citet{Fama-1992}. Every month, we rank the stocks with respect
to their carbon beta, and form five quintile portfolios. Portfolio $Q_{1}$
corresponds to the $20\%$ lowest carbon beta stocks while Portfolio $Q_{5}$
corresponds to the $20\%$ highest carbon beta stocks. The stocks are
equally weighted in each portfolio and the portfolios are rebalanced every
month. In Figure \ref{fig:carima14b} on page \pageref{fig:carima14b}, we have
reported the average carbon beta of the five sorted portfolios for each region.
Since European stocks are the most positively impacted by an unexpected change
in the transition process towards a green economy, we have to compare similar
portfolios between the regions. For instance, Portfolio $Q_{3}$ in the Eurozone
and Portfolio $Q_{1}$ in North America started with an average carbon beta
around minus one. In the Eurozone, the increase in the carbon beta for
Portfolio $Q_{3}$ is much higher than the increase of Portfolio $Q_{1}$ in
North America. The Europe ex EMU region is another example.
Portfolio $Q_{4}$ in this region is comparable with Portfolio $Q_{3}$ in North
America since their average carbon betas started at a similar level.
Nevertheless, the average carbon beta of Portfolio $Q_{4}$ in Europe ex EMU increases
while it decreases for Portfolio $Q_{3}$ in North America. We can deduce that
the increase of the carbon beta in Europe is not due to a sensitivity effect
for the stocks that are the most negatively sensitive to the BMG factor, but to
a geographical effect.
\end{remark}

\begin{figure}[p]
\centering
\caption{Dynamics of the median carbon risk $\beta_{\bmg,\mathcal{S}}\left(t\right)$ by sector}
\label{fig:carima15a}
\includegraphics[width = \figurewidth, height = \figureheight]{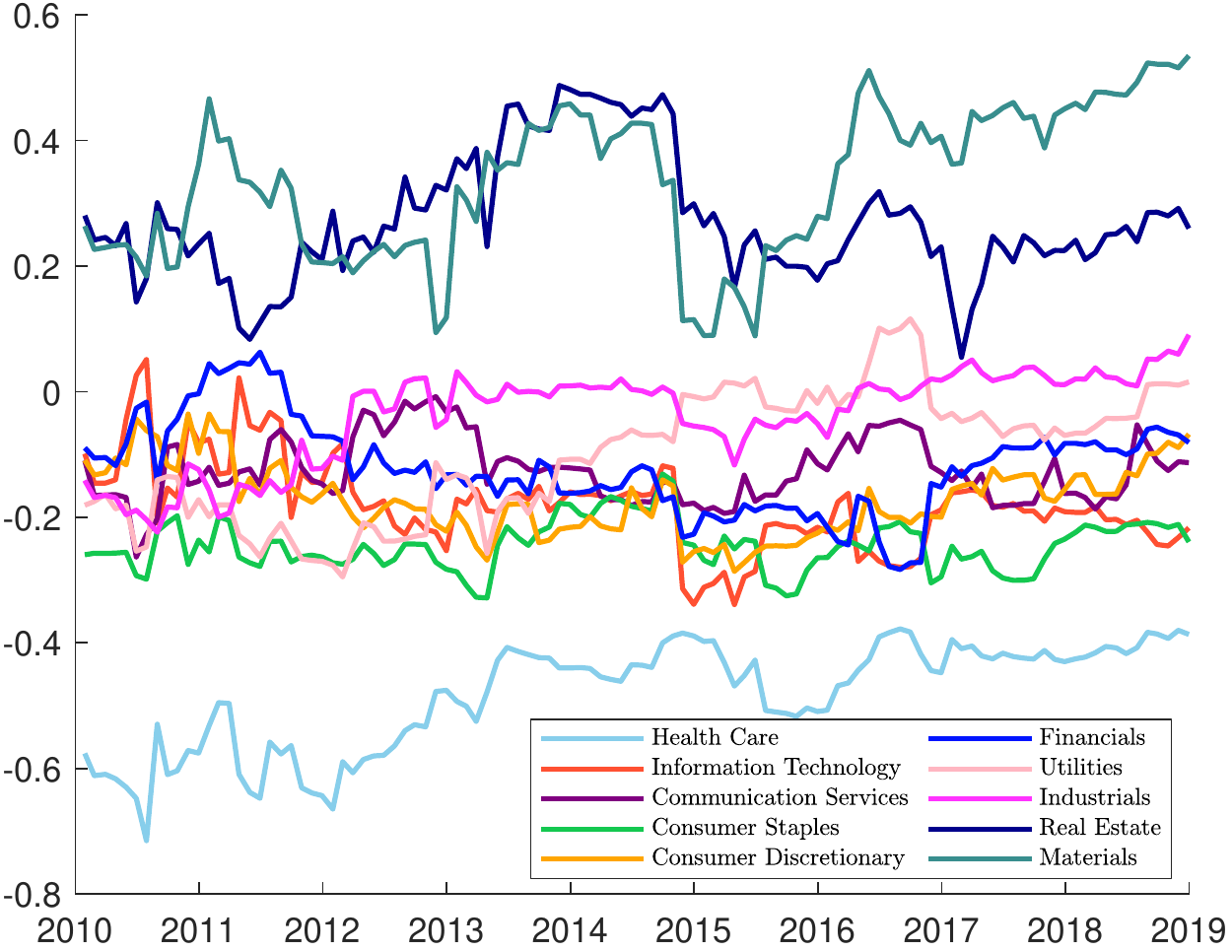}
\end{figure}

\begin{figure}[p]
\centering
\caption{Dynamics of the median carbon risk $\beta_{\bmg,\mathcal{S}}\left(t\right)$ for the energy sector}
\label{fig:carima15b}
\includegraphics[width = \figurewidth, height = \figureheight]{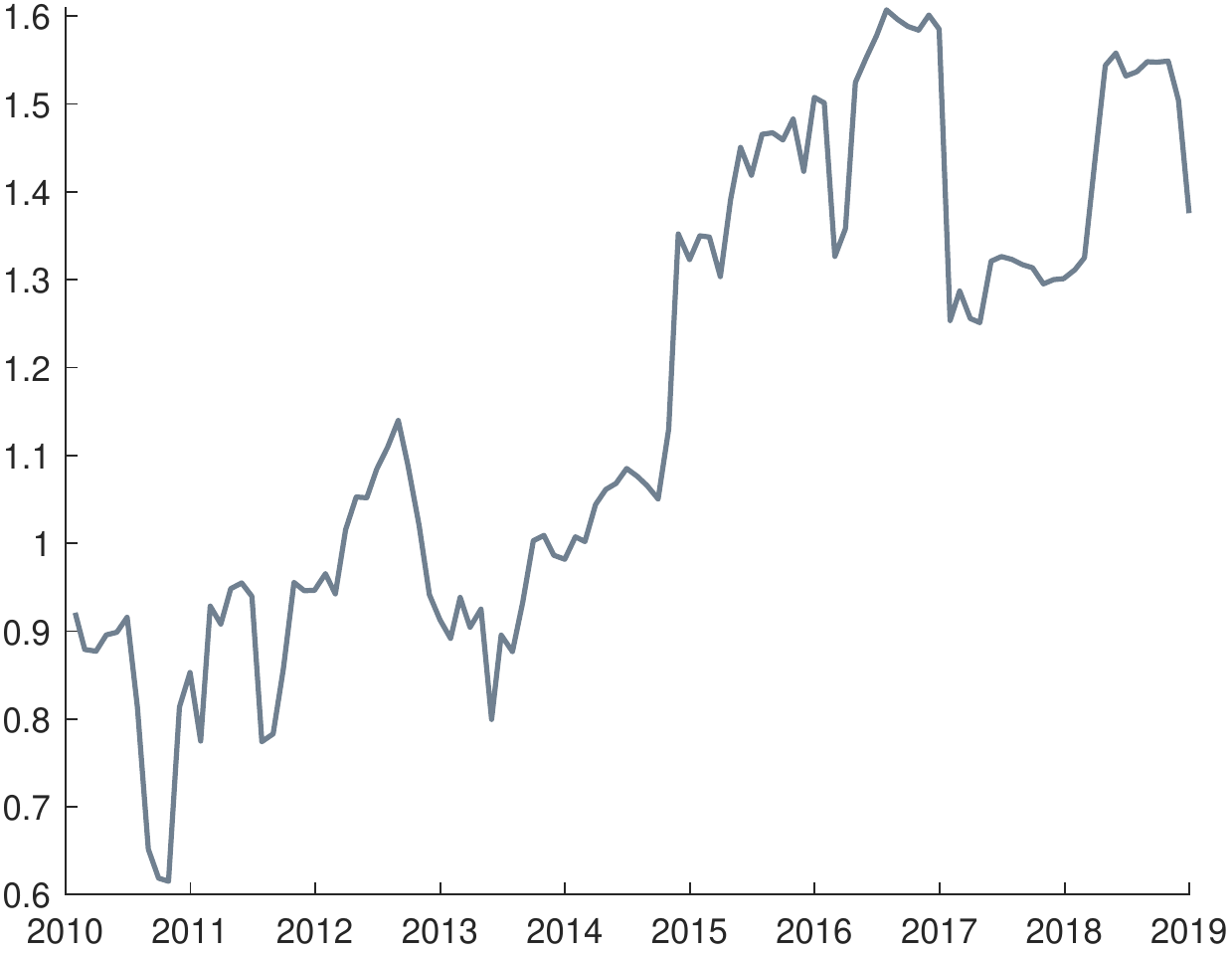}
\end{figure}

Figure \ref{fig:carima10c} provides the dynamics of the average absolute
carbon beta\footnote{It is calculated as follows:
\begin{equation*}
\left\vert \beta\right\vert_{\bmg,\mathcal{R}}\left( t\right) =\frac{%
\sum_{i\in \mathcal{R}}\left\vert \beta_{\bmg,i}\left( t\right) \right\vert
}{\limfunc{card}\mathcal{R}}
\end{equation*}%
} $\left\vert \beta\right\vert _{\bmg,\mathcal{R}}\left( t\right) $ for each
region $\mathcal{R}$\footnote{As we have seen in Table \ref{tab:carima1} on
page \pageref{tab:carima1}, the volatility of the BMG factor is lower than the
volatility of the MKT factor. A variation of the carbon beta can not be
interpreted in an ordinary scale.}. The higher the value of $\left\vert
\beta\right\vert _{\bmg,\mathcal{R}}\left( t\right) $, the greater the impact
(positive or negative) of carbon risk on stock returns. Curiously, we notice
that the integration of carbon risk in the financial market decreases over
time. In particular, there is a substantial decrease in 2012 and then a
stabilization of the global average absolute carbon beta\footnote{The decrease
of $\left\vert \beta\right\vert _{\bmg,\mathcal{R}}\left( t\right) $ in March
2012 is not due to a climate-related policy but to green stocks far
outperforming as we can see in Figure \ref{fig:carima1} on page
\pageref{fig:carima1}. At the same time, the European market declined while the
American market increased. Therefore, the carbon beta considerably increased
for green European stocks, which was driven mostly by the European market's
return rather than their carbon return. In a similar way, the carbon beta
decreased for brown American stocks, which was driven mostly by the American
market's return rather than their carbon return.}. The Eurozone was the region
with the highest sensitivity to carbon risk but this decreased sharply by
almost $44\%$ between 2010 and 2018. However, we observe all regions converging
except Japan\footnote{Carbon risk pricing in Japan is around $25\%$ lower than
globally.}. The convergence of absolute sensitivities between large
geographical regions indicates that investors see carbon risk as a global
issue.\smallskip

We may also be interested in carbon risk trends by sector. In
the static analysis, we recall that the energy, materials and real estate sectors
were the most negatively impacted by an unexpected acceleration in the transition
process, whereas the opposite is true for the health care, information technology
and consumer staples sectors. Figures \ref{fig:carima15a} and
\ref{fig:carima15b} provide the trends in the median carbon beta $\beta
_{\bmg,\mathcal{S}}\left( t\right) $ for the sector $\mathcal{S}$ at time
$t$, which is defined as follows:
\begin{equation*}
\beta_{\bmg,\mathcal{S}}\left(t\right) =\limfunc{median}_{i\in \mathcal{S}%
}\beta_{\bmg,i}\left( t\right)
\end{equation*}%
We distinguish four categories. The first one concerns high positively
sensitive sectors to the carbon factor. This includes only the energy sector.
The stock price of the latter is increasingly negatively influenced by the
movements of the carbon factor. The second category includes the materials
and real estate sectors for which the positive sensitivity of stock price to
carbon factor is much more moderate. The third category includes the sectors with
a neutral or a low negative sensitivity to the carbon factor. This
category is made up of the industrials, utilities, communication services,
consumer discretionary, consumer staples, financials and information
technology sectors. The last category, including only the health care sector,
concerns a moderate negative sensitivity to the carbon factor. However, this
sector is getting closer and closer to the carbon risk-neutral category over time,
even though we continue to observe a gap.\smallskip

Among the second category, we observe that the materials and real
estate sectors started with a similar median carbon beta. However,
the spread has been increasing between the two sectors since 2016, because
the median carbon risk is stable in the case of the real estate
sector whereas it is increasing for the materials sector. This gap may
be persistent in the long run, implying that the materials sector may
be increasingly affected by carbon risk. Concerning the third
category, we observe that the industrials sector was mostly
negatively influenced by the BMG factor, but it has become a carbon
risk-neutral sector\footnote{This increase in the median carbon beta
may push the industrial sector to the second category in the
future.}. Overall, sector differentiation is more
important than geographical breakdown for investors since
market-based carbon risks converge both in absolute and relative values at the geographical level.\smallskip

\begin{remark}
We may also consider the absolute average carbon beta for each sector.
Results are reported in Figure \ref{fig:carima15d} on page \pageref{fig:carima15d}.
In this case, we distinguish two main categories. The first
one corresponds to high carbon pricing. This category includes the energy
and materials sectors. The second one includes the sectors with a low (either
upward or downward) carbon sensitivity to stock prices.
\end{remark}

\begin{figure}[tbph]
\centering
\caption{Density of the carbon risk first difference}
\label{fig:carima13}
\includegraphics[width = \figurewidth, height = \figureheight]{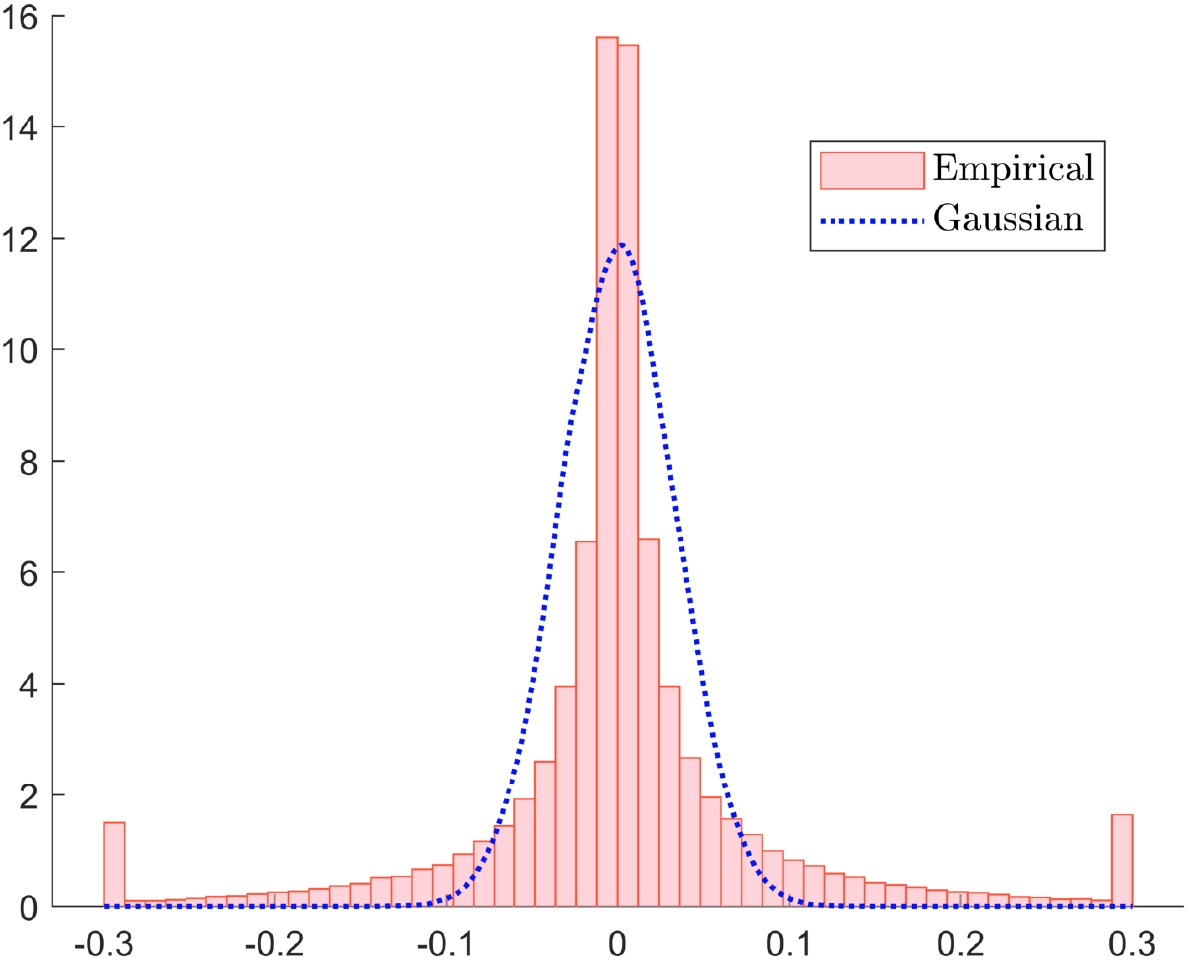}
\end{figure}

The advantage of this dynamic analysis is to assume that common risks are
time-varying. In Figure \ref{fig:carima13}, we have reported the density of the monthly variations
$\beta_{\bmg,i}\left(t\right) - \beta_{\bmg,i}\left( t-1\right)$. We observe that
it is far from being a Gaussian distribution
since we observe fat tails with a significant number of extreme variations.
Hence, we deduce that the time-varying model allows us to take into account some
extreme changes in carbon risk. However, a microeconomic analysis shows
that, in the event of environmental controversies, the model is not able to
substantially change the carbon risk in the short run\footnote{This is for example the
case of some famous controversial events, e.g. Volkswagen, Bayer, etc. Whatever the
variation of the carbon factor, the firm's stock return decreases in the case of
an environmental controversy.}. The extreme changes are more
explained by regional or sector-related effects. This confirms that $\beta
_{\bmg,i}\left( t\right) $ is more a low-frequency systematic measure than a
high-frequency idiosyncratic measure of the carbon risk.\smallskip

\subsection{Alternative measures of the BMG factor}

The brown minus green or BMG factor developed by \citet{Gorgen-2019} is an
aggregation of numerous variables and we wonder what climate
change-related dimensions\footnote{In this section, the dimensions do not
correspond to the dimensions previously introduced (value chain, public
perception and adaptability). When we refer to climate-related
dimensions, it concerns any variables involved in climate change.} are
most priced in by the financial market. Indeed, some variables taken into
account by the Carima project may create some noise without necessarily being
relevant. With a set of more than 50 proxy variables, it is likely
that some variables can include information that is not priced in by the market or
that is not specifically related to carbon risk. For the value
chain and adaptability dimensions, the variables provide climate-relevant information.
Nevertheless, the public perception dimension mainly contains ESG or E pillar scores while these scores do not
incorporate just climate-related information.\smallskip

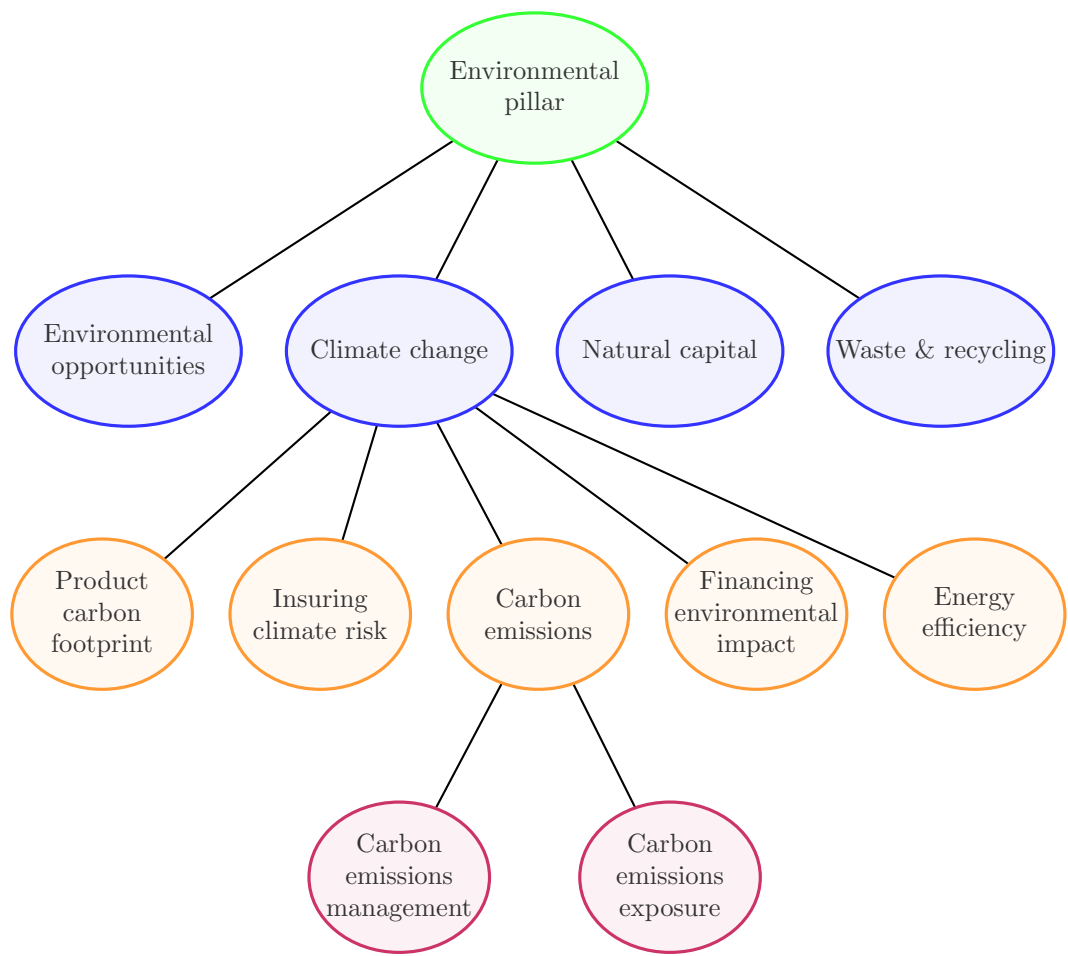
\begin{figure}[h!]
\caption{Dimension hierarchy in the environmental pillar (MSCI methodology)}
\centering
\label{fig:msci_hierarchy}
\begin{tikzpicture}[every text node part/.style={align=center}]

\draw[black, thick] (-0.6,0) -- (-6,-3.5);
\draw[black, thick] (-0.6,0) -- (-2.4,-3.5);
\draw[black, thick] (-0.6,0) -- (1.2,-3.5);
\draw[black, thick] (-0.6,0) -- (4.8,-3.5);

\draw[black, thick] (-2.4,-3.5) -- (-6.35,-7);
\draw[black, thick] (-2.4,-3.5) -- (-3.45,-7);
\draw[black, thick] (-2.4,-3.5) -- (-0.55,-7);
\draw[black, thick] (-2.4,-3.5) -- (2.35,-7);
\draw[black, thick] (-2.4,-3.5) -- (5.25,-7);

\draw[black, thick] (-0.55,-7) -- (-2.4,-10.5);
\draw[black, thick] (-0.55,-7) -- (1.2,-10.5);

\filldraw[color=green!80, fill=green!5, very thick](-0.6,0) ellipse (1.5 and 1) node[color=black!80]{Environmental\\pillar};

\filldraw[color=blue!80, fill=blue!5, very thick](-6,-3.5) ellipse (1.5 and 1) node[color=black!80]{Environmental\\opportunities};
\filldraw[color=blue!80, fill=blue!5, very thick](-2.4,-3.5) ellipse (1.5 and 1) node[color=black!80]{Climate change};
\filldraw[color=blue!80, fill=blue!5, very thick](1.2,-3.5) ellipse (1.5 and 1) node[color=black!80]{Natural capital};
\filldraw[color=blue!80, fill=blue!5, very thick](4.8,-3.5) ellipse (1.5 and 1) node[color=black!80]{Waste \& recycling};

\filldraw[color=orange!80, fill=orange!5, very thick](-6.35,-7) ellipse (1.2 and 1) node[color=black!80]{Product\\carbon\\footprint};
\filldraw[color=orange!80, fill=orange!5, very thick](-3.45,-7) ellipse (1.2 and 1) node[color=black!80]{Insuring\\climate risk};
\filldraw[color=orange!80, fill=orange!5, very thick](-0.55,-7) ellipse (1.2 and 1) node[color=black!80]{Carbon\\emissions};
\filldraw[color=orange!80, fill=orange!5, very thick](2.35,-7) ellipse (1.2 and 1) node[color=black!80]{Financing\\environmental\\impact};
\filldraw[color=orange!80, fill=orange!5, very thick](5.25,-7) ellipse (1.2 and 1) node[color=black!80]{Energy\\efficiency};

\filldraw[color=purple!80, fill=purple!5, very thick](-2.4,-10.5) ellipse (1.2 and 1) node[color=black!80]{Carbon\\emissions\\management};
\filldraw[color=purple!80, fill=purple!5, very thick](1.2,-10.5) ellipse (1.2 and 1) node[color=black!80]{Carbon\\emissions\\exposure};

\end{tikzpicture}
\begin{flushleft}
{\small \textit{Source}: \citet{MSCI-2020}.}
\end{flushleft}
\end{figure}

In what follows, we present some risk factors built on different climate-related
dimensions (Figure \ref{fig:msci_hierarchy}). These factors have long exposure to worst-in-class
green stocks and short exposure to best-in-class green stocks. To obtain
results that are comparable with the carbon risk factor developed for the Carima
project, we use the methodology of \citet{Fama-1992, Fama-1993}. Nevertheless,
the returns of the four portfolios (SG, BG, SB and BB) are
equally weighted\footnote{We have also derived the risk factors with a
capitalization-weighted scheme. The results do not change much. We have a
correlation of around $95\%$ between EW and CW factors. Nevertheless, we obtain
a better explanatory power for the multi-factor regression models when we
consider an equally-weighted scheme.}, the portfolio weights are rebalanced
every month and the stock universe is the MSCI World index. Another difference from the Carima approach is
that we integrate the financial firms into our factors even though
their carbon risk is determined differently from the
other sectors. According to \citet{Gorgen-2019}, their carbon risk is more influenced by their investments
than by their carbon emissions. We retain this sector because
excluding financial firms would result in lower statistical significance of
the factors. Moreover, we think that financial firms are exposed to some
climate-related dimensions. Unless otherwise specified, all the
variables come from the MSCI ESG Ratings dataset.\smallskip

\subsubsection{Exposure to carbon costs}

\begin{figure}[tbph]
\centering
\caption{Cumulative performance of the factors based on the exposure to carbon costs}
\label{fig:factor8a}
\includegraphics[width = \figurewidth, height = \figureheight]{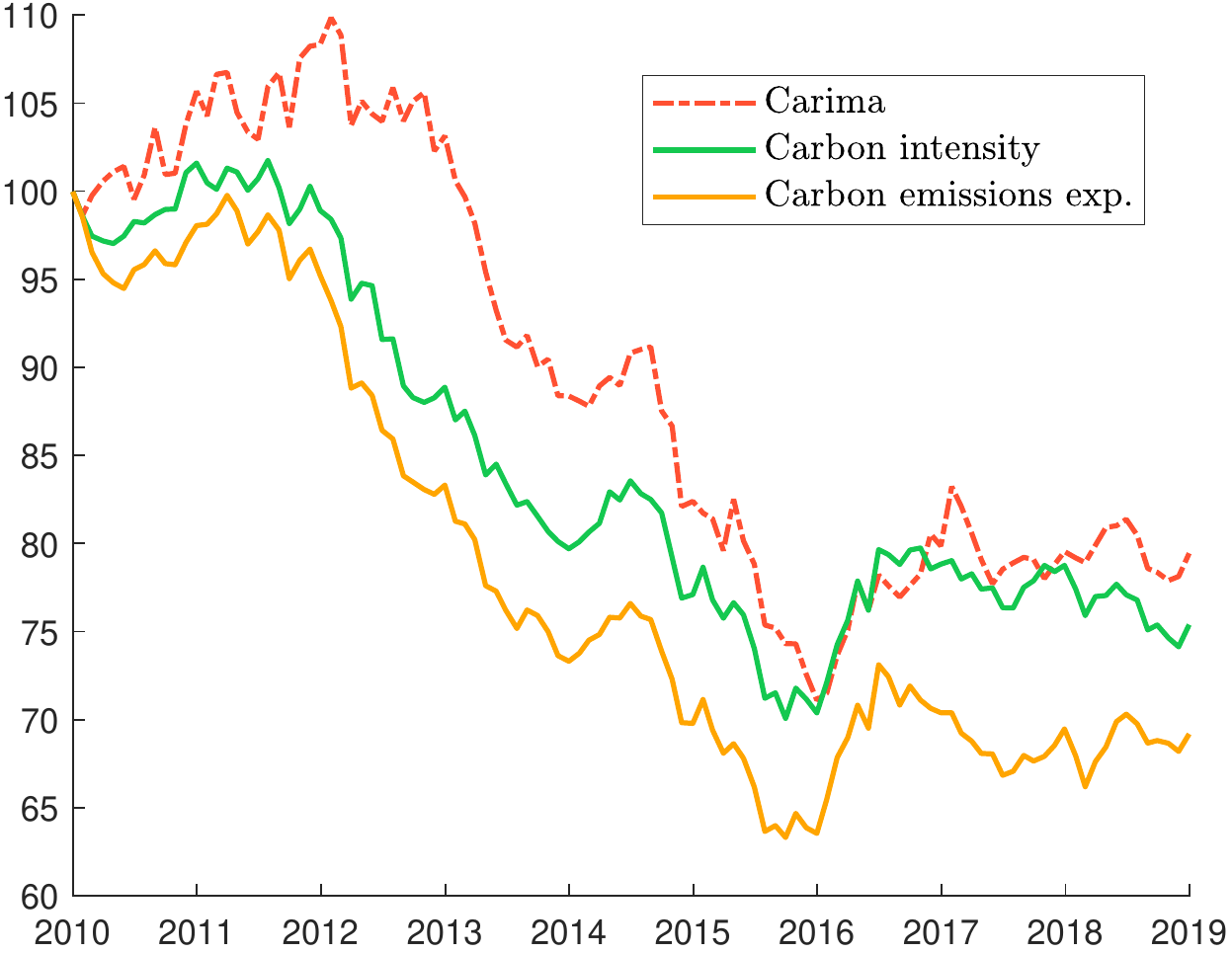}
\end{figure}

The first comparison concerns the factors built on the exposure to
carbon pricing and regulatory caps: (1) the carbon
intensity\footnote{This factor has already been proposed by \citet{In-2017} on a universe of American stocks.}
derived on the three scopes\footnote{The three scopes are available in the Trucost
dataset. Assets are selected every month with a
reporting lag of one year.} and (2) the
carbon emissions exposure score based on the carbon-intensive
business activities and the current or potential future carbon
regulations \citep{MSCI-2020}. In Figure \ref{fig:factor8a}, we have
reported the cumulative performance of these two factors and the
carbon factor developed by \citet{Gorgen-2019}. We observe that the
three factors are very similar. For instance, we have a strong
linear correlation greater than $90\%$ between the carbon intensity
and carbon emissions exposure factors. Because of the greater
similarity with the Carima factor, we wonder if the carbon
intensity risk measure is the only carbon dimension priced in by the
market\footnote{It is obvious that numerous other carbon variables
explain the fluctuations in stock prices but some (or all) of them
increase the explanatory power of a multi-factor model because they
are correlated with the carbon intensity.}. Table \ref{tab:factor4}
on page \pageref{tab:factor4} provides the correlation between the
carbon risk factors and the reference factors. We have a close
correlation between the two current carbon factors and the Carima factor,
but not so much ($58\%$ and $64\%$). In Table \ref{tab:factor7} on
page \pageref{tab:factor7}, we have reported a comparison between
multi-factor models and their nested models. There is a slightly
larger factor exposure for the two current carbon risk
factors with respect to the Carima risk factor. Hence, the carbon
exposure is a dimension widely taken into account by the market to
determine changes in stock prices, but we cannot deduce that this
dimension is the only dimension priced in because our methodology and
our universe of stocks to build factors are different from
the Carima project.

\begin{remark}
We have also built the factors with the capitalization-weighted scheme and
excluding financial firms. In this case, the carbon intensity factor has
a smaller factor exposure than the Carima factor because of a strong
correlation with the market, value and momentum factors. However, this decrease in
explanatory power is mainly due to the exclusion of financials. Indeed, this
sector has little exposure to the potential risk of increased costs linked to
carbon pricing or cap and trade systems. We have seen previously in Figure
\ref{fig:carima5a} on page \pageref{fig:carima5a} that the financial sector
has a negative relative carbon risk. Even though carbon
risk is constructed differently for this sector, excluding it means losing information.
Concerning the CW carbon emissions exposure factor, the results are halfway
between the two EW factors and the Carima factor. Nevertheless, the better results
of explanatory power for this CW factor in comparison with the Carima factor
are too low to conclude that the carbon exposure is the only carbon dimension priced in by the market.
\end{remark}

\subsubsection{Carbon exposure and management pricing}

While prior literature \citep{Semenova-2015} distinguishes between
environmental performance (or management) and environmental risk (or exposure),
we focus on the same distinction but on the carbon dimension. In what follows,
we will provide an answer to whether carbon emissions exposure or carbon
management contributes more to variations in stock prices. The carbon emissions
exposure mainly involves current carbon emissions and factors inherent to the
firm's business whereas the carbon emissions management is about future
potential carbon emissions and measures the efforts to reduce this exposure. If
we consider the carbon axes used by the Carima project, the first one
corresponds to the value chain axis and the second one to the adaptability
axis. In Table \ref{tab:factor7} on page \pageref{tab:factor7}, we remark that
both carbon exposure and carbon management significantly increase the
explanatory power of the common factor models but this variation is greater
with the first factor. With respect to the other carbon risk factors, the
carbon management factor has a lower correlation with the Carima factor as we
can see in Table \ref{tab:factor4} on page \pageref{tab:factor4}. Moreover, the
close correlation of this factor with the Carhart risk factors is a problem in
the case of minimum variance and enhanced index portfolios where the risk
factors are supposed to be uncorrelated. Furthermore, this carbon factor is not
an alternative option to the reference carbon factor.\smallskip

Despite the issues with the carbon management factor, abandoning this dimension
would be unfortunate. The carbon emissions factor built on the carbon emissions
score, which is derived from both the carbon emissions management and exposure
scores, allows us to overcome this problem. The higher the exposure score and
the lower the management score, the lower the carbon emissions score. The
latter assesses the capacity of a firm to handle increasing carbon costs. As we
can see in Table \ref{tab:factor7} on page \pageref{tab:factor7}, aggregating
the two dimensions leads to an increase in explanatory power, which is
significant for almost $19\%$ of the stocks at the threshold of $5\%$.
Moreover, as seen in Table \ref{tab:factor4} on page \pageref{tab:factor4}, the
carbon emissions dimension is the most closely correlated to the Carima risk
factor.

\begin{remark}
The median sector carbon beta coefficients have almost the same ranks for the
carbon exposure and the aggregated carbon factors. Nonetheless, we have very
different results with the carbon management factor. In the case of carbon
management, the carbon betas are very high for the health care sector because
of the lack of environmental performances. For the utilities and materials
sector, the carbon beta has significantly decreased -- the utilities sector has
the lower average carbon beta. These results are not surprising since the more
environmentally responsible firms face greater environmental challenges
\citep{Delmas-2010, Rahman-2012}. We confirm again that the exposure factor is
better than the management factor since the aggregated carbon factor is
associated to a higher adjusted $\mathfrak{R}^{2}$ coefficient while it is very
related to the carbon exposure dimension.
\end{remark}

\subsubsection{Environmental, climate and carbon dimensions}

Let us consider now the three main climate-related dimensions: environment,
climate change and carbon emissions. The carbon dimension is nested into the
climate dimension which is itself nested into the environmental dimension (see
Figure \ref{fig:msci_hierarchy} on page \pageref{fig:msci_hierarchy}). The last
two factors are based respectively on the environmental pillar score and the
climate change theme score available in the MSCI ESG Ratings dataset. The
environmental pillar includes the climate change dimension but also
environmental opportunities, waste and recycling, and natural capital. The
climate change scope includes carbon emissions, environmental risk financing,
climate change vulnerability of insurance companies and the product carbon
footprint \citep{MSCI-2020}. The three new factors are very similar with a
correlation between them around $75\%$ and $80\%$. In Table \ref{tab:factor7}
on page \pageref{tab:factor7}, we notice that the mimicking factor portfolio
for environmental risk is more closely correlated with the factor built by
\citet{Gorgen-2019} than the climate change factor, certainly because the
Carima risk factor doesn't just incorporate carbon emissions variables. One
issue with the environmental factor is that it is significantly negatively
correlated with the market factor\footnote{In a minimum variance portfolio
where the average relative carbon risk is negative, a bear market may imply a
higher loss since the best-in-class green stocks underperform the other
stocks.}.\smallskip

In Figure \ref{fig:factor10b} on page \pageref{fig:factor10b}, we
have reported the sector analysis of the carbon beta
$\hat{\beta}_{\bmg,i}$ estimated with the MKT+BMG model. We notice that the carbon
emissions and environmental factors are very similar while there are
some differences with the climate change factor. For this latter
factor, the median and quantiles are associated with small relative
carbon risk ${\beta}_{\bmg,i}$ for most sectors. These
small carbon risks are mainly offset by the financial sector's higher carbon risk.
While the financial sector has the lower
median carbon beta with the carbon emissions factor, it ranks eighth
with the climate change factor because it takes into account the
vulnerability of insurance companies to insured individuals' physical risks
and the integration of the environmental component into banks' or
asset managers' business models. We also notice that the carbon beta
of the consumer staples and discretionary sectors have overall
increased because the climate change factor takes into account the
product carbon footprint.\smallskip

\begin{figure}[t]
\centering \caption{Dynamics of the average absolute carbon risk
$\left\vert \beta\right\vert_{\bmg,i}\left(t\right)$}
\label{fig:factor12d}
\includegraphics[width = \figurewidth, height = \figureheight]{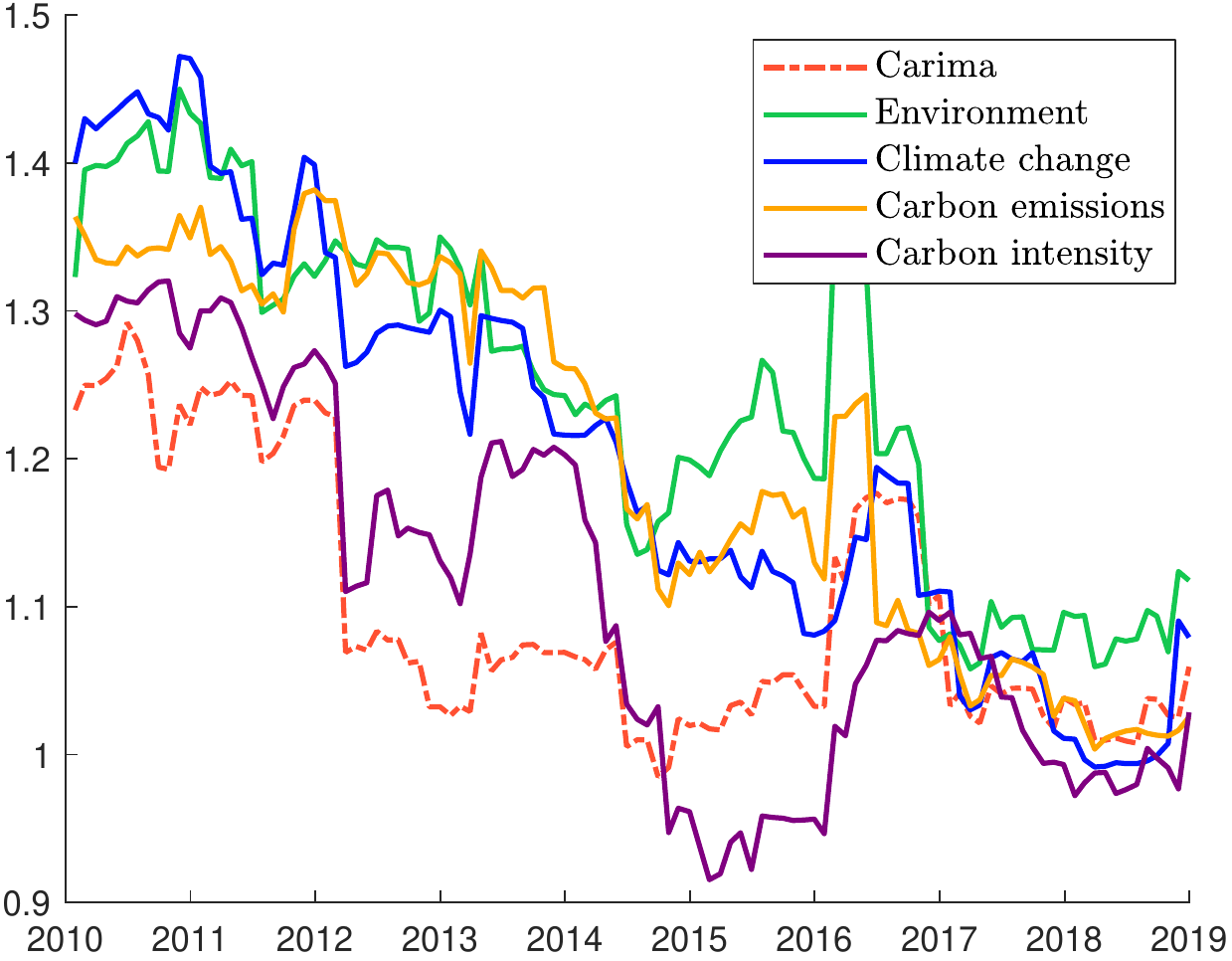}
\end{figure}

\subsubsection{Overview of the factors}

Numerous climate-related dimensions can be used to measure carbon
risk. Among the studied factors, some of them are more appropriate for assessing
stock price fluctuations. Figure \ref{fig:factor12d} provides the
dynamics of the average absolute carbon risk $\left\vert
\beta\right\vert_{\bmg,i}\left(t\right)$ for some climate change-related
dimensions and the Carima factor. In order to have comparable carbon betas,
each carbon factor $\BMG_{j}$ has been standardized so that its monthly
volatility is equal to $1\%$:
\begin{equation*}
\tilde{R}_{\bmg,j}\left(t\right) = \frac{R_{\bmg,j}\left(t\right)}{100 \times \sigma_{\bmg,j}\left(t\right)}
\end{equation*}
where $\tilde{R}_{\bmg,j}\left(t\right)$ is the return of the standardized
carbon factor and $\sigma_{\bmg,j}\left(t\right)$ is the conditional volatility
of the factor $\BMG_{j}$ estimated with a GARCH(1,1) model. In this case, the
factors' volatilities are the same at any time $t$. Overall, the trends of
absolute carbon risk in the environment, climate change and carbon emissions
factors are very similar. Therefore, we think that the carbon emissions
dimension is highly priced in, whereas adding other environmental variables
does not significantly increase the average absolute carbon beta\footnote{Our
results do not mean that the other environmental dimensions are barely priced
in. For a wide range of environmental variables, the associated average
absolute betas are often high but not so much as the carbon emissions
dimension. By taking into account several variables into one factor like the
climate change and environmental factors do, there is some collinearity between
variables and it is difficult, from a statistical point of view, to determine
which dimension is predominant in stock price variations.}. We can observe that
the Carima factor is less priced in than the three main climate change-related
factors. One reason might be the different methodologies between the factors.
Nevertheless, our results do not change much with a capitalization-weighted
scheme or by excluding financial firms. Therefore, we may wonder whether
including a large range of environmental variables is more informative. The
carbon intensity dimension is less priced in by the stock market at the
beginning of the study period in comparison with the three main climate
change-related dimensions, but this gap has been eliminated since
2017.\smallskip

According to Table \ref{tab:factor7} on page \pageref{tab:factor7},
carbon emissions is the better climate change-related factor to explain
stock price fluctuations, followed by the carbon emissions exposure and
carbon intensity factors. By splitting the period into two equal
subperiods\footnote{The first period starts at the beginning of 2010 and ends
in mid-2014 and the second period starts in mid-2014 and ends at the end of
2018.}, we obtain the following adjusted $\mathfrak{R}^{2}$ difference for
the CAPM model against the MKT+BMG model:
\begin{equation*}
\begin{tabular}{ccc}
\hline
 & \nth{1} period & \nth{2} period \\
 & 2010-2014 & 2014-2018 \\    \hline
Carima           & 1.16 & 2.21 \\
Carbon intensity & 1.43 & 2.53 \\
Carbon emissions & 2.18 & 2.39 \\
Climate change   & 1.98 & 1.83 \\
Environment      & 1.35 & 2.17 \\
\hline
\end{tabular}
\end{equation*}
The use of the BMG factors to explain stock price fluctuations is not
consistent over time. According to these results, the carbon intensity
dimension is currently the most important climate change-related axis in the two-factor
model. Concerning the environment and climate change dimensions, the results
and their correlations with the market factor lead to these factors being abandoned when it comes to
managing carbon risk in investment portfolios. Overall, the carbon
intensity and carbon emissions dimensions are the more interesting alternative
factors to the Carima factor.

\begin{remark}
Even though the climate change-related dimensions are less priced in by the stock
market over time, their integration -- except for the climate change dimension
-- in a CAPM model significantly increases the explanatory power during the
second period. This is because the CAPM model has an average adjusted
$\mathfrak{R}^{2}$ equal to $34.31\%$ during the \nth{1} period and $24.15\%$
during the \nth{2} period. Moreover, if we compute a single carbon factor
model, the adjusted $\mathfrak{R}^{2}$ is lower during the \nth{2} period.
Therefore, our results are coherent with the observation that the carbon risk
is less priced in today than before.
\end{remark}

\section{Managing carbon risk}

In what follows, we consider how to manage carbon risk in an
investment portfolio. Three methods are used: the minimum variance strategy,
the enhanced index portfolio and factor investing. For each method, we use an
easily understood example and then we apply the method to the MSCI World index.
In this last case, we take into account the dynamic betas of the MKT+BMG model
estimated by the Kalman filter, implying that we use the beta coefficients and
the weights of the MSCI World index at the end of December 2018.\smallskip

In this section, it is important to keep in mind the distinction between
absolute and relative carbon risks. In the first case, the underlying idea is
to have a neutral exposure to the BMG factor. In other words, we search the
closest carbon exposure to zero. In the second case, the objective is to have a
negative exposure to carbon risk. These two approaches lead us to consider
different objective functions or constraints of the portfolio optimization
program.

\subsection{Application to the minimum variance portfolio}

We consider the global minimum variance (GMV) portfolio, which corresponds to
this optimization program:
\begin{eqnarray}
x^{\star } &=&\arg \min \frac{1}{2}x^{\top }\Sigma x \label{eq:gmv1} \\
&\text{s.t.}&\mathbf{1}_n^{\top }x=1 \notag
\end{eqnarray}%
where $x$ is the vector of portfolio weights and $\Sigma$ is the covariance
matrix of stock returns. The solution is given by the well-known formula:
\begin{equation}
x^{\star }=\frac{\Sigma ^{-1}\mathbf{1}_n}{\mathbf{1}_n^{\top }\Sigma ^{-1}%
\mathbf{1}_n}  \label{eq:gmv2}
\end{equation}
Problem (\ref{eq:gmv1}) can be extended by considering other constraints:%
\begin{eqnarray}
x^{\star } &=&\arg \min \frac{1}{2}x^{\top }\Sigma x \label{eq:mv1} \\
&\text{s.t.}&\left\{
\begin{array}{l}
\mathbf{1}^{\top }x=1 \\
x\in \Omega
\end{array}%
\right.   \notag
\end{eqnarray}%
For instance, the most famous formulation is the long-only optimization
problem where $\Omega =\left[ 0,1\right] ^{n}$ \citep{Jagannathan-2003}.

\subsubsection{The CAPM risk factor model}

In the capital asset pricing model, we recall that:
\begin{equation}
R_{i}\left(t\right) = \alpha _{i} + \beta_{\mkt,i} R_{\mkt}\left(t\right) + \varepsilon _{i}\left(t\right) \label{eq:mv-model1}
\end{equation}%
where $R_{i}\left(t\right)$ is the return of asset $i$,
$R_{\mkt}\left(t\right)$ is the return of the market factor and $\varepsilon
_{i}\left(t\right)$ is the idiosyncratic risk. It follows that the covariance
matrix $\Sigma $ can be decomposed as:
\begin{equation*}
\Sigma =\beta_{\mkt} \beta_{\mkt}^{\top }\sigma _{\mkt}^{2}+D
\end{equation*}%
where $\beta_{\mkt} =\left( \beta_{\mkt,1},\ldots ,\beta_{\mkt,n}\right) $ is
the vector of betas, $\sigma _{\mkt}^{2}$ is the variance of the market
portfolio and $D=\func{diag}\left( \tilde{\sigma}_{1}^{2},\ldots
,\tilde{\sigma}_{n}^{2}\right) $ is the diagonal matrix of specific
variances. Using the Sherman-Morrison-Woodbury formula\footnote{This is provided
in Appendix \ref{appendix:smw-formula} on page
\pageref{appendix:smw-formula}. The expression of $\Sigma ^{-1}$ is obtained
with $A=D$ and $u=v=\sigma _{\mkt}\beta_{\mkt}$.}, we deduce that the inverse
of the covariance matrix is:
\begin{equation*}
\Sigma ^{-1}=D^{-1}-\frac{\sigma_{\mkt}^{2}}{1+\sigma _{\mkt}^{2}\varphi\left(\tilde{\beta}_{\mkt},
\beta_{\mkt}\right) }\tilde{\beta}_{\mkt}\tilde{\beta}_{\mkt}^{\top }
\end{equation*}%
where $\tilde{\beta}_{\mkt,i}=\beta_{\mkt,i}/\tilde{\sigma}_{i}^{2}$ and
$\varphi\left(\tilde{\beta}_{\mkt},\beta_{\mkt}\right)
=\tilde{\beta}_{\mkt}^{\top }\beta_{\mkt} $. Solution (\ref{eq:gmv2})
becomes:
\begin{equation*}
x^{\star }=\sigma ^{2}\left( x^{\star }\right) \left( D^{-1}\mathbf{1}_n-\frac{%
\sigma _{\mkt}^{2}}{1+\sigma _{\mkt}^{2}\varphi\left(\tilde{\beta}_{\mkt},\beta_{\mkt}\right) }
\tilde{\beta}_{\mkt}\tilde{\beta}_{\mkt}^{\top } \mathbf{1}_n\right)
\end{equation*}%
Using this new expression, \citet{Scherer-2011} showed that:
\begin{equation}
x_{i}^{\star }=\frac{\sigma ^{2}\left( x^{\star }\right) }{\tilde{\sigma}%
_{i}^{2}}\left( 1-\frac{\beta_{\mkt,i}}{\beta_{\mkt}^{\star }}\right)   \label{eq:gmv3}
\end{equation}%
where:%
\begin{equation}
\beta_{\mkt}^{\star }=\frac{1+\sigma _{\mkt}^{2}\varphi\left(\tilde{\beta}_{\mkt},\beta_{\mkt}\right) }%
{\sigma _{\mkt}^{2}\tilde{\beta}_{\mkt}^{\top }\mathbf{1}_n}  \label{eq:gmv4}
\end{equation}%
If we consider this formula, we note that the minimum variance portfolio is
exposed to stocks with low volatility and low beta. More precisely, if asset
$i$ has a beta $\beta_{\mkt,i}$ smaller than $\beta_{\mkt}^{\star }$, the
weight of this asset is positive ($x_{i}^{\star }>0$). If
$\beta_{\mkt,i}>\beta_{\mkt} ^{\star }$, then $x_{i}^{\star }<0$.
\citet{Clarke-2011} extended Formula (\ref{eq:gmv3}) to the long-only case
with the threshold $\beta_{\mkt} ^{\star }$ defined as follows:
\begin{equation}
\beta_{\mkt} ^{\star }=\frac{1+\sigma _{\mkt}^{2}\sum_{\beta_{\mkt,i}<\beta_{\mkt} ^{\star }}%
\tilde{\beta}_{\mkt,i}\beta_{\mkt,i}}{\sigma _{\mkt}^{2}\sum_{\beta_{\mkt,i}<\beta_{\mkt} ^{\star }}%
\tilde{\beta}_{\mkt,i}}  \label{eq:mv1}
\end{equation}%
In this case, if $\beta_{\mkt,i}>\beta_{\mkt} ^{\star }$, $x_{i}^{\star }=0$.

\subsubsection{Including the absolute carbon risk}

We consider an extension of the CAPM by including the BMG risk factor:
\begin{equation}
R_{i}\left(t\right)=\alpha _{i}+\beta_{\mkt,i}R_{\mkt}\left(t\right)+\beta_{\bmg,i}R_{\bmg}\left(t\right)+\varepsilon_{i}\left(t\right)
\label{eq:mv-model2}
\end{equation}%
where $R_{\bmg}\left(t\right)$ is the return of the BMG factor and
$\beta_{\bmg,i}$ is the BMG sensitivity (or the carbon beta) of stock $i$.
Moreover, we assume that $R_{\mkt}\left(t\right)$ and $R_{\bmg}\left(t\right)$
are uncorrelated. It follows that the expression of the covariance matrix
becomes:
\begin{equation*}
\Sigma =\beta_{\mkt} \beta_{\mkt} ^{\top }\sigma _{\mkt}^{2}+\beta_{\bmg} \beta_{\bmg}^{\top }\sigma
_{\bmg}^{2}+D
\end{equation*}%
In Appendix \ref{appendix:gmv-bmg} on page \pageref{appendix:gmv-bmg}, we
show that the GMV portfolio is defined as:
\begin{equation}
x_{i}^{\star }=\frac{\sigma ^{2}\left( x^{\star }\right) }{\tilde{\sigma}%
_{i}^{2}}\left( 1-\frac{\beta_{\mkt,i}}{\beta_{\mkt} ^{\star }}-\frac{\beta_{\bmg,i}}{%
\beta_{\bmg}^{\star }}\right)   \label{eq:gmv5}
\end{equation}%
where $\beta_{\mkt}^{\star }$ and $\beta_{\bmg}^{\star }$ are two threshold
values given by Equations (\ref{eq:app-beta-star}) and
(\ref{eq:app-gamma-star}) on page \pageref{eq:app-beta-star}. In the case of
long-only portfolios, we obtain a similar formula:
\begin{equation}
x_{i}^{\star }=\left\{
\begin{array}{ll}
\dfrac{\sigma ^{2}\left( x^{\star }\right) }{\tilde{\sigma}_{i}^{2}}\left( 1-%
\dfrac{\beta_{\mkt,i}}{\beta_{\mkt}^{\star }}-\dfrac{\beta _{\bmg,i}}{\beta_{\bmg}^{\star }}%
\right)  & \text{if }\dfrac{\beta_{\mkt,i}}{\beta_{\mkt} ^{\star }}+\dfrac{\beta _{\bmg,i}}{%
\beta_{\bmg} ^{\star }}\leq 1 \\
0 & \text{otherwise}%
\end{array}%
\right. \label{eq:mv2}
\end{equation}%
but the expressions of the thresholds\footnote{They are given by Equations
(\ref{eq:app-beta-star2}) and (\ref{eq:app-gamma-star2}) on page
\pageref{eq:app-beta-star2}.} $\beta_{\mkt} ^{\star }$ and $\beta_{\bmg}
^{\star } $ are different from those obtained in the GMV case.\smallskip

Contrary to the single-factor model, the impact of sensitivities is more
complex in the two-factor model. Indeed, we know that
$\bar{\beta}_{\mkt}\approx 1$ and $\bar{\beta}_{\bmg}\approx 0$. It follows
that $\beta_{\mkt}^{\star }$ is positive, but $\beta_{\bmg}^{\star }$ may be
positive or negative. We deduce that the ratio
$\dfrac{\beta_{\mkt,i}}{\beta_{\mkt} ^{\star }}$ is an increasing function of
$\beta_{\mkt,i} $. Therefore, the MV portfolio selects assets that present a
low MKT beta value. For the BMG factor, the impact of $\beta_{\bmg,i}$ is more
complex. Let us first compute the volatility of the asset $i$. We have:
\begin{equation*}
\sigma _{i}^{2}=\beta_{\mkt,i}^{2}\sigma _{\mkt}^{2}+\beta_{\bmg,i}^{2}\sigma
_{\bmg}^{2}+\tilde{\sigma}_{i}^{2}
\end{equation*}%
Selecting low volatility assets is then equivalent to considering assets with a
low absolute value $\left\vert \beta _{\bmg,i}\right\vert $. If we calculate
the correlation between assets $i$ and $j$, we obtain:
\begin{equation*}
\rho _{i,j}=\frac{\beta_{\mkt,i}\beta _{\mkt,j}\sigma _{\mkt}^{2}+\beta_{\bmg,i}\beta
_{\bmg,j}\sigma _{\bmg}^{2}}{\sigma _{i}\sigma _{j}}
\end{equation*}%
In practical cases, the cross product $\beta_{\mkt,i}\beta_{\mkt,j}$ is generally
positive, whereas the cross product $\beta_{\bmg,i}\beta_{\bmg,j}$ is positive or
negative. In this context, diversifying a portfolio consists in selecting
assets with low values of $\beta_{\mkt,i}\beta _{\mkt,j}$. We then observe
consistency between low volatility and low correlated assets if we consider
market beta contributions. In terms of BMG sensitivities, diversifying a portfolio
consists in selecting assets with high negative values of $\beta_{\bmg,i}\beta
_{\bmg,j}$. Therefore, we have to choose assets, with high absolute values
$\left\vert \beta_{\bmg,i}\beta_{\bmg,j}\right\vert $, and opposite signs of
$\beta_{\bmg,i}$ and $\beta_{\bmg,j}$. We do not have consistency
between low volatility and low correlated assets if we consider BMG
contributions. This explains that the ratio $\dfrac{\beta
_{\bmg,i}}{\beta_{\bmg} ^{\star }}$ may be an increasing or decreasing function
of $\beta_{\bmg,i}$. The MV portfolio will then overweight assets with a
negative value of $\beta_{\bmg,i}$ only if $\beta_{\bmg}^{\star }$ is positive.
Otherwise, the MV portfolio may prefer assets with a positive sensitivity to
the BMG factor.

\begin{remark}
Let us denote by $x^{\star }\left( \beta_{\mkt},\beta_{\bmg}\right) $ the
minimum variance portfolio that depends on the parameters $\beta_{\mkt} $ and
$\beta_{\bmg}$. We have the following properties:
\begin{equation*}
\left\{
\begin{array}{l}
x^{\star }\left( \beta_{\mkt} ,-\beta_{\bmg} \right) =x^{\star }\left( \beta_{\mkt} ,\beta_{\bmg}\right)  \\
\beta_{\mkt}^{\star }\left(\beta_{\mkt} ,-\beta_{\bmg} \right) =\beta_{\mkt} ^{\star }\left( \beta_{\mkt},\beta_{\bmg} \right)  \\
\beta_{\bmg}^{\star }\left(\beta_{\mkt} ,-\beta_{\bmg} \right) =-\beta_{\bmg}^{\star }\left( \beta_{\mkt},\beta_{\bmg} \right)
\end{array}%
\right.
\end{equation*}%
Changing the BMG sensitivities by their opposite values does not change the
solution\footnote{This result holds for both the GMV portfolio and the
long-only MV portfolio.}.
\end{remark}

\subsubsection{Some examples}

We consider an example\label{label:example-mv} given in \citet[Example 24,
page 168]{Roncalli-2013}. The investment universe is made up of five assets.
Their market beta is respectively equal to $0.9$, $0.8$, $1.2$, $0.7$ and $1.3$
whereas their specific volatility is $4\%$, $12\%$, $5\%$, $8\%$ and $5\%$.
The market portfolio volatility is equal to $25\%$. Using these figures, we
have computed the composition of the minimum variance portfolio. The results
are reported in Table \ref{tab:mv1}. The fourth and fifth columns contain the
weights (in \%) of the unconstrained and long-only minimum variance
portfolios. In the case of the unconstrained portfolio (GMV), we have
$\beta_{\mkt}^{\star } = 1.0972$. We deduce then that long exposures concern
the first, second and fourth assets whereas the short exposures concern the
third and fifth assets. For the long-only portfolio (MV), we obtain
$\beta_{\mkt}^{\star } = 0.8307$. This implies that only the second and
fourth assets are represented in the long-only minimum variance
portfolio.\smallskip

\begin{table}[tbph]
\centering
\caption{Composition of the MV portfolio (parameter set \#1)}
\label{tab:mv1}
\begin{tabular}{|c|cc|rr:rr|}
\hline
\multirow{2}{*}{Asset} & \multirow{2}{*}{$\beta_{\mkt,i}$} & \multirow{2}{*}{$\beta_{\bmg,i}$} &
\multicolumn{2}{c:}{CAPM} & \multicolumn{2}{c|}{MKT+BMG}        \\
    &        &                 & GMV      & MV      & GMV      & MV      \\ \hline
$1$ & $0.90$ &         $-0.50$ & $147.33$ &  $0.00$ & $166.55$ & $33.54$ \\
$2$ & $0.80$ & ${\TsVIII}0.70$ &  $24.67$ &  $9.45$ &  $21.37$ &  $1.46$ \\
$3$ & $1.20$ & ${\TsVIII}0.20$ & $-49.19$ &  $0.00$ & $-58.80$ &  $0.00$ \\
$4$ & $0.70$ & ${\TsVIII}0.90$ &  $74.20$ & $90.55$ &  $65.06$ & $64.99$ \\
$5$ & $1.30$ &         $-0.30$ & $-97.01$ &  $0.00$ & $-94.18$ &  $0.00$ \\
\hline
\end{tabular}
\end{table}

We now consider the impact of the BMG factor. We assume that the BMG
sensitivities are respectively equal to $-0.5$, $0.7$, $0.2$, $0.9$ and
$-0.3$, whereas the volatility of the BMG factor is set to $10\%$. In the
case of the GMV, the thresholds are equal to $\beta_{\mkt}^{\star} = 1.0906$ and
$\beta_{\bmg}^{\star} = 19.7724$. In this case, we obtain the same long and short
exposures with different magnitudes than previously. In the case of the
long-only portfolio, the thresholds become $\beta_{\mkt}^{\star} = 0.8667$ and
$\beta_{\bmg}^{\star} = 9.7394$. Compared to the CAPM solution, we observe that the
weights in the second and fourth assets are reduced, because they have a
positive BMG sensitivity. At the same time, the portfolio is exposed to the
first asset. The reason lies in the negative value of BMG sensitivity.
Indeed, we have:
\begin{equation*}
\frac{\beta_{\mkt,1}}{\beta_{\mkt}^{\star}}=\frac{0.90}{0.8307}>1
\end{equation*}
but:%
\begin{equation*}
\frac{\beta _{\mkt,1}}{\beta_{\mkt} ^{\star }}+\frac{\beta _{\bmg,1}}{\beta_{\bmg}^{\star }}=%
\frac{0.90}{0.8667}+\frac{-0.50}{9.7394}<1
\end{equation*}
Therefore, thanks to the BMG factor, the first asset passes the eligibility
test.\smallskip

\begin{table}[tbph]
\centering
\caption{Composition of the MV portfolio (parameter sets \#2 and \#3)}
\label{tab:mv2}
\begin{tabular}{|c|c|rrr:rrr|}
\hline
\multirow{2}{*}{Asset} & \multirow{2}{*}{$\beta_{\mkt,i}$} &
\multicolumn{3}{c:}{Parameter set \#2} & \multicolumn{3}{c|}{Parameter set \#3}                 \\
    &        & \multicolumn{1}{c}{$\beta_{\bmg,i}$} & \multicolumn{1}{c}{GMV} & \multicolumn{1}{c:}{MV} &
               \multicolumn{1}{c}{$\beta_{\bmg,i}$} & \multicolumn{1}{c}{GMV} & \multicolumn{1}{c|}{MV} \\ \hline
$1$ & $0.90$ &         $-1.50$ &  $105.46$ &   $0.00$ & ${\TsVIII}1.50$ &  $105.46$ &   $0.00$   \\
$2$ & $0.80$ &         $-0.50$ &   $27.88$ &  $19.48$ & ${\TsVIII}0.50$ &   $27.88$ &  $19.48$   \\
$3$ & $1.20$ & ${\TsVIII}3.00$ &   $40.19$ &  $13.61$ &         $-3.00$ &   $40.19$ &  $13.61$   \\
$4$ & $0.70$ &         $-1.20$ &   $76.77$ &  $66.91$ & ${\TsVIII}1.20$ &   $76.77$ &  $66.91$   \\
$5$ & $1.30$ &         $-0.90$ & $-150.30$ &   $0.00$ & ${\TsVIII}0.90$ & $-150.30$ &   $0.00$   \\
\hline
\end{tabular}
\end{table}

We now consider a variant of the previous example. We use the same parameter values,
but different BMG sensitivity values. In the case of the
parameter set \#2, they are equal to $-1.5$, $-0.5$, $3.0$, $-1.2$ and $-0.9$.
For the long/short MV portfolio, we obtain $\beta_{\mkt}^{\star} = 1.0982$ and
$\beta_{\bmg}^{\star} = -19.4470$. For the long-only MV portfolio, the
thresholds become $\beta_{\mkt}^{\star} = 0.9070$ and $\beta_{\bmg}^{\star} =
-9.0718$. We notice that the BMG threshold is negative, whereas it was positive
in the case of the parameter set \#1. Moreover, we observe a positive exposure
on the third asset even though it has a high positive market beta. The reason
is the high magnitude of the BMG sensitivity and the negative correlation with
the BMG sensitivities of the other assets. In the case of the parameter set
\#3, we have only changed the sign of the BMG sensitivities. We obtain the same
composition of the MV portfolio and the same value of $\beta_{\mkt}^{\star}$,
but the value of $\beta_{\bmg}^{\star}$ is different. Indeed, we obtain
$\beta_{\bmg}^{\star} = +19.4470$ for the long/short portfolio and
$\beta_{\bmg}^{\star} = +9.0718$ for the long-only portfolio.\smallskip

\begin{figure}[tbph]
\centering
\caption{Weights of the MV portfolio}
\label{fig:gmv5}
\figureskip
\includegraphics[width = \figurewidth, height = \figureheight]{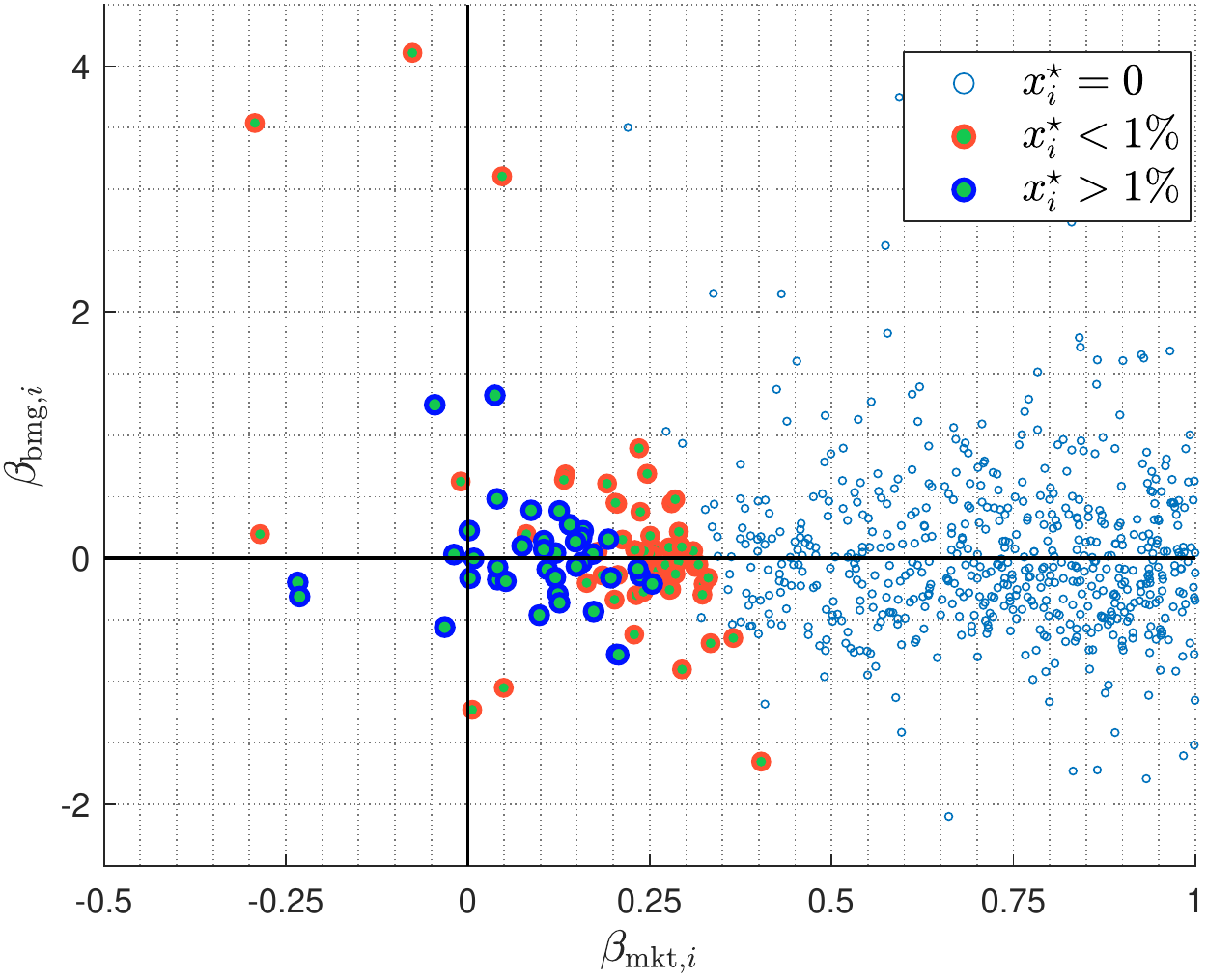}
\end{figure}

We apply the previous framework to the MSCI World index at December 2018. We
have already estimated the MKT+BMG model\footnote{We have
reported the scatter plot of MKT and BMG sensitivities in Figure
\ref{fig:carima4} on page \pageref{fig:carima4}. We observe a low positive
correlation between $\beta_{\mkt,i}$ and $\beta_{\bmg,i}$.} in Section
\ref{section:carima-dynamic} on page \pageref{section:carima-dynamic}. By computing the
long-only MV portfolio, we obtain $\beta^{\star }_{\mkt} = 0.3465$ and $\beta
^{\star }_{\bmg} = 5.3278$. In Figure \ref{fig:gmv5}, we indicate the assets
that make up the MV portfolio with respect to their beta values
$\beta_{\mkt,i}$ and $\beta_{\bmg,i}$. We verify that the most important axis
is the MKT beta. Indeed, the market risk of a stock determines whether the
stock is included in the MV portfolio or not whereas the carbon risk adjusts the
weights of the asset. As we can see, the portfolio overweights assets whose MKT
and BMG sensitivities are both close to zero. This solution is
satisfactory if the original motivation is to reduce the portfolio's absolute carbon risk,
but it is not satisfactory if the objective is to manage the
portfolio's relative carbon risk.

\subsubsection{New formulation of the minimum variance portfolio with relative carbon risk}

In order to circumvent the previous drawback, we can directly add a BMG
constraint in the optimization program:
\begin{eqnarray}
x^{\star } &=&\arg \min \frac{1}{2}x^{\top }\Sigma x  \label{eq:mv3} \\
&\text{s.t.}&\left\{
\begin{array}{l}
\mathbf{1}_{n}^{\top }x=1 \\
\beta _{\bmg}^{\top }x\leq \beta _{\bmg}^{+} \\
x\geq \mathbf{0}_{n}%
\end{array}%
\right.   \notag
\end{eqnarray}%
where $\beta _{\bmg}^{+}$ is the maximum tolerance of the investor
with respect to the relative BMG risk. In this case, the values of
$\beta _{\bmg,i}$ influence both the covariance matrix and the
optimization problem. In this case, it is not possible to obtain an
analytical solution. Nevertheless, we can always analyze the
solution using the framework developed by \citet{Jagannathan-2003}.
Introducing the BMG constraint is equivalent to applying a shrinkage of
the covariance matrix\footnote{The analysis of
\citet{Jagannathan-2003} only includes bound constraints. The
extension to other linear constraints can be found in
\citet{Roncalli-2013}.}:
\begin{equation}
\tilde{\Sigma}=\Sigma +\lambda _{\bmg}\left( \beta _{\bmg}\mathbf{1}%
_{n}^{\top }+\mathbf{1}_{n}\beta _{\bmg}^{\top }\right)   \label{eq:mv4}
\end{equation}%
where $\lambda _{\bmg}\geq 0$ is the Lagrange coefficient associated to the
inequality constraint $\beta _{\bmg}^{\top }x\leq \beta _{\bmg}^{+}$. We deduce
that the shrinkage matrix is equal to:
\begin{equation*}
\tilde{\Sigma}=\sigma _{\mkt}^{2}\left( \beta _{\mkt}\beta _{\mkt}^{\top
}-\left( \frac{\lambda _{\bmg}}{\sigma _{\mkt}\sigma _{\bmg}}\right)
^{2}\right) +\sigma _{\bmg}^{2}\left( \dot{\beta}_{\bmg}\dot{%
\beta}_{\bmg}^{\top }\right) +D
\end{equation*}%
where $\dot{\beta}_{\bmg}=\beta _{\bmg}+\dfrac{\lambda _{\bmg}}{\sigma _{\bmg}^{2}}%
\mathbf{1}_{n}$. By imposing that the MV portfolio has a carbon beta lower than
$\beta _{\bmg}^{+}$, we implicitly introduce two effects:

\begin{enumerate}
\item first, we shift the BMG sensitivities by a positive scalar $\dfrac{
    \lambda _{\bmg}}{\sigma _{\bmg}^{2}}$;

\item second, we reduce the MKT covariance matrix by a uniform parallel shift,
    because of the term $\dfrac{\lambda _{\bmg}}{\sigma _{\mkt}\sigma
    _{\bmg}}$.
\end{enumerate}
Therefore, the BMG constraint $\beta _{\bmg}^{\top }x\leq \beta _{\bmg}^{+}$
can be interpreted as an active view.\smallskip

In order to illustrate the BMG constraint, we consider the previous examples
and impose that the BMG sensitivity of the MV portfolio cannot be positive.
Results are reported in Table \ref{tab:gmv7}. We notice that the invariance
of the BMG sign is broken. For instance, we do not obtain the same solution
between parameter sets \#2 and \#3. Moreover, we verify that the constrained MV
portfolio promotes negative BMG sensitivities\footnote{We recall that this is
not the case of the unconstrained MV portfolio, which better promotes weak BMG
sensitivities.}.\smallskip

\begin{table}[tbph]
\centering
\caption{Composition of the constrained MV portfolio ($\beta_{\bmg}^{+} = 0$)}
\label{tab:gmv7}
\begin{tabular}{|c|c|rr:rr:rr|}
\hline
\multirow{2}{*}{Asset} & \multirow{2}{*}{$\beta_{\mkt,i}$} &
\multicolumn{2}{c:}{Parameter set \#1} & \multicolumn{2}{c:}{Parameter set \#2} &
\multicolumn{2}{c|}{Parameter set \#3} \\
    &        & \multicolumn{1}{c}{$\beta_{\bmg,i}$} & \multicolumn{1}{c:}{MV} &
               \multicolumn{1}{c}{$\beta_{\bmg,i}$} & \multicolumn{1}{c:}{MV} &
               \multicolumn{1}{c}{$\beta_{\bmg,i}$} & \multicolumn{1}{c|}{MV} \\ \hline
$1$ & $0.90$ &         $-0.50$ & $64.29$ &         $-1.50$ &  $0.00$  & ${\TsVIII}1.50$ &  $0.00$   \\
$2$ & $0.80$ & ${\TsVIII}0.70$ &  $0.00$ &         $-0.50$ & $19.48$  & ${\TsVIII}0.50$ & $16.11$   \\
$3$ & $1.20$ & ${\TsVIII}0.20$ &  $0.00$ & ${\TsVIII}3.00$ & $13.61$  &         $-3.00$ & $25.89$   \\
$4$ & $0.70$ & ${\TsVIII}0.90$ & $35.71$ &         $-1.20$ & $66.91$  & ${\TsVIII}1.20$ & $58.00$   \\
$5$ & $1.30$ &         $-0.30$ &  $0.00$ &         $-0.90$ &  $0.00$  & ${\TsVIII}0.90$ &  $0.00$   \\
\hline
\multicolumn{2}{|c|}{$\lambda _{\bmg}$} & \multicolumn{2}{c:}{$65$ bps} &
\multicolumn{2}{c:}{$0$} & \multicolumn{2}{c|}{$56$ bps} \\
\hline
\end{tabular}
\end{table}

\begin{figure}[tbph]
\centering
\caption{Weights of the constrained MV portfolio ($\beta_{\bmg}^{+} = -0.50$)}
\label{fig:gmv6}
\figureskip
\includegraphics[width = \figurewidth, height = \figureheight]{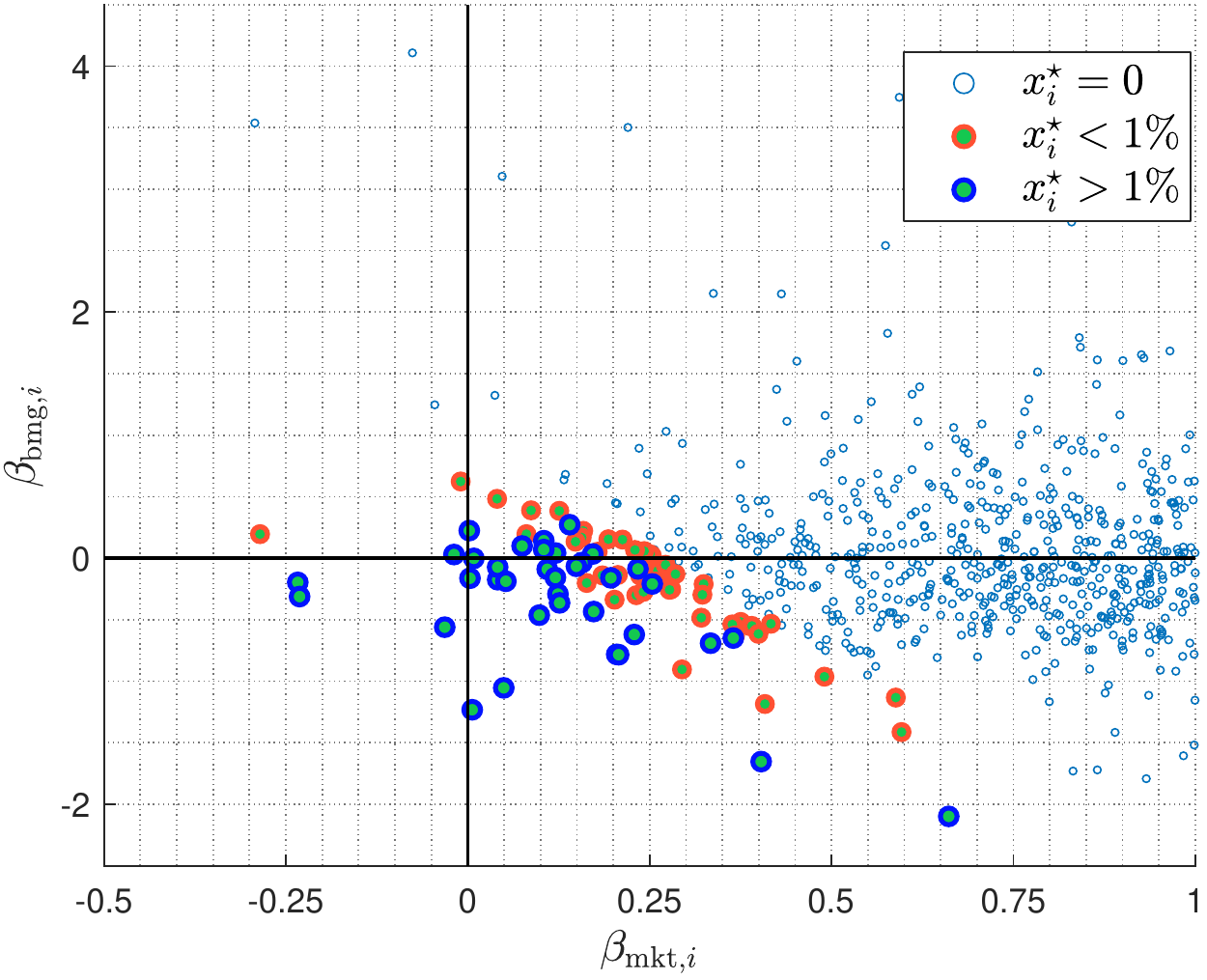}
\end{figure}

\begin{figure}[tbph]
\centering
\caption{Relationship between $\beta_{\bmg}^{+}$ and $\lambda_{\bmg}$}
\label{fig:gmv8}
\figureskip
\includegraphics[width = \figurewidth, height = \figureheight]{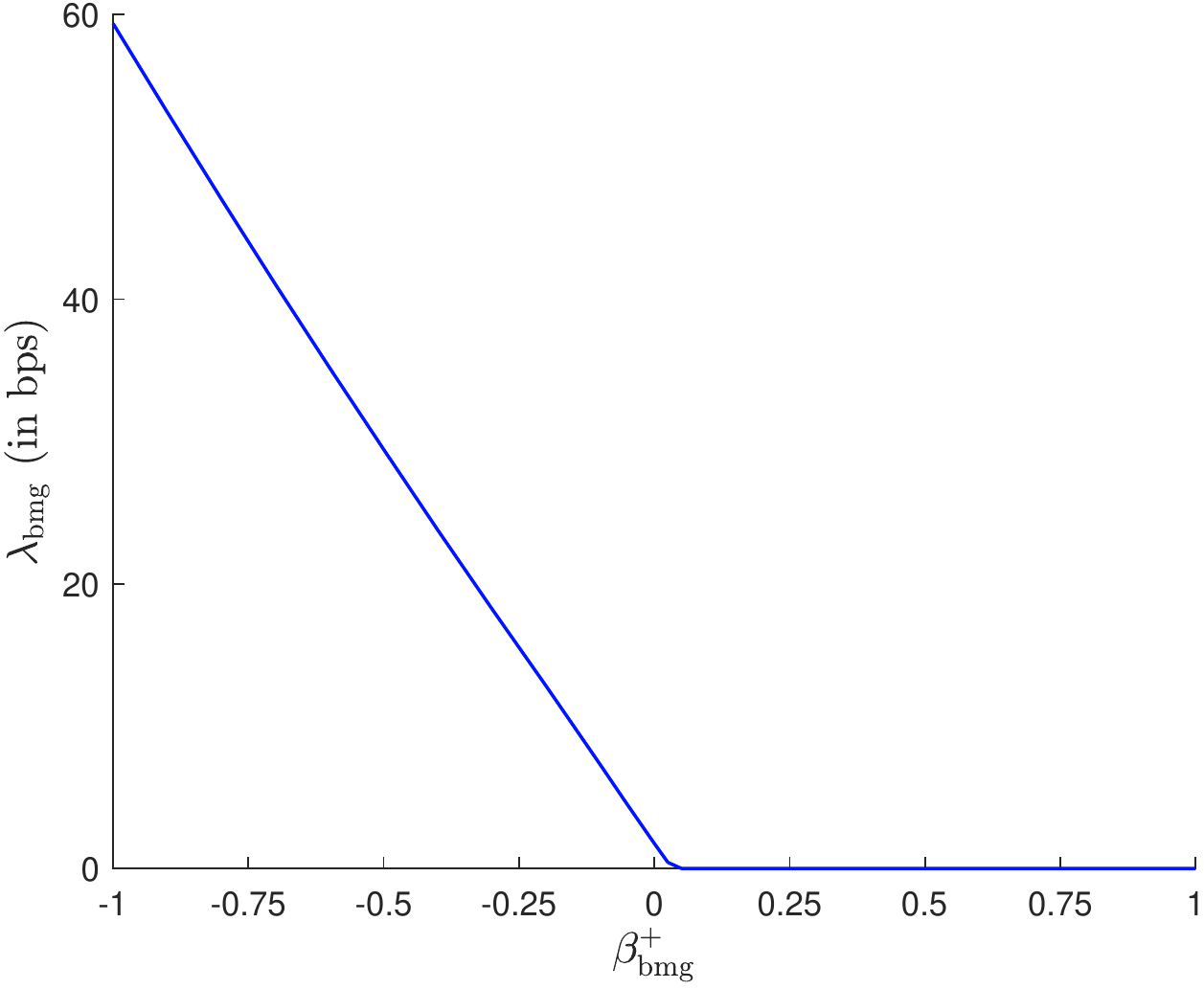}
\end{figure}

We consider again the MSCI World index universe at December
2018. Since asset managers are mostly interested in long-only portfolios, we
only report the long-only MV portfolio. If we would like to impose that the
BMG sensitivity is lower than $-0.5$, we obtain results given in Figure
\ref{fig:gmv6}. The comparison with the previous results (Figure
\ref{fig:gmv5}) shows that the MV portfolio tends to select assets with both
a low MKT beta and a negative BMG sensitivity. Moreover, large weights are
associated with large negative values of $\beta _{\bmg,i}$ on average. These
results can be explained because the Lagrange coefficient $\lambda _{\bmg}$
is equal to 29 bps. Of course, the magnitude of the shrinkage depends on the
value of $\beta_{\bmg}^{+}$. The lower the BMG constraint, the higher the
Lagrange coefficient. For instance, we report the relationship between
$\beta_{\bmg}^{+}$ and $\lambda_{\bmg}$ in Figure \ref{fig:gmv8}. This
trade-off is not free since it will also impact the volatility of the MV
portfolio.\smallskip

\begin{figure}[tbph]
\centering \caption{$\WACI$ of the constrained MV portfolio}
\label{fig:gmv9}
\figureskip
\includegraphics[width = \figurewidth, height = \figureheight]{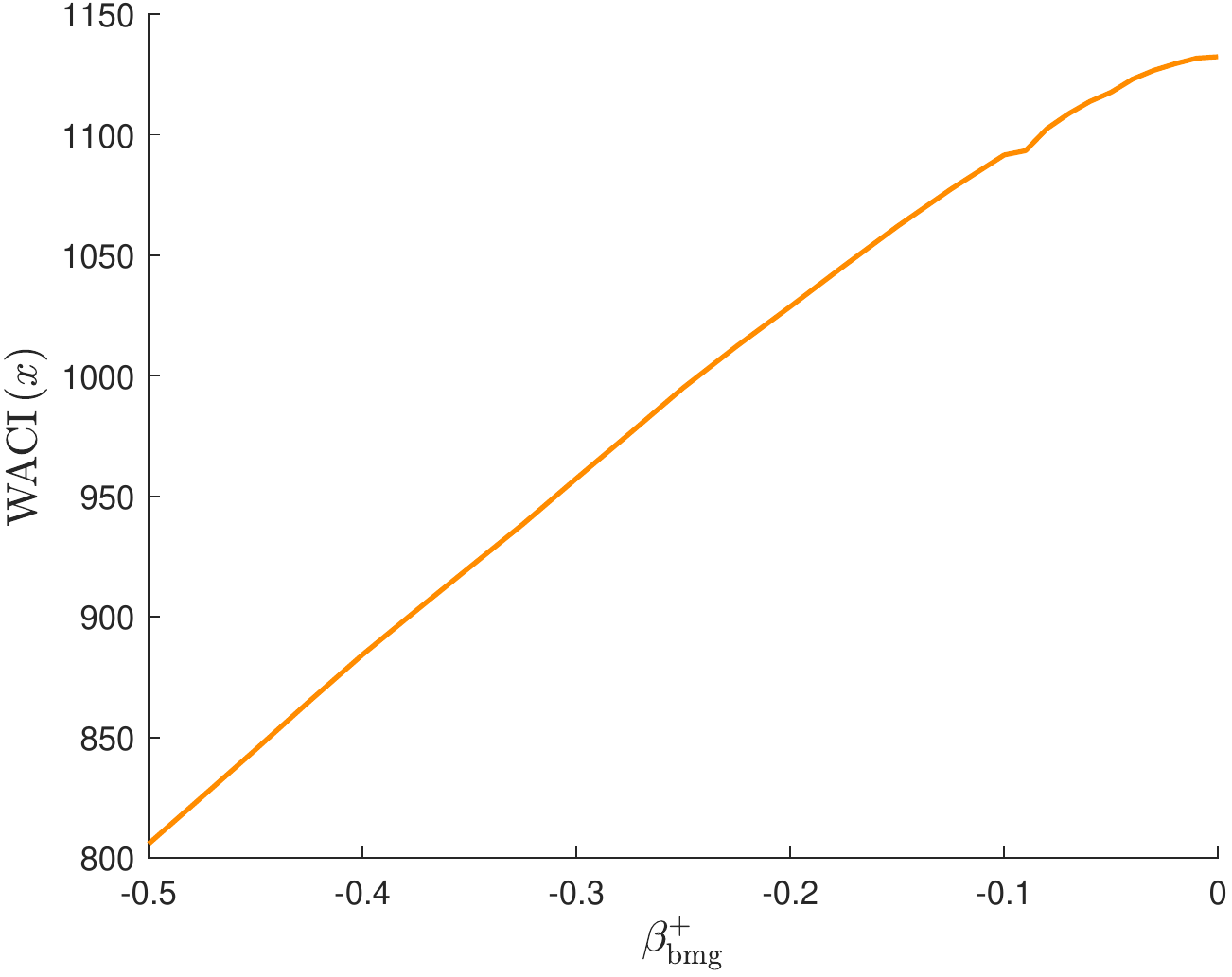}
\end{figure}

For some time now, an important preoccupation of asset managers and asset
owners is about the GHG emissions associated with their investment portfolios.
There are many portfolio carbon footprint metrics but most of them are
not consistent over time because of the equity ownership approach. To overcome
this issue, we use the weighted average carbon intensity (or WACI) recommended
by the Task Force on Climate-related Financial Disclosures \citep{TCFD-2017}:
\begin{equation*}
\WACI\left( x\right) = \sum_{i=1}^{n} x_{i} \cdot \intensity_{i}
\end{equation*}
where $\intensity_{i}$ is the carbon intensity of issuer $i$. We define the
carbon intensity $\intensity$ as the issuer's direct and first-tier indirect
GHG emissions\footnote{The direct emissions correspond to the scope 1 emissions
and the first-tier indirect emissions correspond to the GHG emissions of the
firm's direct suppliers (scope 2 emissions + some upstream scope 3 emissions).}
divided by the revenue. This measure is expressed in tons CO$_{2}$e
per million dollars in revenue. In Figure \ref{fig:gmv9}, we have reported the
relationship between $\WACI$ and $\beta_{\bmg}^{+}$ of the MV portfolio. We
remark that the lower the relative carbon risk threshold, the lower the
portfolio exposure to carbon-intensive companies. Nevertheless, the carbon intensities
related to these different portfolios are very high in comparison
with the equally-weighted portfolio whose $\WACI$ is approximatively equal to
$315$. The reason is that carbon footprint metrics cannot be necessarily
interpreted as risk metrics. Some firms in the energy sector whose stock prices
follow a singular pathway have both non-significant market and carbon betas,
whereas they have a high carbon intensity. These unusual firms, whose
carbon risk is captured in the idiosyncratic risk, contribute to dramatically
increasing the $\WACI$ of the MV portfolio. In order to circumvent
this issue, we can add a constraint to the MV problem:
\begin{eqnarray}
x^{\star } &=&\arg \min \frac{1}{2}x^{\top }\Sigma x  \label{eq:mv5} \\
&\text{s.t.}&\left\{
\begin{array}{ll}
\mathbf{1}_{n}^{\top }x=1 & \\
\beta _{\bmg}^{\top }x\leq \beta _{\bmg}^{+} & \\
x_{i} = 0 & \text{if } \intensity_{i} > \intensity^{+} \\
x\geq \mathbf{0}_{n} &
\end{array}
\right.   \notag
\end{eqnarray}
where $\intensity^{+}$ is a maximum carbon intensity threshold. We have
reported in Figure \ref{fig:gmv11} on page \pageref{fig:gmv11} the relationship
between $\WACI$ and $\beta_{\bmg}^{+}$ of the MV portfolio when the carbon
intensity threshold $\intensity^{+}$ is equal to the $\WACI$ of the EW
portfolio. In this case, we considerably reduce the $\WACI$ regardless of the value
of the portfolio carbon risk threshold $\beta_{\bmg}^{+}$. Even if we impose
$\beta_{\bmg}^{+} = 0$ and $\intensity^{+} = 2\,000$, we obtain a $\WACI$
around $225$, which is very low in comparison with the MV portfolio without
a carbon intensity threshold. Figure \ref{fig:gmv13} on page \pageref{fig:gmv13}
provides the tradeoff between the portfolio volatility $\sigma\left(x\right)$,
the carbon risk threshold $\beta_{\bmg}^{+}$ and the carbon intensity threshold
$\intensity^{+}$. This latter constraint has no substantial impact on portfolio
volatility. Therefore, it is possible to reduce the weighted average carbon
intensity without substantially increasing volatility.

\begin{remark}
In order to highlight the difference between a market measure of carbon risk
and a fundamental measure of carbon risk, we have reported in Figure
\ref{fig:factor20} on page \pageref{fig:factor20} the relationship between
$\intensity_{i}$ and $\beta_{\bmg,i}$. On average, the linear correlation is
equal to $17.5\%$ for the Carima factor. It slightly increases if we consider
the carbon intensity factor, but it remains lower than $30\%$.
\end{remark}

\subsection{Enhanced index portfolio}
\label{section:enhanced-index}

\subsubsection{Analysis of the optimization problem}

Enhanced index portfolios may be obtained by considering the
portfolio optimization method in the presence of a benchmark \citep{Roncalli-2013}.
For that, we define $b=\left( b_{1},\ldots ,b_{n}\right) $ and $x=\left(
x_{1},\ldots ,x_{n}\right) $ as the asset weights in the benchmark and the
portfolio. The tracking error between the active portfolio $x$ and its
benchmark $b$ is the difference between the portfolio's return and the
benchmark's return:
\begin{eqnarray*}
R\left( x\mid b\right)  &=&R\left( x\right) -R\left( b\right)  \\
&=&\left( x-b\right) ^{\top }R
\end{eqnarray*}%
where $R=\left( R_{1},\ldots ,R_{n}\right) $ is the vector of asset returns.
The volatility of the tracking error $R\left( x\mid b\right) $ corresponds to
the standard deviation of $R\left( x\right) -R\left( b\right) $:
\begin{equation*}
\sigma\left( x\mid b\right) =\sqrt{\left( x-b\right) ^{\top }\Sigma \left(
x-b\right)}
\end{equation*}%
The optimization problem of enhanced index portfolios consists in replacing
the portfolio's volatility with the portfolio's tracking error volatility in a
minimum variance framework and imposing long-only weights:
\begin{eqnarray}
x^{\star } &=&\arg \min \frac{1}{2}\left( x-b\right) ^{\top }\Sigma \left(
x-b\right)   \label{eq:te1} \\
&\text{s.t.}&\left\{
\begin{array}{l}
\mathbf{1}_{n}^{\top }x=1 \\
x\geq \mathbf{0}_{n} \\
x\in \Omega
\end{array}%
\right.   \notag
\end{eqnarray}%
If no other constraint is added ($\Omega =\mathbb{R}^{n}$), the optimal
solution $x^{\star }$ is the benchmark $b$. But this framework only makes
sense if we impose a second objective using the restriction $x\in
\Omega $. For instance, we can impose that the optimal portfolio has a carbon
beta less than a threshold as in the case of the minimum variance problem:
\begin{equation}
\Omega =\left\{ x\in \mathbb{R}^{n}:\beta _{\bmg}^{\top }x\leq \beta
_{\bmg}^{+}\right\}   \label{eq:te2}
\end{equation}%
Another approach consists in excluding the first $m$ stocks that present the
largest carbon risk beta:
\begin{equation}
\Omega =\left\{ x\in \mathbb{R}^{n}:x_{i}=0\text{ if }\beta _{\bmg%
,i}\geq \beta _{\bmg}^{\left( m,n\right) }\right\}   \label{eq:te3}
\end{equation}%
where $\beta _{\bmg}^{\left( m,n\right) }=\beta _{\bmg,n-m+1:n}$ is the
$\left( n-m+1\right) $\textit{-th} order statistic of $\left( \beta
_{\bmg,1},\ldots ,\beta _{\bmg,n}\right) $.\smallskip

These two approaches are similar to the ones proposed by \citet{Andersson-2016}.
The difference comes from the fact that we use a market measure to estimate
the carbon risk, whereas \citet{Andersson-2016} measured the carbon
risk directly using the carbon intensity. The two methods have their own advantages and
drawbacks. It is obvious that the method of \citet{Andersson-2016} is more
objective than our method, because it directly uses the carbon footprint of the
issuer. However, it is difficult to know if the stock price is sensitive to
this carbon footprint measure, especially since there are
several carbon intensity measures\footnote{We generally distinguish scope 1, 2
and 3 carbon emissions. According to the \citet{GHG-Protocol-2013}, scope 1
corresponds to all direct emissions from the firm's activities, scope 2
includes indirect emissions from electricity purchased and used by the firm,
whereas scope 3 measures all other indirect emissions from the firm's activities.}.
We understand that the carbon footprint is an ecological environment
risk for planet Earth. But it is less obvious that it corresponds to the
carbon market risk which is priced in by the stock market at the security level.
If this were the case, it would mean that two corporate firms with the same
carbon footprint present the same carbon beta, regardless of the firms' other
characteristics. Our method is less objective since the carbon
risk is estimated through the dynamics of stock prices, and also depends on the
methodology to build the carbon risk factor. Nevertheless, it is more relevant
from a financial point of view, because we consider the carbon risk directly
priced in by the stock market. In a sense, the first method has
a longer-term horizon, whereas the second method is short-term by construction.

\begin{remark}
We can replace the absolute threshold $\beta_{\bmg}^{+}$ with a relative
threshold. Indeed, imposing a reduction of the carbon risk with respect to
the benchmark, e.g. $\beta_{\bmg}^{\top }\left(x-b\right) \leq
-\Delta_{\bmg}$, is equivalent to using an absolute threshold, e.g.
$\beta_{\bmg}^{+} = \beta _{\bmg}^{\top }b - \Delta_{\bmg}$ where
$\Delta_{\bmg}$ corresponds to the relative or absolute difference between
the benchmark's carbon risk and the threshold value of relative carbon
risk.
\end{remark}

The mathematical analysis of the optimization problem (\ref{eq:te1}) with the
constraint (\ref{eq:te2}) is given in Appendix \ref{appendix:te-bmg} on page
\pageref{appendix:te-bmg}. We show that $\Delta _{i}=x_{i}^{\star }-b_{i}$ is
a decreasing function of the scaled BMG sensitivity $\breve{\beta}_{\bmg,i}$,
which is equal to $\left( \Sigma ^{-1}\beta _{\bmg}\right) _{i}$. To
illustrate this property, we consider the example used on page
\pageref{label:example-mv} (parameter set \#1) and assume that the benchmark
is the equally-weighted (EW) portfolio. Moreover, we impose that the relative carbon
risk of the optimized portfolio is less than zero -- $\beta _{\bmg}^{+} =
0$. Results are reported in Table \ref{tab:te3}. We verify that underweights
and overweights depend on the sign of $\breve{\beta}_{\bmg,i}$. If
$\breve{\beta}_{\bmg,i}$ is negative, $\Delta_i$ is positive and the asset is
overweighted with respect to the benchmark. Otherwise, the asset is
underweighted if $\breve{\beta}_{\bmg,i}$ is positive.\smallskip

\begin{table}[tbph]
\centering
\caption{Enhanced index portfolio}
\label{tab:te3}
\begin{tabular}{cccccc}
\hline
Asset & $b_i$ & $x_i^{\star}$ & $\Delta_i$ & $\beta_{\bmg,i}$ & $\breve{\beta}_{\bmg,i}$ \\ \hline
$1$ & $20.00\%$ & $36.77$ & ${\TsIII}16.77\%$ &         $-0.5$ &         $-56.38$  \\
$2$ & $20.00\%$ & $17.12$ &         $-2.88\%$ & ${\TsVIII}0.7$ & ${\TsVIII}12.22$  \\
$3$ & $20.00\%$ & $11.61$ &         $-8.39\%$ & ${\TsVIII}0.2$ & ${\TsVIII}29.46$  \\
$4$ & $20.00\%$ & $12.03$ &         $-7.97\%$ & ${\TsVIII}0.9$ & ${\TsVIII}34.10$  \\
$5$ & $20.00\%$ & $22.48$ & ${\TsVIII}2.48\%$ &         $-0.3$ &         $-14.33$  \\
\hline
\end{tabular}
\end{table}

\subsubsection{Application to the MSCI World index}

\begin{figure}[p]
\centering
\caption{Relationship between $\beta _{\bmg,i}$ and $\Delta _{i}=x_{i}^{\star }-b_{i}$ for the EW benchmark}
\label{fig:te4}
\includegraphics[width = \figurewidth, height = \figureheight]{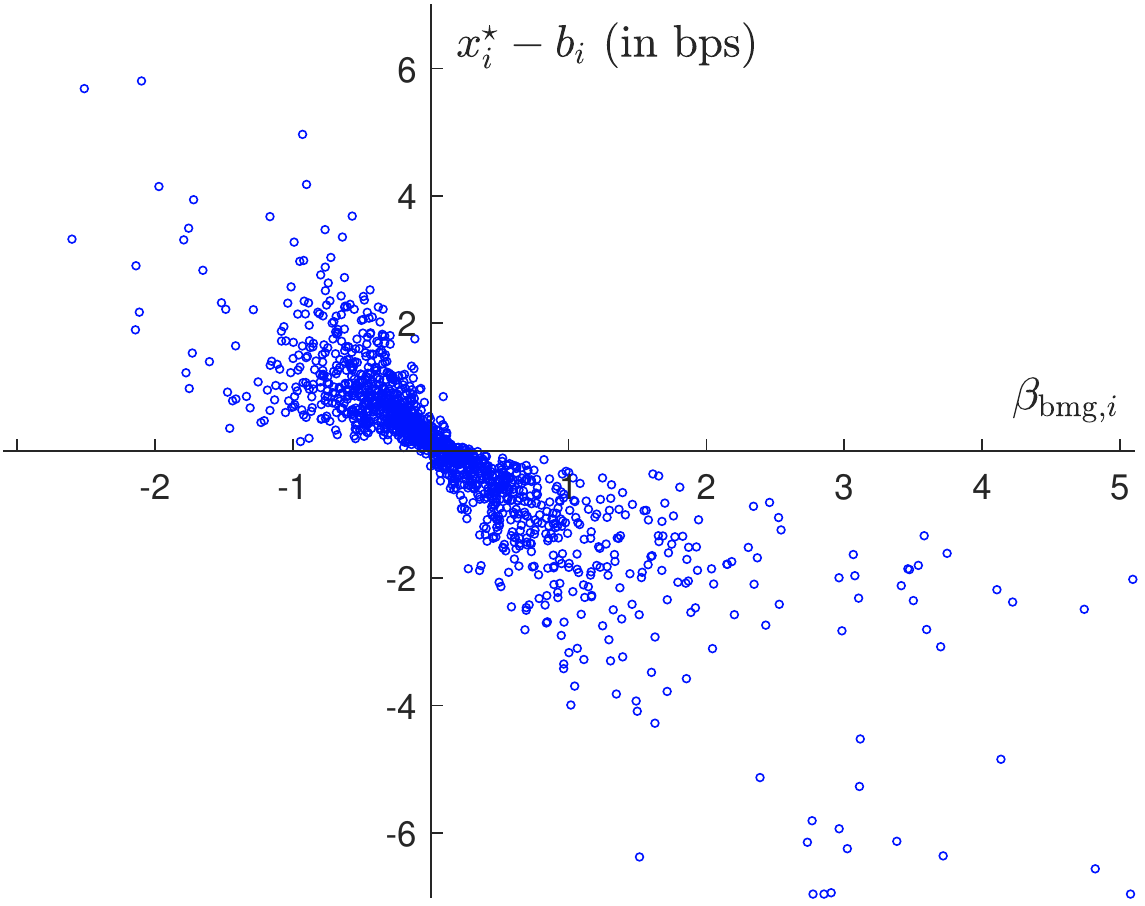}
\end{figure}

\begin{figure}[p]
\centering
\caption{Relationship between $\breve{\beta}_{\bmg,i}$ and $\Delta _{i}=x_{i}^{\star}-b_{i}$ for the EW benchmark}
\label{fig:te5}
\includegraphics[width = \figurewidth, height = \figureheight]{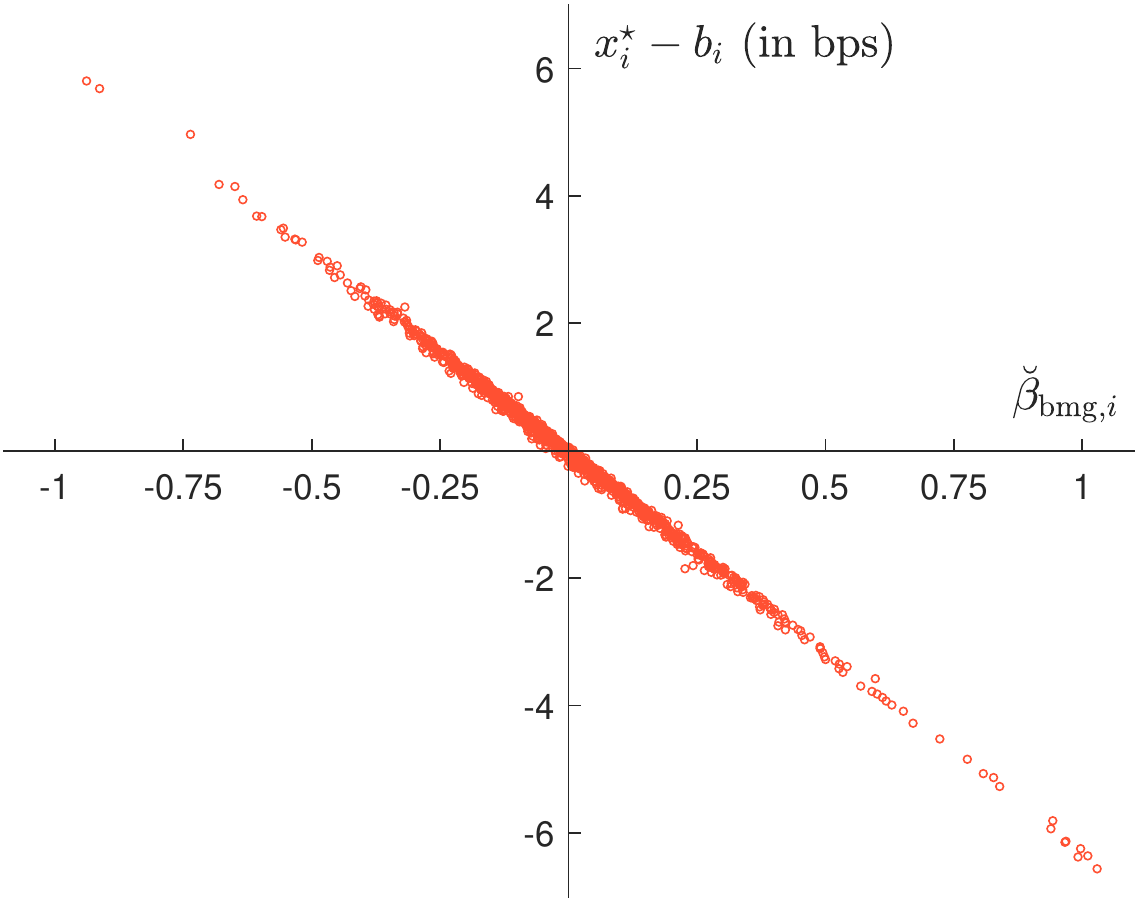}
\end{figure}

The previous example gives the impression that underweights and overweights
can also be predicted thanks to the BMG sensitivity $\beta_{\bmg,i}$. In this
example, $\beta_{\bmg,i}$ and $\breve{\beta}_{\bmg,i}$ have the same sign. In
order to illustrate that the statistic $\beta_{\bmg,i}$ is less relevant than
$\breve{\beta}_{\bmg,i}$, we apply the previous framework to
the MSCI World index universe. We consider that the benchmark is the EW portfolio, and
we impose that the BMG sensitivity is less than zero -- $\beta _{\bmg}^{+} =
0$. In Figures \ref{fig:te4} and \ref{fig:te5}, we have reported the
relationships between $\beta_{\bmg,i}$, $\breve{\beta}_{\bmg,i}$ and $\Delta
_{i}=x_{i}^{\star}-b_{i}$. We verify that $\breve{\beta}_{\bmg,i}$ is a
better statistic than $\beta_{\bmg,i}$ for estimating the weighting
direction. Indeed, the relationship between $\beta_{\bmg,i}$ and $\Delta
_{i}=x_{i}^{\star}-b_{i}$ is noisier than the relationship between
$\breve{\beta}_{\bmg,i}$ and $\Delta _{i}=x_{i}^{\star}-b_{i}$. Moreover, in
this last case, the relationship is almost linear.\smallskip

In what follows, we always consider that the benchmark is the
capitalization-weighted (CW) portfolio. We have again reported the
relationships between $\beta_{\bmg,i}$, $\breve{\beta}_{\bmg,i}$ and
$\Delta_{i}=x_{i}^{\star}-b_{i}$ in Figures \ref{fig:te6} and \ref{fig:te7} on
page \pageref{fig:te6}. In this case, we impose a relative carbon risk of the
portfolio\footnote{We impose a smaller threshold value of carbon risk when we
consider the CW portfolio as the benchmark because the relative carbon risk is
equal to $0.1491$ for the EW portfolio whereas it is equal to $-0.0851$ for the
CW portfolio. Indeed, Figure \ref{fig:te8} on page \pageref{fig:te8} shows that
the higher the market capitalisation of a firm, the lower its relative carbon risk. If
we set $\beta_{\bmg}^{+} = 0$ in the case where the benchmark is the CW
portfolio, the optimized portfolio $x^{\star}$ is then equal to the benchmark
$b$.} less than $-0.3$ -- $\beta_{\bmg}^{+} = -0.3$. We can notice that
some assets are not in line with the linear relationship between
$\breve{\beta}_{\bmg,i}$ and $\Delta_{i} = x_{i}^{\star}-b_{i}$. Since we
consider a long-only portfolio, these assets are excluded in the optimized
portfolio. The slope of the relationship between $\breve{\beta}_{\bmg,i}$ and
$\Delta_{i}$ is steeper in the current case than in the case where the
benchmark is the EW portfolio for two reasons. The first one is that we
have set a higher $\Delta_{\bmg}$ and the second one is that a steeper slope
allows us to offset the assets whose weights $x_{i}^{\star}$ have already
reached the value of zero.

\begin{remark}
Since the relationship between $\beta_{\bmg,i}$ and $\Delta
_{i}=x_{i}^{\star}-b_{i}$ is not a monotonically decreasing function, the
order-statistic optimization problem defined by the restriction
(\ref{eq:te3}) is not equivalent to the max-threshold optimization problem
defined by the inequality (\ref{eq:te2}).
\end{remark}

\begin{table}[tbph]
\centering
\caption{Regional composition of the portfolio (in \%)}
\label{tab:te14}
\begin{tabular}{cccc}
\hline
Region        & $b_{\mathcal{R}}$ & $x_{\mathcal{R}}^{\star}$ & $\Delta_{\mathcal{R}}$ \\ \hline
Eurozone      & $10.89$           & $12.77$                   & ${\TsVIII}1.88$        \\
Europe ex EMU & $10.83$           & $10.73$                   & $-0.09$                \\
North America & $65.16$           & $64.97$                   & $-0.19$                \\
Japan         & ${\TsV}8.70$      & ${\TsV}8.90$              & ${\TsVIII}0.19$        \\
Others        & ${\TsV}4.41$      & ${\TsV}2.63$              & $-1.79$                \\
\hline
\end{tabular}
\end{table}

Table \ref{tab:te14} reports some results about the portfolio's regional
exposure. Since the exposures of the capitalization-weighted and optimized
portfolios to a region $\mathcal{R}$ are respectively equal to $b_{\mathcal{R}}
= \sum_{i\in \mathcal{R}} b_{i}$ and $x_{\mathcal{R}}^{\star} = \sum_{i\in
\mathcal{R}} x_{i}^{\star}$, the long/short regional exposure
$\Delta_{\mathcal{R}}$ of the optimized portfolio with respect to the benchmark
is equal to $\Delta_{\mathcal{R}} = x_{\mathcal{R}}^{\star} - b_{\mathcal{R}}$.
The long/short exposure for the Eurozone and Japan is positive while it is
negative for Europe ex EMU and North America. We notice that these results are
consistent with the regional trend analysis provided by Figure
\ref{fig:carima10b} on page \pageref{fig:carima10b}. Indeed, the higher a
region's relative carbon risk, the lower the long/short exposure to a region.
Long/short exposure to the Eurozone is high in absolute and relative
values. Indeed, the optimized portfolio has a long exposure to the Eurozone of
almost $17.2\%$ higher than the benchmark. Nevertheless, the exposure to
European ex EMU and North American stocks has not changed significantly with
respect to the benchmark. This result is quite surprising for North America
since the average relative carbon risk $\beta_{\bmg,\mathcal{R}}\left(
t\right)$ was high in December 2018. For the rest of the world,
long/short exposure has significantly decreased by almost $40.5\%$ with respect
to the benchmark. We obtain such a result because of the very high average
relative carbon risk $\beta_{\bmg,\mathcal{R}}\left( t\right)$ of the other
regions which is around $0.75$ at the end of December 2018.\smallskip

\begin{figure}[tbph]
\centering
\caption{Long/short sector exposure $\Delta_{\mathcal{R}}$ of the portfolio (in \%)}
\label{fig:te13}
\includegraphics[width = \figurewidth, height = \figureheight]{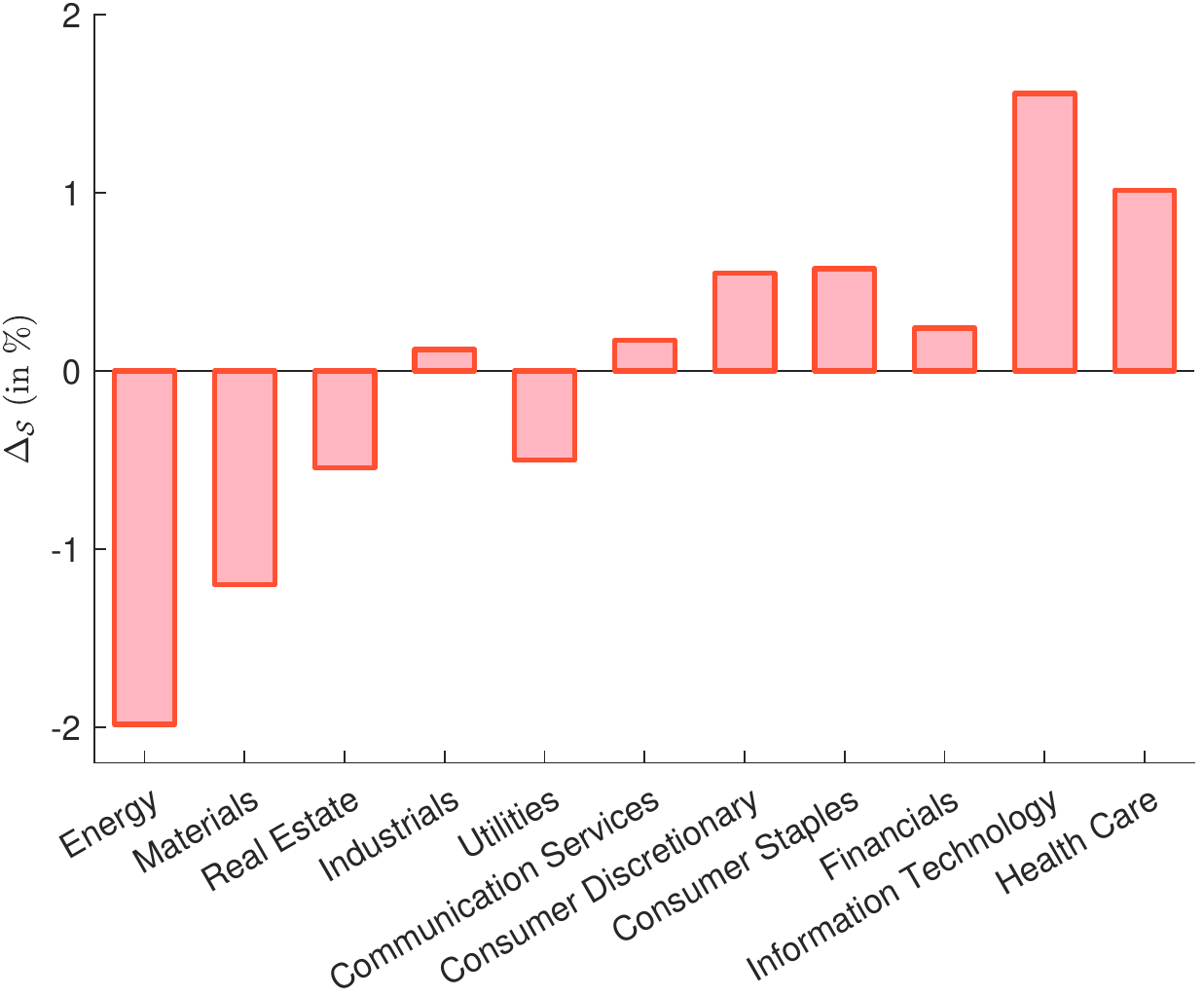}
\end{figure}

We may also be interested in the sector composition of the optimized
portfolio\footnote{We recall that $\beta_{\bmg}^{+} = -0.3$.}. Figure
\ref{fig:te13} provides the portfolio's long/short exposure
$\Delta_{\mathcal{S}}$ with respect to the benchmark, which is defined as
$\Delta_{\mathcal{S}} = \sum_{i \in \mathcal{S}} \left(x_{i}^{\star} -
b_{i}\right)$ where $\mathcal{S}$ is the sector. In this case, the difference
(both in absolute and relative terms) between the CW and optimized portfolios
is high. The energy sector is the most impacted. Indeed, the weight of the
energy sector is $6.00\%$ in the benchmark while it is equal to $4.02\%$ in the
optimized portfolio which represents a decrease of $33\%$. The materials sector
closely follows the energy sector with a decrease of its exposure from $4.58\%$
to $3.38\%$. The real estate and utilities sectors also see decreases in their
weights, of $17.42\%$ and $14.80\%$, respectively. In contrast, the sectors
whose exposure has increased are mainly information technology and health care.
Indeed, the weight increases from $14.95\%$ to $16.51\%$ for the information
technology sector and $13.32\%$ to $14.34\%$ for the health care sector. The
weight for the other sectors has changed slightly. Overall, the results are
consistent with the results obtained in Figures \ref{fig:carima5a},
\ref{fig:carima15a} and \ref{fig:carima15b}. In comparison with the benchmark,
the optimized portfolio is more exposed to the sectors which are positively
impacted by an unexpected acceleration in the transition process towards a
green economy.\smallskip

\begin{figure}[tbph]
\centering
\caption{Solution of the max-threshold optimization problem}
\label{fig:te9}
\includegraphics[width = \figurewidth, height = \figureheight]{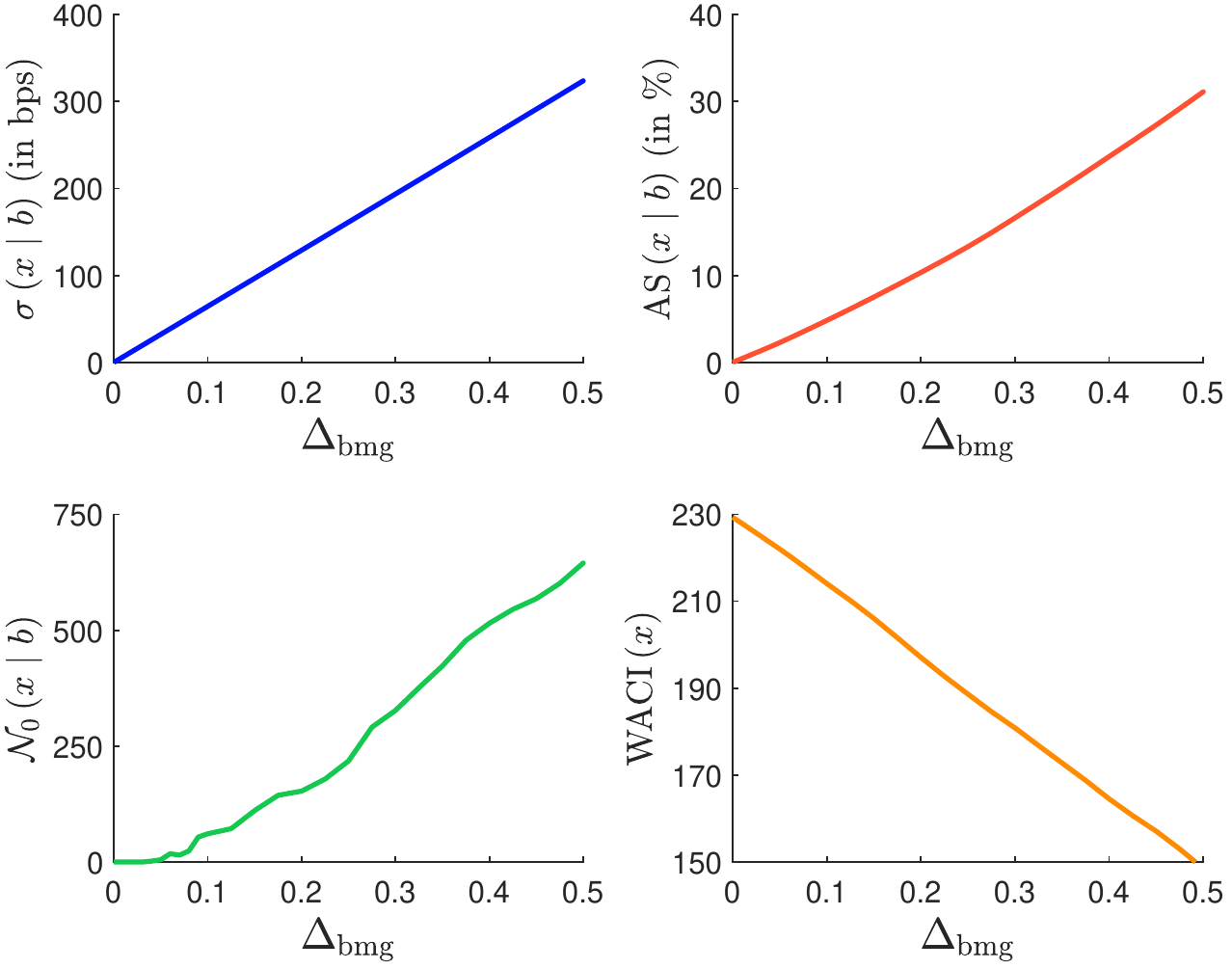}
\end{figure}

In what follows, we again consider that the benchmark is the CW portfolio.
Figure \ref{fig:te9} provides the relationships between the difference between
the benchmark's carbon risk and the portfolio's carbon risk $\Delta_{\bmg}$, the
tracking error $\sigma\left(x \mid b\right)$, the active share $\AS\left(x
\mid b\right)$, the number of excluding stocks $\mathcal{N}_{0}\left(x
\mid b\right)$ and the weighted average carbon intensity $\WACI$ for the max-threshold optimization problem. We notice that the
relationship between $\Delta_{\bmg}$ and $\sigma\left(x \mid b\right)$ is
linear. Indeed, we can demonstrate that\footnote{The semi-formal proof is
given in Appendix \ref{appendix:te-approximation} on page
\pageref{appendix:te-approximation}.}:
\begin{equation*}
\sigma\left( x^{\star }\mid b\right) \approx c\Delta_{\bmg}
\end{equation*}
By decreasing the relative carbon risk of the portfolio by $0.1$, the tracking
error increases by almost $65$ bps whatever the initial value of the
portfolio's carbon risk. In the current optimization problem, the active share
remains relatively low for any value of $\Delta_{\bmg}$. Moreover, we verify that the higher the $\Delta_{\bmg}$, the lower the $\WACI$.\smallskip

Figure \ref{fig:te11} on page \pageref{fig:te11} provides the solution to the
order-statistic optimization problem defined by the constraint (\ref{eq:te3}).
In this case, we solve a standard tracking error problem by having zero
exposure to the first $m$ stocks with higher carbon betas. We can notice
that for a same level of relative carbon risk tolerance, the tracking error and
the active share are always larger in the case of the order-statistic problem
in comparison with the max-threshold problem. By excluding the assets with the
higher relative carbon risk, the optimization problem excludes itself the
assets with a low relative carbon risk in order to have a carbon exposure
similar to the benchmark. Hence, we must exclude a large number of stocks to
move from $\Delta_{\bmg} = 0$ to $\Delta_{\bmg} = 0.1$. This implies both high
active share and tracking error. For instance, the tracking error is almost
twice that of the max-threshold optimization problem when $\Delta_{\bmg}$ is
equal to $0.1$.\smallskip

Some variants can be used to define the enhanced index portfolio problem.
For instance, we can use the following constraint:
\begin{equation}
\Omega = \left\{x\in \mathbb{R}^{n} : x_{i} = 0 \text{ if }b_{i}\beta_{\bmg,i}
\geq \left(b \odot \beta_{\bmg}\right)^{\left(m,n\right)}\right\}   \label{eq:te4}
\end{equation}
where $\left(b \odot \beta_{\bmg}\right)^{\left(m,n\right)} = \left(b \odot
\beta_{\bmg}\right)_{n-m+1:n}$ is the $\left( n-m+1\right) $\textit{-th}
order statistic of the vector $\left(b_{1} \beta_{\bmg,1},\ldots ,b_{n} \beta
_{\bmg,n}\right)$. In this case, we exclude the assets with both high weight
and high carbon beta in order to considerably reduce the portfolio's relative
carbon risk. Figure \ref{fig:te12} on page \pageref{fig:te12} provides the
solution to this optimization problem. Surprising, the outcomes are slightly
less successful than those of the standard order-statistic problem.
Therefore, we can deduce that excluding the assets with both high weight and
high relative carbon risk implies that the tracking error is more impacted.

\subsubsection{New formulation of the optimization problem with absolute carbon risk}

For the minimum variance portfolio, we have seen that managing the relative or
absolute carbon risk leads to two different portfolio optimization programs. In
the previous section, we considered the case where the fund manager would
like to reduce exposure to the brownest stocks. In order to decrease the BMG
sensitivity, the optimized portfolio increases its exposure to the greenest stocks.
This is normal if the investor's moral values are to fight climate
risk. However, this implies taking a bet particularly in the short run.
Indeed, if the investor chooses a negative value of $\beta _{\bmg}^{+}$ (e.g.
$\beta _{\bmg}^{+}=-30\%$), his optimized investment portfolio may substantially
underperform the benchmark if the BMG factor posts a positive performance.
Therefore, by choosing a negative BMG sensitivity, the investor expects that
the performance of the BMG factor will be negative in the future.\smallskip

Another enhanced index management approach is to limit the
exposure to absolute carbon risk. In this case, we obtain the
same optimization problem (\ref{eq:te1}), but with another system of
constraints:
\begin{equation*}
\Omega =\left\{ x\in \mathbb{R}^{n}:\left\vert \beta _{\bmg}^{\top
}x\right\vert \leq \left\vert \beta \right\vert _{\bmg}^{+}\right\}
\end{equation*}%
where $\left\vert \beta \right\vert _{\bmg}^{+}$ is the maximum sensitivity to
absolute carbon risk. This is equivalent to imposing the following inequality
constraints\footnote{Indeed, we have:
\begin{equation*}
\left\vert \beta _{\bmg}^{\top }x\right\vert \leq \left\vert \beta
\right\vert _{\bmg}^{+}\Leftrightarrow -\left\vert \beta \right\vert _{\bmg%
}^{+}\leq \beta _{\bmg}^{\top }x\leq +\left\vert \beta \right\vert _{\bmg%
}^{+}\Leftrightarrow \left\{
\begin{array}{l}
\beta _{\bmg}^{\top }x\leq \left\vert \beta \right\vert _{\bmg}^{+} \\
\left\vert \beta \right\vert _{\bmg}^{+}\geq -\beta _{\bmg}^{\top }x%
\end{array}%
\right.
\end{equation*}
}:
\begin{equation*}
\left(
\begin{array}{cc}
\beta _{\bmg} & -\beta _{\bmg}%
\end{array}%
\right) ^{\top }x\leq \left(
\begin{array}{c}
\left\vert \beta \right\vert _{\bmg}^{+} \\
\left\vert \beta \right\vert _{\bmg}^{+}%
\end{array}%
\right)
\end{equation*}%
Again, we obtain a QP problem, which is easy to solve. The special
case $\left\vert \beta \right\vert _{\bmg}^{+}=0$ corresponds to the
neutral exposure to the absolute carbon risk\footnote{In this case,
the inequality constraints are replaced by the equality constraint
$\beta _{\bmg}^{\top }x=0$, and the optimization program remains a
QP problem.}. From the viewpoint of passive management, imposing
that $\beta _{\bmg}^{\top }x=0$ can be justified because the
objective of passive management is to implement no active bets.

\begin{remark}
While we have two specific optimization programs in the case of the minimum
variance portfolios, the boundary between relative and absolute carbon risk is
not obvious when we consider tracking error optimization problems. Indeed,
imposing a neutral absolute carbon risk is equivalent to using an
inequality constraint on the relative carbon risk\footnote{%
If $\beta _{\bmg}^{\top }b>0$ (resp. $\beta _{\bmg}^{\top }b<0$), a neutral
absolute carbon risk is achieved using the constraint $\beta _{\bmg}^{\top
}x\leq 0$ (resp. $\beta _{\bmg}^{\top }x\geq 0$).}.
\end{remark}

\subsection{A factor investing perspective}

\citet{Bennani-2018} and \citet{Drei-2019} discussed the relationships between
ESG investing and factor investing. Among the different results, they showed
that ESG may be considered as a beta strategy in the Eurozone while it
continues to remain an alpha strategy in North America \citep{Roncalli-2020b}.
We may wonder if the carbon risk can play a similar role from a factor
investing point of view. Indeed, the results obtained in Section Two show that
it helps to explain the cross-section of stock returns.\smallskip

We recall that the objective of factor investing is to build a well-diversified
portfolio of risk factors to better capture the equity risk premium.
The underlying idea is that there is not only one risk premium, but several
risk premia, implying that the equity risk premium cannot be reduced to the
CAPM risk factor. The traditional approach is then to capture the equity risk
premia with five risk factors: size, value, momentum, low-volatility and
quality.\smallskip

Based on the previous analysis, it is tempting to include a carbon
risk factor as a sixth risk premium. However, we think that carbon
risk is too specific and cannot be considered as a risk premium.
First, it is obviously not a skewness risk premium
\citep{Roncalli-2017}. Second, it may be a market anomaly, but we
must be cautious with the \textquoteleft\textit{factor
zoo}\textquoteright\ \citep{Cochrane-2011}. Indeed, we think that
the ESG risk factor does make more sense than the carbon risk factor
in a factor investing framework, because it represents a broader
investment type. Moreover, ESG investing is today a big investment
topic of institutional investors, whereas carbon risk is embedded in
the Environment pillar of ESG. Carbon risk is more a risk management
subject than an investing approach. This is why we speak about ESG
investing, but not about carbon investing.\smallskip

Nevertheless, as shown previously, carbon risk is a financial risk
and is interesting to manage. Since equity factor
investing is benchmarked against capitalization-weighted indices, we
can use the framework developed in Section
\ref{section:enhanced-index} on page
\pageref{section:enhanced-index}. Instead of using a bottom-up
approach to define a carbon risk factor, it is easier to implement
an overlay or a top-down approach that controls the relative or
absolute carbon risk of the factor investing portfolio with respect
to its benchmark.


\section{Conclusion}

This paper studies the methodology proposed by \citet{Gorgen-2019} for
measuring carbon risk in investment portfolios. We confirm the results of
these authors, that showed that carbon risk is priced in at the stock level and is
relevant in a cross-sectional multi-factor analysis. By considering a dynamic
framework, we highlight several stylized facts. First, we notice that
carbon risk was priced in more at the beginning of the 2010s than it is today.
Nevertheless, this is mainly due to the Eurozone. Second, we observe a
convergence of absolute carbon risk pricing among the different regions,
except in Japan. If we focus on relative carbon risk, we confirm another
transatlantic divide that we generally observe in ESG investing
\citep{Bennani-2018, Drei-2019}. On average, European stocks have a negative
carbon beta, whereas it is positive for North America. We also observe some
differences between sectors. For instance, there is clearly a difference in the
dynamics of the carbon beta between the materials and energy sectors, and the
other sectors.\smallskip

Because the brown-minus-green (BMG) factor developed by \citet{Gorgen-2019} is
based on more than $50$ proxy variables, investors may have some difficulty
understanding which risk dimension is priced in by the market. For each stock,
these authors calculate a brown green score, which is based on three
dimensions: value chain, public perception and adaptability. Each dimension
is the result of mixing several sub-dimensions. In this article, we have
preferred to consider very basic dimensions based on the Trucost and MSCI
databases. We have focused our analysis on carbon intensity data provided by
Trucost, and data on carbon emissions exposure, carbon management, the climate
change score and the environment pillar provided by MSCI. All these dimensions
have an explanatory power that is in line with the Carima factor of
\citet{Gorgen-2019}. However, at first order, we find that carbon intensity and
carbon emissions exposure are the best alternative approaches to the Carima
factor.\smallskip

ESG rating agencies and other NGOs have developed many fundamental measures and
scores to assess a firm's carbon risk. Carbon beta is not a new measure.
However, the approach of \citet{Gorgen-2019} is definitively original since it
is a market-based measure and not a fundamental-based measure. The carbon beta
of a stock corresponds to the carbon risk of the stock priced in by the
financial market. Therefore, we may observe wide discrepancies between the
market perception of the carbon risk, and for instance the direct value of the
carbon intensity. Our analysis shows that they are weakly correlated (less than
$30\%$). We can draw a parallel with the value risk factor. Indeed, the value
beta of a stock may be related to its book-to-price, but they are two different
measures. And the financial market may consider that a stock with a high
book-to-price is not necessarily a value stock, but a growth stock.\smallskip

The carbon beta therefore constitutes useful information for managing carbon risk in
investment portfolios. In this article, we have mainly studied two investment
strategies: the minimum variance portfolio and the enhanced index portfolio.
Our results highlight the difference between managing absolute carbon risk
and relative carbon risk. Managing absolute carbon risk
implies having zero exposure to the BMG factor, whereas managing
relative carbon risk implies having negative exposure to the BMG factor. In
the first case, the objective is to propose an immunization-hedging investment
strategy against carbon risk. In the second case, we explicitly take an active
management bet by overweighting green stocks and underweighting brown stocks.
This second approach is certainly the most frequently observed, even in
passive management, because of investors' moral values. We show that the two
approaches led us to consider different objective functions or constraints of the
portfolio optimization program, implying that we obtain very different
solutions. Another finding is that managing market-based carbon risk
and fundamental-based carbon risk\footnote{By considering for instance
a direct measure of carbon risk such as carbon intensity.} does not give
the same solution, even though we observe similar properties between the optimized
portfolios. This is particularly true in the case of enhanced index portfolios.
Finally, we discuss whether the BMG factor can be considered as a new
factor or not, alongside traditional factors (size, value, momentum, etc.). Our
conviction is that carbon risk is a risk management subject, and not an
investment style such as ESG investing. This is why we consider that carbon
risk is more appropriate for better defining a minimum variance portfolio than
improving the diversification of a factor investing portfolio. Nevertheless,
all our results confirm the initial findings of \citet{Gorgen-2019}. Investors
must be aware that carbon risk is priced in by the stock market. This is why they
must measure and manage this risk, especially when it is too high or when it
is incompatible with the fiduciary duties of their investment portfolios.

\clearpage

\bibliographystyle{apalike}

\clearpage

\appendix

\section*{Appendix}

\section{Mathematical results}

\subsection{Time-varying estimation with Kalman filter}
\label{appendix:kalman-filter}

The time-varying risk factor model can be written as a state space model:%
\begin{equation}
\left\{
\begin{array}{l}
y\left( t\right) =x\left( t\right) ^{\top }\beta \left( t\right)
+\varepsilon \left( t\right)  \\
\beta \left( t\right) =\beta \left( t-1\right) +\eta \left( t\right)
\end{array}%
\right.   \label{eq:app-ssm1}
\end{equation}%
where $\varepsilon \left( t\right) \sim \mathcal{N}\left( 0,\sigma
_{\varepsilon }^{2}\right) $, $\eta \left( t\right) \sim \mathcal{N}\left(
\mathbf{0}_{K+1},\Sigma _{\beta }\right) $ and $K$ is the number of risk
factors. For instance, in the case of the MKT+BMG model, $y\left( t\right) $
corresponds to the asset return $R_{i}\left( t\right) $, $x\left( t\right) $
is a $3\times 1$ vector, whose elements are $1$, $R_{\mathrm{mkt}}\left(
t\right) $ and $R_{\mathrm{bmg}}\left( t\right) $ and:%
\begin{equation}
\beta \left( t\right) =\left(
\begin{array}{c}
\alpha _{i}\left( t\right)  \\
\beta _{\mathrm{mkt},i}\left( t\right)  \\
\beta _{\mathrm{bmg},i}\left( t\right)
\end{array}%
\right)   \label{eq:app-ssm2}
\end{equation}%
It follows that the variable $y_{t}$ is observable, but this is not the case for
the state vector $\beta \left( t\right) $. The Kalman filter is a statistical
tool to estimate the distribution function of $\beta \left( t\right) $. Let
$\beta \left( 0\right) \sim \mathcal{N}\left( \beta
_{0},P_{0}\right) $ be the initial position of the state vector. We note $%
\hat{\beta}\left( t\mid t-1\right) =\mathbb{E}\left[ \beta \left( t\right)
\mid \mathcal{F}\left( t-1\right) \right] $ and $\hat{\beta}\left( t\mid
t\right) =\mathbb{E}\left[ \beta \left( t\right) \mid \mathcal{F}\left(
t\right) \right] $ as the optimal estimators of $\beta \left( t\right) $ given
the available information until time $t-1$ and $t$. $P\left( t\mid t-1\right)
$ and $P\left( t\mid t\right) $ are the covariance matrices associated with
$\hat{\beta}\left( t\mid t-1\right) $ and $\hat{\beta}\left(
t\mid t\right) $. Since the estimate of $y\left( t\right) $ is equal to $%
\hat{y}\left( t\mid t-1\right) =x\left( t\right) ^{\top }\hat{\beta}\left(
t\mid t-1\right) $, we can compute the variance $F\left( t\right) $ of the
innovation process $v\left( t\right) =y\left( t\right) -\hat{y}\left( t\mid
t-1\right) $. These different quantities can be calculated thanks to the
Kalman filter, which consists in the following recursive algorithm\footnote{%
The algorithm is initialized with values $\hat{\beta}\left( 0\mid 0\right)
=\beta _{0}$ and $P\left( 0\mid 0\right) =P_{0}$.} \citep[page
654]{Roncalli-2020}:
\begin{equation}
\left\{
\begin{array}{l}
\hat{\beta}\left( t\mid t-1\right) =\hat{\beta}\left( t-1\mid t-1\right)  \\
P\left( t\mid t-1\right) =P\left( t-1\mid t-1\right) +\Sigma _{\beta } \\
v\left( t\right) =y\left( t\right) -x\left( t\right) ^{\top }\hat{\beta}%
\left( t\mid t-1\right)  \\
F\left( t\right) =x\left( t\right) ^{\top }P\left( t\mid t-1\right) x\left(
t\right) +\sigma _{\varepsilon }^{2} \\
\hat{\beta}\left( t\mid t\right) =\hat{\beta}\left( t\mid t-1\right) +\left(
\dfrac{P\left( t\mid t-1\right) }{F\left( t\right) }\right) x\left( t\right)
v\left( t\right)  \\
P\left( t\mid t\right) =\left( I_{K+1}-\left( \dfrac{P\left( t\mid t-1\right)
}{F\left( t\right) }\right) x\left( t\right) x\left( t\right) ^{\top
}\right) P\left( t\mid t-1\right)
\end{array}%
\right.   \label{eq:app-ssm3}
\end{equation}%
In this model, the parameters $\sigma _{u}^{2}$ and $\Sigma _{\beta }$ are
unknown and can be estimated by the method of maximum likelihood. Since $%
v\left( t\right) \sim \mathcal{N}\left( 0,F\left( t\right) \right) $, the
log-likelihood function is equal to:
\begin{equation}
\LogL\left( \theta \right) =-\frac{T}{2}\ln \left( 2\pi \right) -\frac{1}{2}%
\sum_{t=1}^{T}\left( \ln F\left( t\right) +\frac{v^{2}\left( t\right) }{%
F\left( t\right) }\right)   \label{eq:app-ssm4}
\end{equation}%
where $\theta =\left( \sigma ^{2},\Sigma \right) $. Maximizing the
log-likelihood function requires specifying the initial conditions $\beta
_{0}$ and $P_{0}$, which are not necessarily known. In this case, we use the
linear regression $y\left( t\right) =x\left( t\right) ^{\top }\beta
+\varepsilon \left( t\right) $, and the OLS estimates $\hat{\beta}_{\mathrm{%
ols}}$ and $\hat{\sigma}_{\varepsilon }^{2}\left( X^{\top }X\right) ^{-1}$ to
initialize $\beta _{0}$ and $P_{0}$.

\subsection{Sherman-Morrison-Woodbury formula}
\label{appendix:smw-formula}

Suppose $u$ and $v$ are two $n\times 1$ vectors and $A$ is an invertible $%
n\times n$ matrix. We can show that \citep{Golub-2013}:
\begin{equation}
\left( A+uv^{\top }\right) ^{-1}=A^{-1}-\frac{1}{1+v^{\top }A^{-1}u}%
A^{-1}uv^{\top }A^{-1}  \label{eq:smw1}
\end{equation}%
\citet{Batista-2008} and \citet{Batista-2009} extended the SMW formula when
the outer product is a
sum:%
\begin{equation}
\left( A+\sum_{k=1}^{m}u_{k}v_{k}^{\top }\right)
^{-1}=A^{-1}-A^{-1}US^{-1}V^{\top }A^{-1}  \label{eq:smw2}
\end{equation}%
where $U=\left(\begin{array}{ccc} u_{1} & \cdots & u_{m}
\end{array}\right) $ and $V=\left(\begin{array}{ccc} v_{1} & \cdots & v_{m}
\end{array}\right) $ are two $n\times m$ matrices, and $S=I+T$ and $T=\left(
T_{i,j}\right) $ are two $m\times m$ matrices where $T_{i,j}=v_{i}^{\top
}A^{-1}u_{j}$.\smallskip

In the case $m=2$, the SMW formula becomes:%
\begin{equation}
\left( A+u_{1}v_{1}^{\top }+u_{2}v_{2}^{\top }\right)
^{-1}=A^{-1}-A^{-1}US^{-1}V^{\top }A^{-1}  \label{eq:smw3}
\end{equation}%
where:%
\begin{equation*}
S=\left(
\begin{array}{cc}
1+v_{1}^{\top }A^{-1}u_{1} & v_{1}^{\top }A^{-1}u_{2} \\
v_{2}^{\top }A^{-1}u_{1} & 1+v_{2}^{\top }A^{-1}u_{2}%
\end{array}%
\right)
\end{equation*}%
Since we have:%
\begin{equation*}
S^{-1}=\frac{1}{\left\vert S\right\vert }\left(
\begin{array}{cc}
1+v_{2}^{\top }A^{-1}u_{2} & -v_{1}^{\top }A^{-1}u_{2} \\
-v_{2}^{\top }A^{-1}u_{1} & 1+v_{1}^{\top }A^{-1}u_{1}%
\end{array}%
\right)
\end{equation*}%
where
\begin{eqnarray*}
\left\vert S\right\vert  &=&\left( 1+v_{1}^{\top }A^{-1}u_{1}\right) \left(
1+v_{2}^{\top }A^{-1}u_{2}\right) -v_{2}^{\top }A^{-1}u_{1}v_{1}^{\top
}A^{-1}u_{2} \\
&=&1+v_{1}^{\top }A^{-1}u_{1}+v_{2}^{\top }A^{-1}u_{2}+v_{1}^{\top
}A^{-1}u_{1}v_{2}^{\top }A^{-1}u_{2}-v_{2}^{\top }A^{-1}u_{1}v_{1}^{\top
}A^{-1}u_{2}
\end{eqnarray*}%
We deduce that:%
\begin{eqnarray*}
\left\vert S\right\vert \cdot US^{-1}V^{\top } &=&\left(
\begin{array}{cc}
u_{1} & u_{2}%
\end{array}%
\right) \left(
\begin{array}{cc}
1+v_{2}^{\top }A^{-1}u_{2} & -v_{1}^{\top }A^{-1}u_{2} \\
-v_{2}^{\top }A^{-1}u_{1} & 1+v_{1}^{\top }A^{-1}u_{1}%
\end{array}%
\right) \left(
\begin{array}{c}
v_{1}^{\top } \\
v_{2}^{\top }%
\end{array}%
\right)  \\
&=&u_{1}v_{1}^{\top }+u_{1}v_{2}^{\top }A^{-1}u_{2}v_{1}^{\top
}-u_{2}v_{2}^{\top }A^{-1}u_{1}v_{1}^{\top }- \\
&&u_{1}v_{1}^{\top }A^{-1}u_{2}v_{2}^{\top }+u_{2}v_{2}^{\top
}+u_{2}v_{1}^{\top }A^{-1}u_{1}v_{2}^{\top }
\end{eqnarray*}%
\smallskip

If $A$ is a diagonal matrix, we can simplify the previous SMW formula
(\ref{eq:smw3}). Indeed, we have:
\begin{eqnarray}
\left\vert S\right\vert  &=&1+\sum_{s=1}^{n}\frac{%
u_{1,s}v_{1,s}+u_{2,s}v_{2,s}}{a_{s,s}}+\sum_{s=1}^{n}\frac{u_{1,s}v_{1,s}}{%
a_{s,s}}\sum_{s=1}^{n}\frac{u_{2,s}v_{2,s}}{a_{s,s}}-  \notag \\
&&\sum_{s=1}^{n}\frac{u_{1,s}v_{2,s}}{a_{s,s}}\sum_{s=1}^{n}\frac{%
u_{2,s}v_{1,s}}{a_{s,s}}  \label{eq:smw4a}
\end{eqnarray}%
and\footnote{%
Let $A=\left( a_{i,j}\right) $ and $C=\left( c_{i,j}\right) $ be two $%
n\times n$ matrices and $b$ a vector of dimension $n$. We recall that:%
\begin{equation*}
\left( A\limfunc{diag}\left( b\right) C\right)
_{i,j}=\sum_{s=1}^{n}a_{i,s}b_{s}c_{s,j}
\end{equation*}%
We deduce that:%
\begin{equation*}
\left( a_{1}a_{2}^{\top }\limfunc{diag}\left( b\right) c_{1}c_{2}^{\top
}\right) _{i,j}=a_{1,i}c_{2,j}\sum_{s=1}^{n}a_{2,s}b_{s}c_{1,s}
\end{equation*}%
where $a_{1}$, $a_{2}$, $c_{1}$and $a_{2}$ are $n\times 1$ vectors.}:%
\begin{eqnarray}
\left\vert S\right\vert \cdot US^{-1}V^{\top } &=&\left( 1+\sum_{s=1}^{n}%
\frac{u_{2,s}v_{2,s}}{a_{s,s}}\right) u_{1}v_{1}^{\top }+\left(
1+\sum_{s=1}^{n}\frac{u_{1,s}v_{1,s}}{a_{s,s}}\right) u_{2}v_{2}^{\top }-
\notag \\
&&\left( \sum_{s=1}^{n}\frac{u_{1,s}v_{2,s}}{a_{s,s}}\right)
u_{2}v_{1}^{\top }-\left( \sum_{s=1}^{n}\frac{u_{2,s}v_{1,s}}{a_{s,s}}%
\right) u_{1}v_{2}^{\top }  \label{eq:smw4b}
\end{eqnarray}%
If we assume that $u_{1}$ is uncorrelated to $v_{2}$, $u_{2}$ is uncorrelated
to $v_{1}$, $u_{2}$ and $v_{2}$ are centered around $0$, we
obtain:%
\begin{equation}
\left\vert S\right\vert \approx 1+\sum_{s=1}^{n}\frac{%
u_{1,s}v_{1,s}+u_{2,s}v_{2,s}}{a_{s,s}}+\sum_{s=1}^{n}\frac{u_{1,s}v_{1,s}}{%
a_{s,s}}\sum_{s=1}^{n}\frac{u_{2,s}v_{2,s}}{a_{s,s}}  \label{eq:smw5a}
\end{equation}%
and:%
\begin{equation}
\left\vert S\right\vert \cdot US^{-1}V^{\top }\approx \left( 1+\sum_{s=1}^{n}%
\frac{u_{2,s}v_{2,s}}{a_{s,s}}\right) u_{1}v_{1}^{\top }+\left(
1+\sum_{s=1}^{n}\frac{u_{1,s}v_{1,s}}{a_{s,s}}\right) u_{2}v_{2}^{\top }
\label{eq:smw5b}
\end{equation}

\subsection{Minimum variance portfolio in the MKT+BMG model}
\label{appendix:gmv-bmg}

\begin{remark}
In what follows, we use the notations $\beta_i$ and $\gamma_i$ instead of
$\beta_{\mkt,i}$ and $\beta_{\bmg,i}$ to simplify the notations.
\end{remark}

We have:%
\begin{equation*}
R_{i}\left(t\right)=\alpha _{i}+\beta _{i}R_{\mkt}\left(t\right)+\gamma _{i}R_{\bmg}\left(t\right)+\varepsilon _{i}\left(t\right)
\end{equation*}%
It follows that the covariance matrix is:%
\begin{equation*}
\Sigma =\beta \beta ^{\top }\sigma _{\mkt}^{2}+\gamma \gamma ^{\top }\sigma
_{\bmg}^{2}+D
\end{equation*}%
where $\beta =\left( \beta _{1},\ldots ,\beta _{n}\right) $ is the vector of
MKT betas, $\gamma =\left( \gamma _{1},\ldots ,\gamma _{n}\right) $ is the
vector of BMG betas, $\sigma _{\mkt}^{2}$ is the variance of the market
portfolio, $\sigma _{\bmg}^{2}$ is the variance of the BMG factor and $D=%
\func{diag}\left( \tilde{\sigma}_{1}^{2},\ldots ,\tilde{\sigma}%
_{n}^{2}\right) $ is the diagonal matrix of specific variances. We recall
that the GMV portfolio is equal to:%
\begin{eqnarray}
x^{\star } &=&\frac{\Sigma ^{-1}\mathbf{1}_{n}}{\mathbf{1}_{n}^{\top }\Sigma
^{-1}\mathbf{1}_{n}}  \notag \\
&=&\sigma ^{2}\left( x^{\star }\right) \cdot \Sigma ^{-1}\mathbf{1}_{n}
\label{eq:app-gmv1}
\end{eqnarray}%
because we have:%
\begin{eqnarray*}
\sigma ^{2}\left( x^{\star }\right)  &=&x^{\star ^{\top }}\Sigma x^{\star }
\\
&=&\frac{\mathbf{1}_{n}^{\top }\Sigma ^{-1}}{\mathbf{1}_{n}^{\top }\Sigma
^{-1}\mathbf{1}_{n}}\Sigma \frac{\Sigma ^{-1}\mathbf{1}_{n}}{\mathbf{1}%
_{n}^{\top }\Sigma ^{-1}\mathbf{1}_{n}} \\
&=&\frac{1}{\mathbf{1}_{n}^{\top }\Sigma ^{-1}\mathbf{1}_{n}}
\end{eqnarray*}

\subsubsection{General formula}

We use the generalized Sherman-Morrison-Woodbury with $A=D$, $%
u_{1}=v_{1}=\sigma _{\mkt}\beta $ and $u_{2}=v_{2}=\sigma _{\bmg}\gamma $. It
follows that the inverse of the covariance matrix is equal to:
\begin{equation*}
\Sigma ^{-1}=D^{-1}-D^{-1}US^{-1}V^{\top }D^{-1}
\end{equation*}%
where $U=V=\left( \begin{array}{cc} \sigma _{\mkt}\beta & \sigma
_{\bmg}\gamma\end{array}\right) $ and:%
\begin{equation*}
S=\left(
\begin{array}{cc}
1+\sigma _{\mkt}^{2}\beta ^{\top }D^{-1}\beta  & \sigma _{\mkt}\sigma _{\bmg}\beta
^{\top }D^{-1}\gamma  \\
\sigma _{\mkt}\sigma _{\bmg}\beta ^{\top }D^{-1}\gamma  & 1+\sigma
_{\bmg}^{2}\gamma ^{\top }D^{-1}\gamma
\end{array}%
\right)
\end{equation*}%
We notice that:%
\begin{eqnarray*}
\sum_{s=1}^{n}\frac{u_{1,s}v_{1,s}}{a_{s,s}} &=&\sigma _{\mkt}^{2}\sum_{s=1}^{n}%
\frac{\beta _{s}^{2}}{\tilde{\sigma}_{s}^{2}}=\sigma _{\mkt}^{2}\varphi \left(
\tilde{\beta},\beta \right)  \\
\sum_{s=1}^{n}\frac{u_{2,s}v_{2,s}}{a_{s,s}} &=&\sigma
_{\bmg}^{2}\sum_{s=1}^{n}\frac{\gamma _{s}^{2}}{\tilde{\sigma}_{s}^{2}}%
=\sigma _{\bmg}^{2}\varphi \left( \tilde{\gamma},\gamma \right)  \\
\sum_{s=1}^{n}\frac{u_{1,s}v_{2,s}}{a_{s,s}} &=&\sum_{s=1}^{n}\frac{%
u_{2,s}v_{1,s}}{a_{s,s}}=\sigma _{\mkt}\sigma _{\bmg}\sum_{s=1}^{n}\frac{\beta
_{s}\gamma _{s}}{\tilde{\sigma}_{s}^{2}}=\sigma _{\mkt}\sigma _{\bmg}\varphi
\left( \tilde{\beta},\gamma \right)
\end{eqnarray*}%
where $\tilde{\beta}$ and $\tilde{\gamma}$ are the standardized vectors of $%
\beta $ and $\gamma $ by the idiosyncratic variances and $\varphi \left(
x,y\right) =x\circ y$ is the outer product. Using Equation (\ref{eq:smw4a}),
we obtain:%
\begin{equation*}
\left\vert S\right\vert =1+\sigma _{\mkt}^{2}\varphi \left( \tilde{\beta},\beta
\right) +\sigma _{\bmg}^{2}\varphi \left( \tilde{\gamma},\gamma \right)
+\sigma _{\mkt}^{2}\sigma _{\bmg}^{2}\left( \varphi \left( \tilde{\beta},\beta
\right) \varphi \left( \tilde{\gamma},\gamma \right) -\varphi ^{2}\left(
\tilde{\beta},\gamma \right) \right)
\end{equation*}%
Equation (\ref{eq:smw4b}) becomes:%
\begin{eqnarray*}
\left\vert S\right\vert \cdot US^{-1}V^{\top } &=&\sigma _{\mkt}^{2}\left(
1+\sigma _{\bmg}^{2}\varphi \left( \tilde{\gamma},\gamma \right) \right)
\beta \beta ^{\top }+\sigma _{\bmg}^{2}\left( 1+\sigma _{\mkt}^{2}\varphi \left(
\tilde{\beta},\beta \right) \right) \gamma \gamma ^{\top }- \\
&&\sigma _{\mkt}^{2}\sigma _{\bmg}^{2}\varphi \left( \tilde{\beta},\gamma
\right) \left( \gamma \beta ^{\top }+\beta \gamma ^{\top }\right)
\end{eqnarray*}%
Finally, the inverse of the covariance matrix has the following expression:%
\begin{equation*}
\Sigma ^{-1}=D^{-1}-M^{-1}
\end{equation*}%
where:%
\begin{equation*}
M^{-1}=\omega _{1}\tilde{\beta}\tilde{\beta}^{\top }+\omega _{2}\tilde{\gamma%
}\tilde{\gamma}^{\top }-\omega _{3}\left( \tilde{\gamma}\tilde{\beta}^{\top
}+\tilde{\beta}\tilde{\gamma}^{\top }\right)
\end{equation*}%
and:%
\begin{eqnarray*}
\omega _{0} &=&\left\vert S\right\vert  \\
\omega _{1} &=&\omega _{0}^{-1}\cdot \sigma _{\mkt}^{2}\left( 1+\sigma
_{\bmg}^{2}\varphi \left( \tilde{\gamma},\gamma \right) \right)  \\
\omega _{2} &=&\omega _{0}^{-1}\cdot \sigma _{\bmg}^{2}\left( 1+\sigma
_{\mkt}^{2}\varphi \left( \tilde{\beta},\beta \right) \right)  \\
\omega _{3} &=&\omega _{0}^{-1}\cdot \sigma _{\mkt}^{2}\sigma _{\bmg}^{2}\varphi
\left( \tilde{\beta},\gamma \right)
\end{eqnarray*}%
Therefore, the solution (\ref{eq:app-gmv1}) becomes:%
\begin{equation*}
x^{\star }=\sigma ^{2}\left( x^{\star }\right) \left( D^{-1}\mathbf{1}%
_{n}-M^{-1}\mathbf{1}_{n}\right)
\end{equation*}%
It follows that:%
\begin{equation}
x^{\star }=\sigma ^{2}\left( x^{\star }\right) \left( \xi -\tilde{\omega}_{1}%
\tilde{\beta}-\tilde{\omega}_{2}\tilde{\gamma}\right)   \label{eq:app-gmv2}
\end{equation}%
where $\xi =\left( \tilde{\sigma}_{1}^{-2},\ldots ,\tilde{\sigma}%
_{n}^{-2}\right) $, $\tilde{\omega}_{1}=\omega _{1}\tilde{\beta}^{\top }%
\mathbf{1}_{n}-\omega _{3}\tilde{\gamma}^{\top }\mathbf{1}_{n}$ and $\tilde{%
\omega}_{2}=\omega _{2}\tilde{\gamma}^{\top }\mathbf{1}_{n}-\omega _{3}%
\tilde{\beta}^{\top }\mathbf{1}_{n}$. We conclude that:%
\begin{equation}
x_{i}^{\star }=\frac{\sigma ^{2}\left( x^{\star }\right) }{\tilde{\sigma}%
_{i}^{2}}\left( 1-\frac{\beta _{i}}{\beta ^{\star }}-\frac{\gamma _{i}}{%
\gamma ^{\star }}\right)   \label{eq:app-gmv3}
\end{equation}%
where $\beta ^{\star }=\tilde{\omega}_{1}^{-1}$ and $\gamma ^{\star }=\tilde{%
\omega}_{2}^{-1}$.

\begin{remark}
If we develop the expression of $\beta ^{\star }$ and $\gamma ^{\star }$, we
obtain:%
\begin{equation}
\beta ^{\star }=\frac{1+\sigma _{\mkt}^{2}\varphi \left( \tilde{\beta},\beta
\right) +\sigma _{\bmg}^{2}\varphi \left( \tilde{\gamma},\gamma \right)
+\sigma _{\mkt}^{2}\sigma _{\bmg}^{2}\left( \varphi \left( \tilde{\beta},\beta
\right) \varphi \left( \tilde{\gamma},\gamma \right) -\varphi ^{2}\left(
\tilde{\beta},\gamma \right) \right) }{\sigma _{\mkt}^{2}\left( \tilde{\beta}%
^{\top }\mathbf{1}_{n}+\sigma _{\bmg}^{2}\left( \varphi \left( \tilde{\gamma}%
,\gamma \right) \tilde{\beta}^{\top }\mathbf{1}_{n}-\varphi \left( \tilde{%
\beta},\gamma \right) \tilde{\gamma}^{\top }\mathbf{1}_{n}\right) \right) }
\label{eq:app-beta-star}
\end{equation}%
and:%
\begin{equation}
\gamma ^{\star }=\frac{1+\sigma _{\mkt}^{2}\varphi \left( \tilde{\beta},\beta
\right) +\sigma _{\bmg}^{2}\varphi \left( \tilde{\gamma},\gamma \right)
+\sigma _{\mkt}^{2}\sigma _{\bmg}^{2}\left( \varphi \left( \tilde{\beta},\beta
\right) \varphi \left( \tilde{\gamma},\gamma \right) -\varphi ^{2}\left(
\tilde{\beta},\gamma \right) \right) }{\sigma _{\bmg}^{2}\left( \tilde{\gamma}%
^{\top }\mathbf{1}_{n}+\sigma _{\mkt}^{2}\left( \varphi \left( \tilde{\beta}%
,\beta \right) \tilde{\gamma}^{\top }\mathbf{1}_{n}-\varphi \left( \tilde{%
\beta},\gamma \right) \tilde{\beta}^{\top }\mathbf{1}_{n}\right) \right) }
\label{eq:app-gamma-star}
\end{equation}
\end{remark}

\subsubsection{Special cases}

Let us assume that stock returns are not sensitive to the BMG risk factor,
i.e. $\gamma _{i}=0$. It follows that:%
\begin{equation*}
x_{i}^{\star }=\frac{\sigma ^{2}\left( x^{\star }\right) }{\tilde{\sigma}%
_{i}^{2}}\left( 1-\frac{\beta _{i}}{\beta ^{\star }}\right)
\end{equation*}%
and:%
\begin{equation*}
\beta ^{\star }=\frac{1+\sigma _{\mkt}^{2}\varphi \left( \tilde{\beta},\beta
\right) }{\sigma _{\mkt}^{2}\tilde{\beta}^{\top }\mathbf{1}_{n}}
\end{equation*}%
We retrieve Equations (\ref{eq:gmv3}) and (\ref{eq:gmv4}) that have been
formulated by \citet{Scherer-2011}.\smallskip

On average, we observe that the sensitivities $\beta _{i}$ are distributed
around $1.0$, whereas the sensitivities $\gamma _{i}$ are distributed around
$0.0$. Then, we can assume the hypothesis $\mathcal{H}_{0}$ that $%
\sum_{s=1}^{n}w_{s}\gamma _{s}$ is equal to zero when $w$ is a vector that
is not correlated to $\gamma $. Under $\mathcal{H}_{0}$, we deduce that:%
\begin{equation*}
x_{i}^{\star }=\frac{\sigma ^{2}\left( x^{\star }\right) }{\tilde{\sigma}%
_{i}^{2}}\left( 1-\frac{\beta _{i}}{\beta ^{\star }}-\frac{\gamma _{i}}{%
\gamma ^{\star }}\right)
\end{equation*}%
and:%
\begin{equation*}
\beta ^{\star }=\frac{1+\sigma _{\mkt}^{2}\varphi \left( \tilde{\beta},\beta
\right) +\sigma _{\bmg}^{2}\varphi \left( \tilde{\gamma},\gamma \right)
+\sigma _{\mkt}^{2}\sigma _{\bmg}^{2}\varphi \left( \tilde{\beta},\beta \right)
\varphi \left( \tilde{\gamma},\gamma \right) }{\sigma _{\mkt}^{2}\left(
1+\sigma _{\bmg}^{2}\varphi \left( \tilde{\gamma},\gamma \right) \right)
\tilde{\beta}^{\top }\mathbf{1}_{n}}
\end{equation*}%
and:%
\begin{equation*}
\gamma ^{\star }=\frac{1+\sigma _{\mkt}^{2}\varphi \left( \tilde{\beta},\beta
\right) +\sigma _{\bmg}^{2}\varphi \left( \tilde{\gamma},\gamma \right)
+\sigma _{\mkt}^{2}\sigma _{\bmg}^{2}\varphi \left( \tilde{\beta},\beta \right)
\varphi \left( \tilde{\gamma},\gamma \right) }{\sigma _{\bmg}^{2}\left(
1+\sigma _{\mkt}^{2}\varphi \left( \tilde{\beta},\beta \right) \right) \tilde{%
\gamma}^{\top }\mathbf{1}_{n}}
\end{equation*}

\subsubsection{Extension to the long-only minimum variance portfolio}

The extension to the long-only case follows the semi-formal proof formulated
by \citet{Clarke-2011}. We note $\mathcal{I}_{u}=\left\{ 1,\ldots ,n\right\} $
as the investment universe. We consider the long-only minimum variance (MV)
portfolio, which corresponds to the optimization program:
\begin{eqnarray}
x^{\star } &=&\arg \min \frac{1}{2}x^{\top }\Sigma x \label{eq:app-mv1} \\
&\text{s.t.}&\left\{
\begin{array}{l}
\mathbf{1}_{n}^{\top }x=1 \\
x\geq \mathbf{0}_{n}%
\end{array}%
\right.   \notag
\end{eqnarray}%
The associated Lagrange function is then equal to:%
\begin{equation*}
\mathcal{L}\left( x;\lambda _{0},\lambda \right) =\frac{1}{2}x^{\top }\Sigma
x-\lambda _{0}\left( \mathbf{1}_{n}^{\top }x-1\right) -\lambda ^{\top
}\left( x-\mathbf{0}_{n}\right)
\end{equation*}%
The first-order condition is:%
\begin{equation*}
\frac{\partial \,\mathcal{L}\left( x;\lambda _{0},\lambda \right) }{\partial
\,x}=\Sigma x-\lambda _{0}\mathbf{1}_{n}-\lambda =\mathbf{0}_{n}
\end{equation*}%
whereas the Kuhn-Tucker conditions are $\min \left( \lambda _{i},x_{i}\right)
=0$ for $i=1,\ldots ,n$. The optimal solution is given by:
\begin{equation*}
x^{\star }=\Sigma ^{-1}\left( \lambda _{0}\mathbf{1}_{n}+\lambda \right)
\end{equation*}%
We deduce that\footnote{%
Because we have:%
\begin{eqnarray*}
\mathbf{1}_{n}^{\top }x=1 &\Leftrightarrow &\lambda _{0}\mathbf{1}_{n}^{\top
}\Sigma ^{-1}\mathbf{1}_{n}+\mathbf{1}_{n}^{\top }\Sigma ^{-1}\lambda =1 \\
&\Leftrightarrow &\lambda _{0}=\frac{1-\mathbf{1}_{n}^{\top }\Sigma
^{-1}\lambda }{\mathbf{1}_{n}^{\top }\Sigma ^{-1}\mathbf{1}_{n}}
\end{eqnarray*}%
}:%
\begin{equation*}
x_{mv}=x_{gmv}+\delta ^{+}
\end{equation*}%
where:%
\begin{equation*}
x_{gmv}=\frac{\Sigma ^{-1}\mathbf{1}_{n}}{\mathbf{1}_{n}^{\top }\Sigma ^{-1}%
\mathbf{1}_{n}}
\end{equation*}%
and:%
\begin{equation*}
\delta ^{+}=\left( I_{n}-\frac{\Sigma ^{-1}\mathbf{1}_{n}\mathbf{1}%
_{n}^{\top }}{\mathbf{1}_{n}^{\top }\Sigma ^{-1}\mathbf{1}_{n}}\right)
\Sigma ^{-1}\lambda
\end{equation*}%
Therefore, imposing the long-only constraint is equivalent to adding a
correction term $\delta ^{+}$ to the GMV portfolio $x_{gmv}^{\star }$ in
order to satisfy the constraint $x_{i}\geq 0$.\smallskip

Let us assume that we know the assets that make up the long-only MV
portfolio. We note $\mathcal{I}_{u}^{\prime }\subset \mathcal{I}_{u}$ as this
constrained investment universe. If we restrict the analysis to the set
$\mathcal{I}_{u}^{\prime }$, the GMV portfolio is equal to:
\begin{equation*}
\tilde{x}_{gmv}=\frac{\tilde{\Sigma}^{-1}\mathbf{1}_{n^{\prime }}}{\mathbf{1}%
_{n^{\prime }}^{\top }\tilde{\Sigma}^{-1}\mathbf{1}_{n^{\prime }}}
\end{equation*}%
where $\tilde{\Sigma}$ is the submatrix of $\Sigma $ corresponding to the
restricted universe and $n^{\prime }=\limfunc{card}\mathcal{I}_{u}^{\prime }$
is the number of assets that belong to the long-only MV portfolio. By
construction, we have the equivalence between long-only MV and restricted GMV
portfolios:
\begin{equation*}
x_{mv}\equiv \tilde{x}_{gmv}
\end{equation*}%
It follows that:%
\begin{equation}
x_{i}^{\star }=\left\{
\begin{array}{ll}
\dfrac{\sigma ^{2}\left( x^{\star }\right) }{\tilde{\sigma}_{i}^{2}}\left( 1-%
\dfrac{\beta _{i}}{\beta ^{\star }}-\dfrac{\gamma _{i}}{\gamma ^{\star }}%
\right)  & \text{if }\dfrac{\beta _{i}}{\beta ^{\star }}+\dfrac{\gamma _{i}}{%
\gamma ^{\star }}\leq 1 \\
0 & \text{otherwise}%
\end{array}%
\right. \label{eq:app-mv2}
\end{equation}%
However, contrary to the GMV case, the threshold values are endogenous and
not exogenous:
\begin{equation}
\beta ^{\star }=\frac{1+\sum_{i\in \mathcal{I}_{u}^{\prime }}\left( \sigma
_{\mkt}^{2}\tilde{\beta}_{i}\beta _{i}+\sigma _{\bmg}^{2}\tilde{\gamma}%
_{i}\gamma _{i}\right) +\sigma _{\mkt}^{2}\sigma _{\bmg}^{2}\left( \sum_{i\in
\mathcal{I}_{u}^{\prime }}\tilde{\beta}_{i}\beta _{i}\sum_{i\in \mathcal{I}%
_{u}^{\prime }}\tilde{\gamma}_{i}\gamma _{i}-\left( \sum_{i\in \mathcal{I}%
_{u}^{\prime }}\tilde{\beta}_{i}\gamma _{i}\right) ^{2}\right) }{\sigma
_{\mkt}^{2}\left( \sum_{i\in \mathcal{I}_{u}^{\prime }}\tilde{\beta}_{i}+\sigma
_{\bmg}^{2}\left( \sum_{i\in \mathcal{I}_{u}^{\prime }}\tilde{\gamma}%
_{i}\gamma _{i}\sum_{i\in \mathcal{I}_{u}^{\prime }}\tilde{\beta}%
_{i}-\sum_{i\in \mathcal{I}_{u}^{\prime }}\tilde{\beta}_{i}\gamma
_{i}\sum_{i\in \mathcal{I}_{u}^{\prime }}\tilde{\gamma}_{i}\right) \right) }
\label{eq:app-beta-star2}
\end{equation}%
and:%
\begin{equation}
\gamma ^{\star }=\frac{1+\sum_{i\in \mathcal{I}_{u}^{\prime }}\left( \sigma
_{\mkt}^{2}\tilde{\beta}_{i}\beta _{i}+\sigma _{\bmg}^{2}\tilde{\gamma}%
_{i}\gamma _{i}\right) +\sigma _{\mkt}^{2}\sigma _{\bmg}^{2}\left( \sum_{i\in
\mathcal{I}_{u}^{\prime }}\tilde{\beta}_{i}\beta _{i}\sum_{i\in \mathcal{I}%
_{u}^{\prime }}\tilde{\gamma}_{i}\gamma _{i}-\left( \sum_{i\in \mathcal{I}%
_{u}^{\prime }}\tilde{\beta}_{i}\gamma _{i}\right) ^{2}\right) }{\sigma
_{\bmg}^{2}\left( \sum_{i\in \mathcal{I}_{u}^{\prime }}\tilde{\gamma}%
_{i}+\sigma _{\mkt}^{2}\left( \sum_{i\in \mathcal{I}_{u}^{\prime }}\tilde{\beta}%
_{i}\beta _{i}\sum_{i\in \mathcal{I}_{u}^{\prime }}\tilde{\gamma}%
_{i}-\sum_{i\in \mathcal{I}_{u}^{\prime }}\tilde{\beta}_{i}\gamma
_{i}\sum_{i\in \mathcal{I}_{u}^{\prime }}\tilde{\beta}_{i}\right) \right) }
\label{eq:app-gamma-star2}
\end{equation}%
Indeed, we first need to determine $\mathcal{I}_{u}^{\prime }$ in order to
calculate $\beta ^{\star }$ and $\gamma ^{\star }$.

\subsection{Analysis of the tracking error optimization problem}
\label{appendix:te-bmg}

\subsubsection{General formulation of the optimization program}

We consider the following optimization program:
\begin{eqnarray}
x^{\star } &=&\arg \min \frac{1}{2}\left( x-b\right) ^{\top }\Sigma
\left(
x-b\right)   \label{eq:app-te1} \\
&\text{s.t.}&\mathbf{1}_{n}^{\top }x=1  \notag
\end{eqnarray}%
The associated Lagrange function is then equal to:%
\begin{eqnarray*}
\mathcal{L}\left( x;\lambda _{0}\right)  &=&\frac{1}{2}x^{\top
}\Sigma
x-x^{\top }\Sigma b-\lambda _{0}\left( \mathbf{1}_{n}^{\top }x-1\right)  \\
&=&\frac{1}{2}x^{\top }\Sigma x-x^{\top }\left( \Sigma b+\lambda _{0}\mathbf{%
1}_{n}\right) +\lambda _{0}
\end{eqnarray*}%
We deduce that the first-order condition is:%
\begin{equation*}
\frac{\partial \,\mathcal{L}\left( x;\lambda _{0}\right) }{\partial \,x}%
=\Sigma x-\left( \Sigma b+\lambda _{0}\mathbf{1}_{n}\right)
=\mathbf{0}_{n}
\end{equation*}%
Since we have $x=\Sigma ^{-1}\left( \Sigma b+\lambda
_{0}\mathbf{1}_{n}\right) $ and $\mathbf{1}_{n}^{\top }x=1$, it follows that:
\begin{eqnarray*}
\mathbf{1}_{n}^{\top }x=1 &\Leftrightarrow &\mathbf{1}_{n}^{\top
}\Sigma
^{-1}\left( \Sigma b+\lambda _{0}\mathbf{1}_{n}\right) =1 \\
&\Leftrightarrow &1+\lambda _{0}\left( \mathbf{1}_{n}^{\top }\Sigma ^{-1}%
\mathbf{1}_{n}\right) =1 \\
&\Leftrightarrow &\lambda _{0}=0
\end{eqnarray*}%
and we obtain the trivial solution $x^{\star }=b$. We notice that this
solution remains valid if we introduce long-only constraints $x\geq
\mathbf{0}_{n}$ because the Kuhn-Tucker conditions $\min \left( \lambda
_{i},x_{i}\right) =0$ are already satisfied.\smallskip

We now consider the optimization program:%
\begin{eqnarray}
x^{\star } &=&\arg \min \frac{1}{2}\left( x-b\right) ^{\top }\Sigma
\left(
x-b\right)   \label{eq:app-te2} \\
&\text{s.t.}&\left\{
\begin{array}{l}
\mathbf{1}_{n}^{\top }x=1 \\
x\geq \mathbf{0}_{n} \\
\beta _{\bmg}^{\top }x\leq \beta _{\bmg}^{+}%
\end{array}%
\right.   \notag
\end{eqnarray}%
The associated Lagrange function is then equal to:%
\begin{eqnarray*}
\mathcal{L}\left( x;\lambda _{0}, \lambda, \lambda _{\bmg}\right)  &=&\frac{1}{2}x^{\top
}\Sigma x-x^{\top }\Sigma b-\lambda _{0}\left( \mathbf{1}_{n}^{\top
}x-1\right) -\lambda ^{\top }x+\lambda _{\bmg}\left( \beta
_{\bmg}^{\top}x-\beta _{\bmg}^{+}\right)  \\
&=&\frac{1}{2}x^{\top }\Sigma x-x^{\top }\left( \Sigma b+\lambda _{0}\mathbf{%
1}_{n}+\lambda -\lambda _{\bmg}\beta _{\bmg}\right) +\lambda
_{0}-\lambda _{\bmg}\beta _{\bmg}^{+}
\end{eqnarray*}%
where $\lambda _{0}\geq 0$ is the Lagrange multiplier of the equality
constraint $\mathbf{1}_{n}^{\top }x=1$, $\lambda \geq \mathbf{0}_{n}$ is the
vector of Lagrange multipliers of the bound constraints $x\geq \mathbf{0}_{n}
$ and $\lambda _{\bmg}\geq 0$ is the Lagrange multiplier of the inequality
constraint $\beta _{\bmg}^{\top }x\leq \beta _{\bmg}^{+}$. We deduce that the
first-order condition is:
\begin{equation*}
\frac{\partial \,\mathcal{L}\left( x;\lambda _{0}, \lambda, \lambda _{\bmg}\right) }{\partial \,x}%
=\Sigma x-\left( \Sigma b+\lambda _{0}\mathbf{1}_{n}+\lambda -\lambda _{%
\bmg}\beta _{\bmg}\right) =\mathbf{0}_{n}
\end{equation*}%
implying that:%
\begin{eqnarray}
x^{\star } &=&\Sigma ^{-1}\left( \Sigma b+\lambda
_{0}\mathbf{1}_{n}+\lambda
-\lambda _{\bmg}\beta _{\bmg}\right)   \notag \\
&=&b+\lambda _{0}\Sigma ^{-1}\mathbf{1}_{n}+\Sigma ^{-1}\lambda
-\lambda _{\bmg}\Sigma ^{-1}\beta _{\bmg}  \notag \\
&=&x^{\star }\left( b\right) +x^{\star }\left( \beta _{\bmg}\right)
\label{eq:app-te3}
\end{eqnarray}%
where\ $x^{\star }\left( b\right) =b+\lambda _{0}\Sigma ^{-1}\mathbf{1}_{n}$
and $x^{\star }\left( \beta _{\bmg}\right) =\Sigma ^{-1}\lambda -\lambda
_{\bmg}\Sigma ^{-1}\beta _{\bmg}$. We notice that Portfolio $x^{\star }\left(
b\right) $ only depends on the benchmark $b$, the covariance matrix $\Sigma $
and the Lagrange multiplier $\lambda _{0}$. It does not depend on the BMG
sensitivities, and may be considered as a constant term. Moreover, we can set
$\lambda _{0}=0$, because the constraint $\mathbf{1}_{n}^{\top }x=1$ has no
impact on the solution $x^{\star }$, but only scales the optimal portfolio
such as the weights sum up to 100\%. In this case, we have $x^{\star }\left(
b\right) =b$ and $x^{\star }\left( \beta _{\bmg}\right) $ is the long/short
portfolio $x^{\star }-b$ that depends on several parameters:
\begin{equation}
\Delta = x^{\star }-b \approx \Sigma ^{-1}\lambda -\lambda _{\bmg}\Sigma
^{-1}\beta _{\bmg}  \label{eq:app-te4}
\end{equation}%
At first sight, these parameters are the covariance matrix $\Sigma $, the
vector $\lambda $ of Lagrange coefficients, the Lagrange coefficient $\lambda
_{\bmg}$ and the vector $\beta _{\bmg}$ of carbon risk sensitivities. In
fact, this interpretation of Equation (\ref{eq:app-te4}) is misleading. The
two important quantities are the scaled BMG beta vector
$\breve{\beta}_{\bmg}=\Sigma ^{-1}\beta _{\bmg}$ and the threshold value
$\beta _{\bmg}^{+}$, since $\lambda $ and $\lambda _{\bmg}$ are
endogenous\footnote{The vector $\lambda $ and the scalar $\lambda _{\bmg}$
are related to these two quantities. They do not add more degrees of freedom
to the optimization problem.}.\smallskip

\subsubsection{Special cases}

Let us assume that $\Sigma =\limfunc{diag}\left( \sigma _{1}^{2},\ldots
,\sigma _{n}^{2}\right) $ is a diagonal matrix. We have%
\begin{equation}
\Delta _{i}=\frac{\lambda _{i}}{\sigma _{i}^{2}}-\lambda _{\bmg}\frac{\beta
_{\bmg,i}}{\sigma _{i}^{2}} \label{eq:te-cst1}
\end{equation}%
We notice that $\Delta _{i}$ is a decreasing function of $\beta _{\bmg,i}$.
Moreover, the Kuhn-Tucker conditions implies the following property:
\begin{equation}
x_{i}>0\Rightarrow \lambda _{i}=0\Rightarrow \left\{
\begin{array}{cc}
\Delta _{i}>0 & \text{if }\beta _{\bmg,i}<0 \\
\Delta _{i}<0 & \text{if }\beta _{\bmg,i}>0%
\end{array}%
\right. \label{eq:te-cst2}
\end{equation}

We now assume a constant correlation matrix. We have $\Sigma =\sigma \sigma
^{\top }\odot R$ where $R=C_{n}\left( \rho \right) $. \citet{Maillard-2010}
showed that $\Sigma ^{-1}=\Gamma \odot R^{-1}$ where $\Gamma
_{i,j}=\frac{1}{\sigma _{i}\sigma _{j}}$ and:
\begin{equation*}
R^{-1}=\frac{\rho \mathbf{1}_{n}\mathbf{1}_{n}^{\top }-\left( \left(
n-1\right) \rho +1\right) I_{n}}{\left( n-1\right) \rho ^{2}-\left(
n-2\right) \rho -1}
\end{equation*}%
It follows that:%
\begin{eqnarray*}
\Sigma ^{-1} &=&\frac{\rho \Gamma \odot \left( \mathbf{1}_{n}\mathbf{1}%
_{n}^{\top }\right) -\left( \left( n-1\right) \rho +1\right) \Gamma \odot
I_{n}}{\left( n-1\right) \rho ^{2}-\left( n-2\right) \rho -1} \\
&=&\rho \xi \Gamma -\left( \left( n-1\right) \rho +1\right) \xi \tilde{\Gamma%
}
\end{eqnarray*}%
where $\tilde{\Gamma}=\limfunc{diag}\left( \Gamma \right) $ is the diagonal
matrix with $\tilde{\Gamma}_{i,i}=\Gamma _{i,i}$, and:%
\begin{equation*}
\xi =\frac{1}{\left( n-1\right) \rho ^{2}-\left( n-2\right) \rho -1}
\end{equation*}
Let $u\in \mathbb{R}^{n}$ be a vector. We deduce that:%
\begin{eqnarray*}
v &=&\Sigma ^{-1}u \\
&=&\rho \xi \Gamma u-\left( \left( n-1\right) \rho +1\right) \xi \tilde{u}
\end{eqnarray*}%
where $\tilde{u}$ is a vector with $\tilde{u}_{i}=\sigma _{i}^{-2}u_{i}$.
Finally, we obtain:%
\begin{eqnarray}
v_{i} &=&\rho \xi \sum_{j=1}^{n}\Gamma _{i,j}u_{j}-\left( \left( n-1\right)
\rho +1\right) \xi \tilde{u}_{i}  \notag \\
&=&\frac{n\rho \xi }{\sigma _{i}}\bar{s}-\frac{\left( \left( n-1\right) \rho
+1\right) \xi }{\sigma _{i}}s_{i}  \label{eq:te-ccm1}
\end{eqnarray}%
where $s_{i}=\sigma _{i}^{-1}u_{i}$ and $\bar{s}=n^{-1}\sum_{j=1}^{n}s_{j}$.
If $n$ is very large and we assume that $\rho >0$, we have:
\begin{equation*}
\xi \approx \frac{1}{n\rho \left( \rho -1\right) }
\end{equation*}%
and Equation (\ref{eq:te-ccm1}) reduces to:%
\begin{equation}
v_{i}\approx \frac{1}{1-\rho }\left( \frac{s_{i}-\bar{s}}{\sigma _{i}}%
\right)   \label{eq:te-ccm2}
\end{equation}%
We only consider this case in order to simplify the expression of $\Delta
_{i}$, which is otherwise complex and does not help to interpret the impact
of the BMG constraint. We obtain:
\begin{eqnarray}
\Delta _{i} &\approx &\frac{1}{\left( 1-\rho \right) \sigma _{i}}\left(
\frac{\lambda _{i}}{\sigma _{i}}-\overline{\frac{\lambda }{\sigma }}\right) +%
\frac{\lambda _{\bmg}}{\left( 1-\rho \right) \sigma _{i}}\left( \overline{%
\frac{\beta _{\bmg}}{\sigma }}-\frac{\beta _{\bmg,i}}{\sigma _{i}}\right)
\notag \\
&=&\Delta _{i}^{1}+\Delta _{i}^{2}  \label{eq:te-ccm3}
\end{eqnarray}%
We notice that the overweight or underweight of an asset will depend on the
relative position of the statistic $\sigma _{i}^{-1}\beta _{\bmg,i}$ with
respect to its mean. If it is below the mean, the second term $\Delta
_{i}^{2}$ is positive. Generally, we would observe an overweight of asset
$i$.\smallskip

In the general case, we have:%
\begin{equation}
\Delta _{i}=\left( \Sigma ^{-1}\lambda \right) _{i}-\lambda _{\bmg}\breve{%
\beta}_{\bmg,i}  \label{q:te-gen1}
\end{equation}%
If we omit the impact of the lower bound, $\Delta _{i}$ is positive if $%
\breve{\beta}_{\bmg,i}$ is negative.

\subsubsection{Approximation of the tracking error volatility}
\label{appendix:te-approximation}

\begin{figure}[tbh]
\centering
\caption{Relationship between $\Delta _{\bmg}$ and $\lambda_{\bmg}$}
\label{fig:te10}
\figureskip
\includegraphics[width = \figurewidth, height = \figureheight]{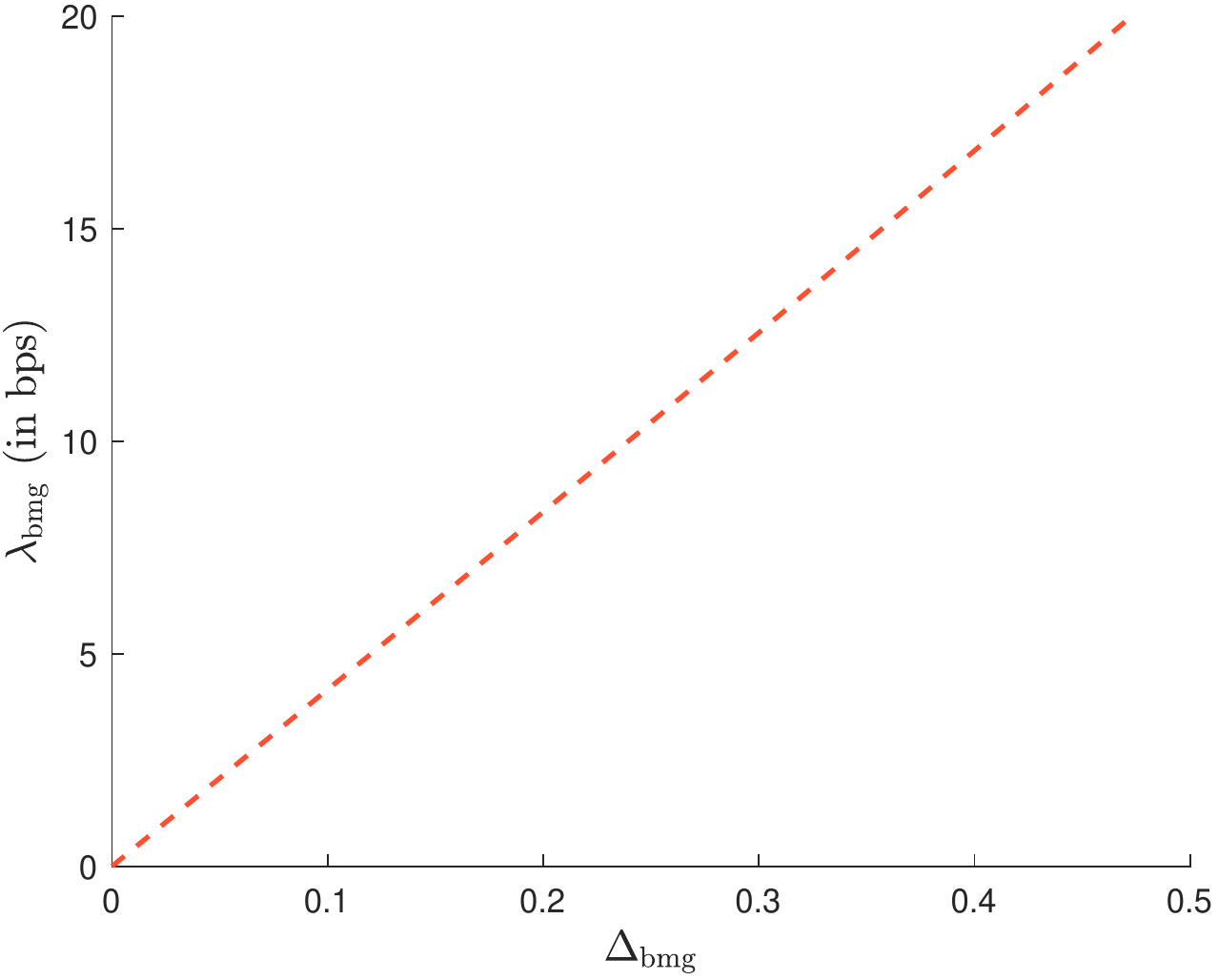}
\end{figure}

We recall that the first-order condition is\footnote{See Equation
(\ref{eq:app-te3}) on page \pageref{eq:app-te3}.} $\Sigma \left( x^{\star
}-b\right) =\lambda _{0}\mathbf{1}_{n}+\lambda -\lambda _{\bmg}\beta _{\bmg}$
where $\lambda _{0}$, $\lambda $ and $\lambda _{\bmg}$ are the Lagrange
coefficients associated with the constraints $\mathbf{1}_{n}^{\top }x=1$, $%
x\geq \mathbf{0}_{n}$ and $\beta _{\bmg}^{\top }x\leq \beta _{\bmg}^{+}$. We
deduce that:
\begin{eqnarray*}
\sigma ^{2}\left( x^{\star }\mid b\right)  &=&\left( x^{\star }-b\right)
^{\top }\Sigma \left( x^{\star }-b\right)  \\
&=&\left( x^{\star }-b\right) ^{\top }\left( \lambda _{0}\mathbf{1}%
_{n}+\lambda -\lambda _{\bmg}\beta _{\bmg}\right)  \\
&=&\left( x^{\star }-b\right) ^{\top }\lambda -\lambda _{\bmg}\left(
x^{\star }-b\right) ^{\top }\beta _{\bmg} \\
&=&\left( x^{\star }-b\right) ^{\top }\lambda +\lambda _{\bmg}\Delta _{\bmg} \\
&\approx &c\Delta _{\bmg}^{2}
\end{eqnarray*}
The last result is obtained because we have $\left( x^{\star }-b\right)
^{\top }\lambda \approx 0$ and we also notice that the Lagrange coefficient
$\lambda _{\bmg}$ is almost a linear function of $\Delta _{\bmg}$ for
reasonable values of $\Delta _{\bmg}$. For instance, we report in Figure
\ref{fig:te10} the relationship between $\Delta _{\bmg}$ and $\lambda_{\bmg}$
when the benchmark corresponds to the CW index. This explains that the
tracking error volatility is approximatively a linear function of $\Delta
_{\bmg}$.

\clearpage

\section{Additional results}

\begin{table}[h]
\centering
\caption{Statistics of factor returns}
\label{tab:carima1}
\scalebox{0.925}{
\begin{tabular}{lccccccc}
\hline
Factor & $\mu $ & $\sigma $ & $\gamma _{1}$ & $\gamma _{2}$ & $\mathcal{MDD}$ & $\mathcal{BM}$ & $\mathcal{WM}$ \\
\hline
MKT                    & ${\TsVIII}7.63\%$ &      $13.23\%$ &         $-0.34$ & $3.63$ &        $-20.4\%$ &      $10.0\%$ & $-9.5\%$ \\
SMB                    &         $-0.10\%$ & ${\TsV}4.65\%$ & ${\TsVIII}0.14$ & $2.82$ &        $-13.6\%$ & ${\TsV}3.9\%$ & $-3.1\%$ \\
HML                    &         $-1.82\%$ & ${\TsV}5.92\%$ & ${\TsVIII}0.25$ & $2.96$ &        $-20.0\%$ & ${\TsV}4.3\%$ & $-4.5\%$ \\
WML                    & ${\TsVIII}6.74\%$ & ${\TsV}8.24\%$ &         $-0.14$ & $2.92$ &        $-11.1\%$ & ${\TsV}6.3\%$ & $-6.8\%$ \\
BMG                    &         $-2.52\%$ & ${\TsV}6.41\%$ &         $-0.06$ & $3.08$ &        $-35.3\%$ & ${\TsV}4.2\%$ & $-5.3\%$ \\
\hdashline
Carbon intensity       &         $-3.09\%$ & ${\TsV}5.13\%$ & ${\TsVIII}0.14$ & $3.38$ &        $-31.1\%$ & ${\TsV}4.5\%$ & $-3.8\%$ \\
Carbon emissions exp.  &         $-4.01\%$ & ${\TsV}5.47\%$ & ${\TsVIII}0.36$ & $3.58$ &        $-36.7\%$ & ${\TsV}5.2\%$ & $-3.8\%$ \\
Carbon emissions mgmt. & ${\TsVIII}4.15\%$ & ${\TsV}3.71\%$ & ${\TsVIII}0.20$ & $2.93$ & ${\TsV}$$-4.7\%$ & ${\TsV}3.1\%$ & $-2.2\%$ \\
Carbon emissions       & ${\TsVIII}0.05\%$ & ${\TsV}4.68\%$ & ${\TsVIII}0.31$ & $4.22$ &        $-15.8\%$ & ${\TsV}5.0\%$ & $-3.6\%$ \\
Climate change         &         $-0.78\%$ & ${\TsV}4.56\%$ & ${\TsVIII}0.25$ & $2.69$ &        $-20.8\%$ & ${\TsV}3.7\%$ & $-2.7\%$ \\
Environment            &         $-0.26\%$ & ${\TsV}4.63\%$ & ${\TsVIII}0.48$ & $3.24$ &        $-17.8\%$ & ${\TsV}4.1\%$ & $-3.1\%$ \\
\hline
\end{tabular}}
\begin{flushleft}
\justify \small
The statistics $\mu $ and $\sigma $ are the annualized return and volatility, $\gamma_{1}$
and $\gamma_{2}$ are the skewness and kurtosis coefficients, $\mathcal{MDD}$
is the observed maximum drawdown, $\mathcal{BM}$ and $\mathcal{WM}$ are the
best and worst monthly returns. BMG corresponds to the Carima factor developed by \citet{Gorgen-2019}.
\end{flushleft}
\vspace*{17pt}
\centering
\caption{Correlation matrix of factor returns (in \%)}
\label{tab:factor4}
\scalebox{0.925}{
\begin{tabular}{l|*{5}{d{3.4}}}
Factor     & \multicolumn{1}{c}{MKT}  &  \multicolumn{1}{c}{SMB} &  \multicolumn{1}{c}{HML}  &  \multicolumn{1}{c}{WML}  & \multicolumn{1}{c}{BMG} \\
\hline
 Carbon intensity       &  -6.46       & 13.71        &   8.71      & -3.04      & 58.40^{***} \\
 Carbon emissions exp.  &  -6.71       & 14.95        &   4.03      & -4.03      & 64.02^{***} \\
 Carbon emissions mgmt. & -17.93^{*}   & 24.16^{**}   & -20.91^{**} & 20.93^{**} & 38.66^{***} \\
 Carbon emissions       &   1.22       & 25.85^{***}  &  -0.23      &  5.15      & 72.36^{***} \\
 Climate change         & -15.02       & 16.30^{*}    &  11.43      &  2.07      & 61.11^{***} \\
 Environment            & -28.20^{***} & 21.16^{**}   &  -0.33      &  3.70      & 68.53^{***} \\
\end{tabular}}
\begin{flushleft}
\justify \small
BMG corresponds to the Carima factor developed by \citet{Gorgen-2019}.
\end{flushleft}
\vspace*{20pt}
\centering
\caption{Comparison of cross-section regressions (in \%)}
\label{tab:factor7}
\scalebox{0.925}{
\begin{tabular}{lc*{3}{d{2.1}}:c*{3}{d{2.1}}}
\hline
 & Adjusted $\mathfrak{R}^{2}$ & \multicolumn{3}{c:}{$F$-test}
 & Adjusted $\mathfrak{R}^{2}$ & \multicolumn{3}{c}{$F$-test} \\
 & difference                  & \multicolumn{1}{c}{$10\%$} & \multicolumn{1}{c}{$5\%$} & \multicolumn{1}{c:}{$1\%$}
 & difference                  & \multicolumn{1}{c}{$10\%$} & \multicolumn{1}{c}{$5\%$} & \multicolumn{1}{c}{$1\%$} \\
\hline
& \multicolumn{4}{c:}{CAPM vs MKT+BMG} & \multicolumn{4}{c}{FF vs FF+BMG} \\
Carima                 & 1.74 & 21.2 & 15.6 &  9.2 & 1.73 & 22.5 & 17.5 &  9.7 \\
Carbon intensity       & 1.77 & 22.0 & 16.1 &  9.0 & 1.80 & 23.3 & 17.5 & 10.6 \\
Carbon emissions exp.  & 1.79 & 21.2 & 16.3 &  9.8 & 1.85 & 23.0 & 18.2 & 10.5 \\
Carbon emissions mgmt. & 1.64 & 23.3 & 17.8 &  9.5 & 1.64 & 25.1 & 19.4 & 10.7 \\
Carbon emissions       & 2.00 & 24.2 & 18.8 & 11.7 & 2.06 & 25.7 & 20.1 & 12.7 \\
Climate change         & 1.58 & 20.9 & 15.9 &  8.6 & 1.48 & 21.2 & 16.5 &  9.4 \\
Environment            & 1.63 & 21.9 & 17.1 & 10.5 & 1.60 & 23.0 & 18.1 & 10.2 \\
\hline
\end{tabular}}
\end{table}


%
\begin{figure}[tbph]
\centering
\caption{Cumulative performance of the BMG factor}
\label{fig:carima1}
\includegraphics[width = \figurewidth, height = \figureheight]{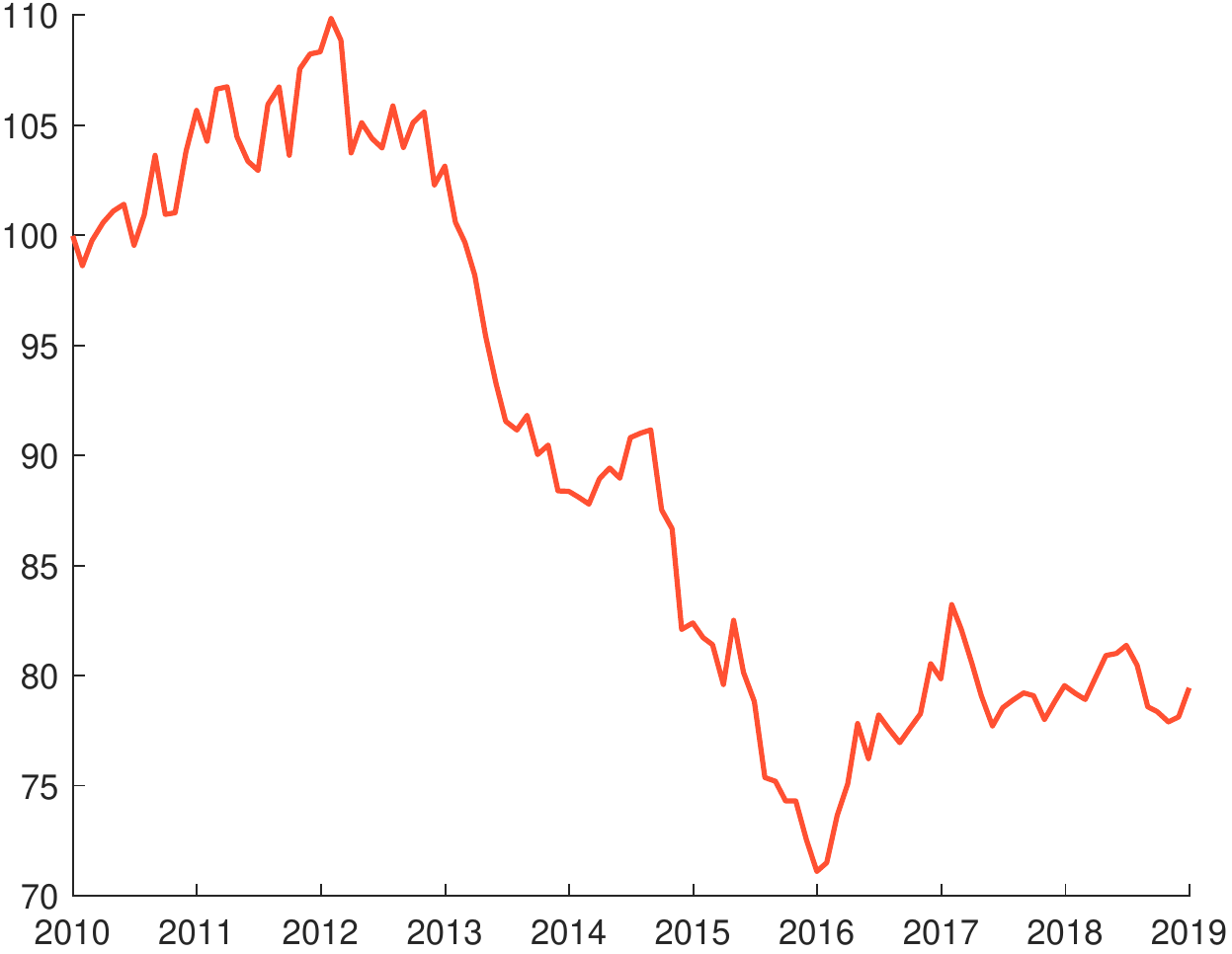}
\begin{flushleft}
{\small \textit{Source}: \citet{Gorgen-2019}.}
\end{flushleft}
\end{figure}

\begin{figure}[tbph]
\centering
\caption{Box plots of the carbon sensitivities by sector and region}
\label{fig:carima5b}
\includegraphics[width = \figurewidth, height = \figureheight]{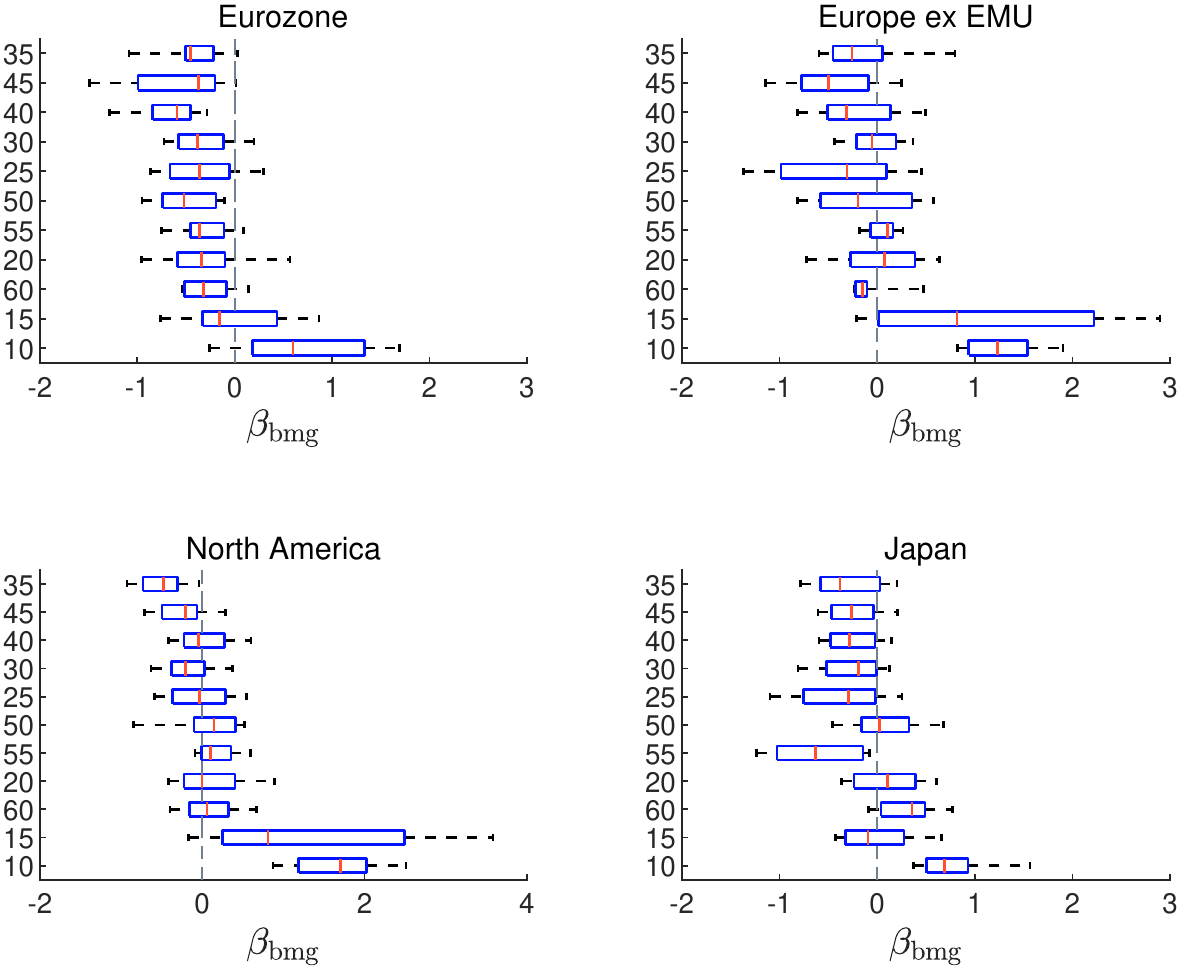}
\justify \small The sorted GICS classification is: energy (10), materials (15), industrials (20),
consumer discretionary (25), consumer staples (30), health care (35), financials (40),
information technology (45), communication services (50), utilities (55) and real estate (60). The box
plots provide the median, the quartiles and the $10\%$ and $90\%$ quantiles of the carbon beta.
\end{figure}

\begin{figure}[tbph]
\centering
\caption{Dynamics of the carbon beta of sorted-portfolios}
\label{fig:carima14b}
\includegraphics[width = \figurewidth, height = \figureheight]{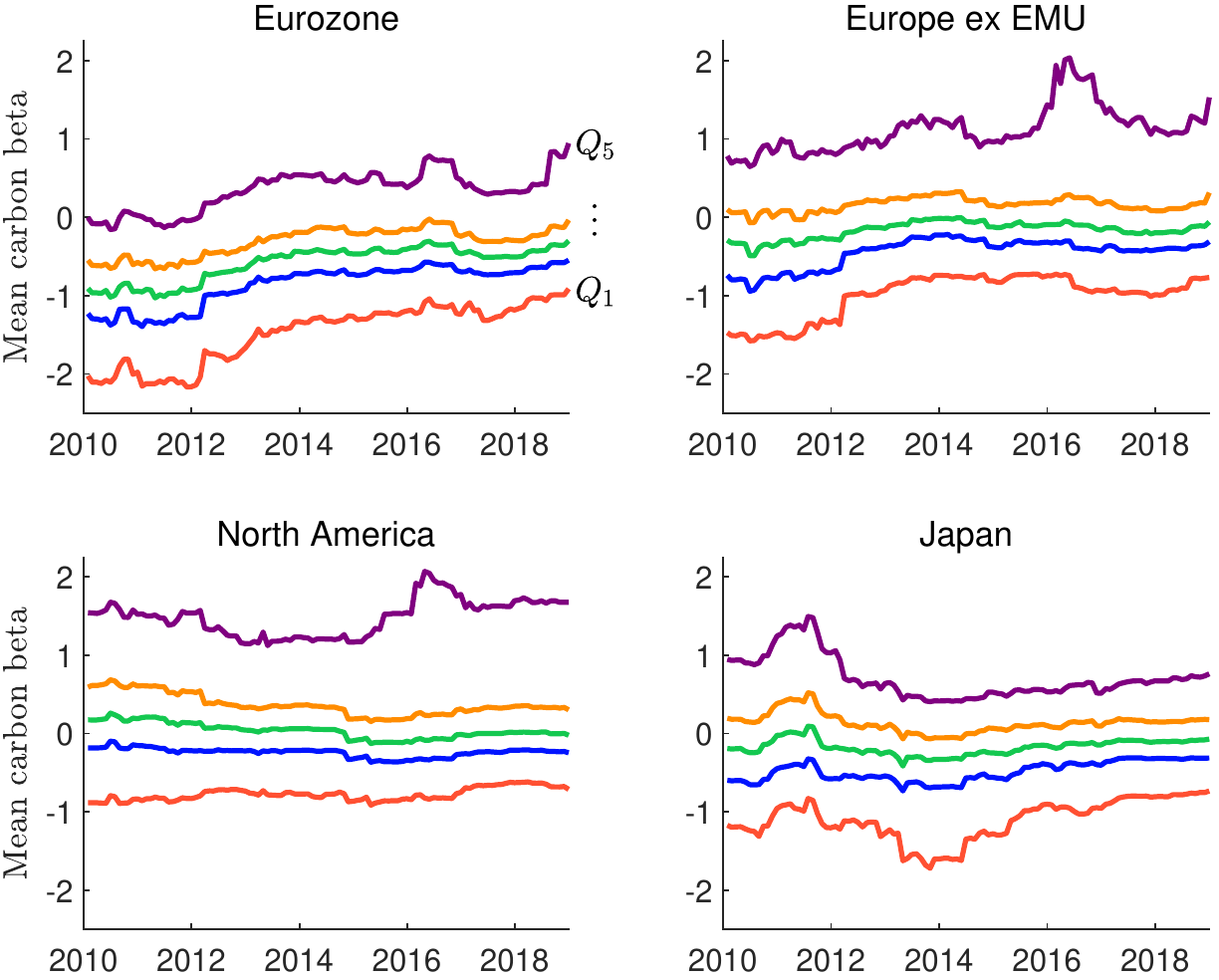}
\end{figure}

\begin{figure}[tbph]
\centering
\caption{Dynamics of the average absolute carbon risk $\left\vert \beta\right\vert_{\bmg,\mathcal{S}}\left(t\right)$ by sector}
\label{fig:carima15d}
\includegraphics[width = \figurewidth, height = \figureheight]{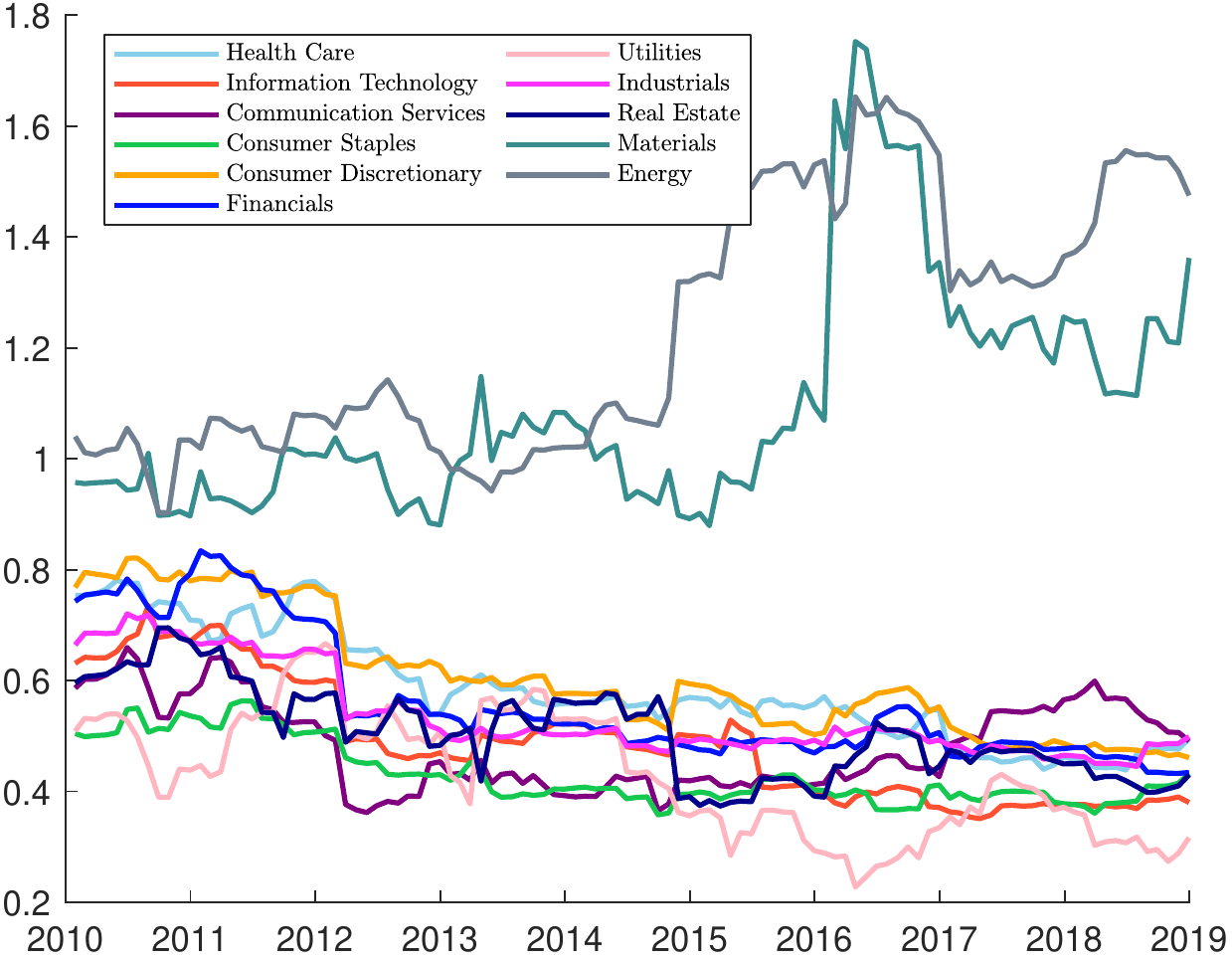}
\end{figure}

\begin{figure}[tbph]
\centering
\caption{Scatter plot of MKT and BMG sensitivities}
\label{fig:carima4}
\figureskip
\includegraphics[width = \figurewidth, height = \figureheight]{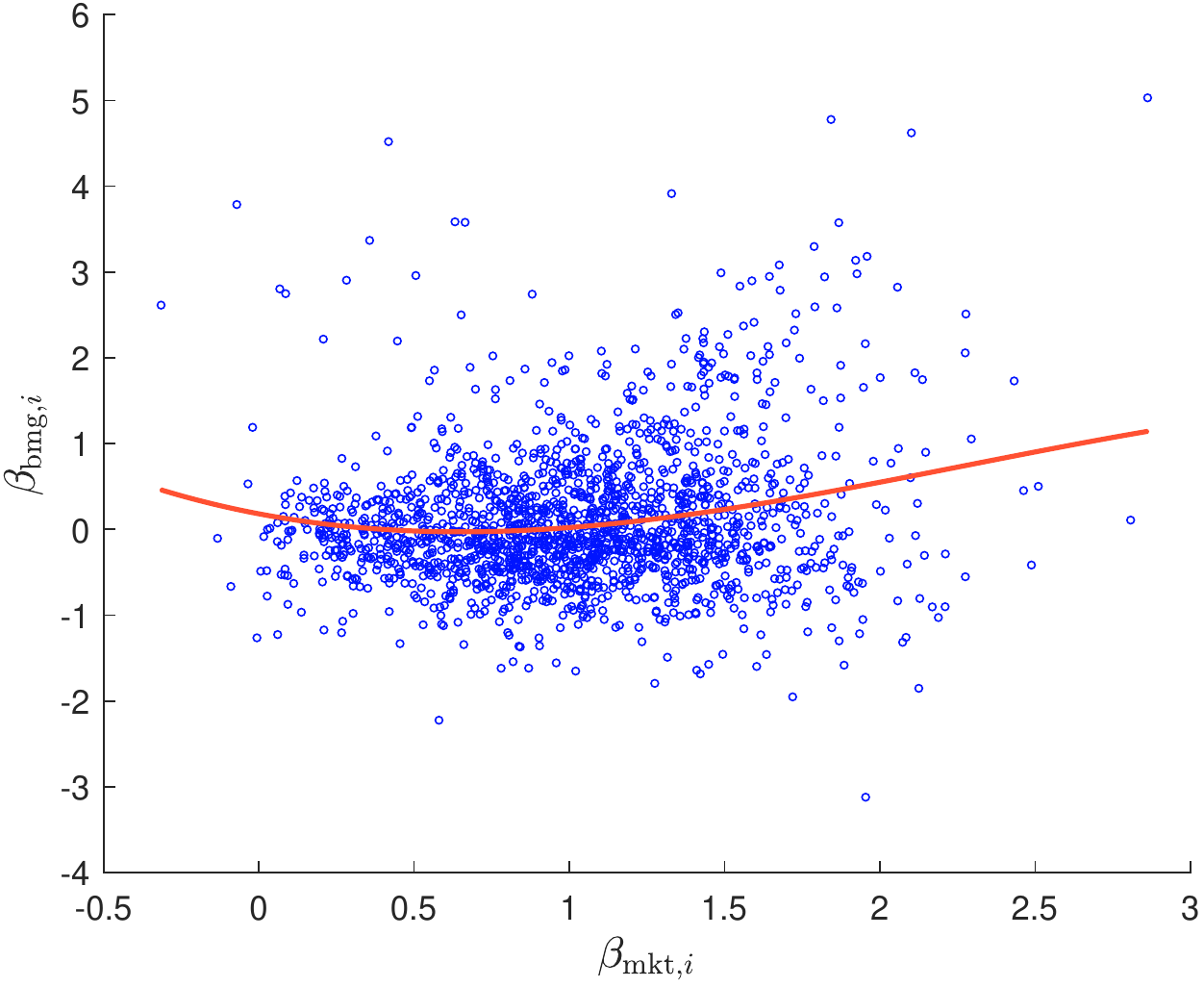}
\end{figure}

\begin{figure}[tbph]
\centering
\caption{Box plots of the carbon sensitivities by sector and factor}
\label{fig:factor10b}
\includegraphics[width = \figurewidth, height = \figureheight]{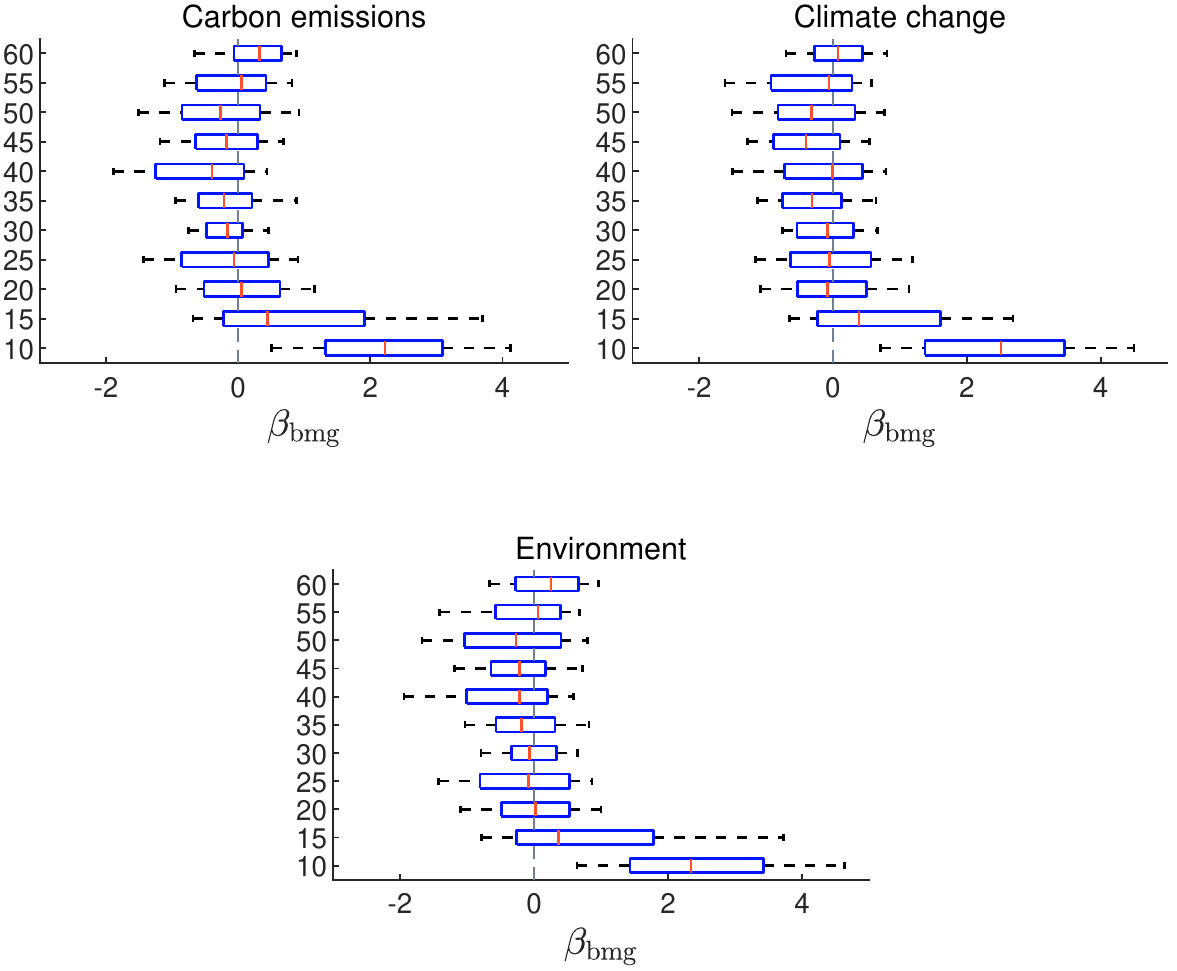}
\justify \small The sorted GICS classification is: energy (10), materials (15), industrials (20),
consumer discretionary (25), consumer staples (30), health care (35), financials (40),
information technology (45), communication services (50), utilities (55) and real estate (60).  The box
plots provide the median, the quartiles and the $10\%$ and $90\%$ quantiles of the carbon beta.
\end{figure}

\begin{figure}[tbph]
\centering
\caption{$\WACI$ of the constrained MV portfolios ($\intensity^{+} = 315$)}
\label{fig:gmv11}
\figureskip
\includegraphics[width = \figurewidth, height = \figureheight]{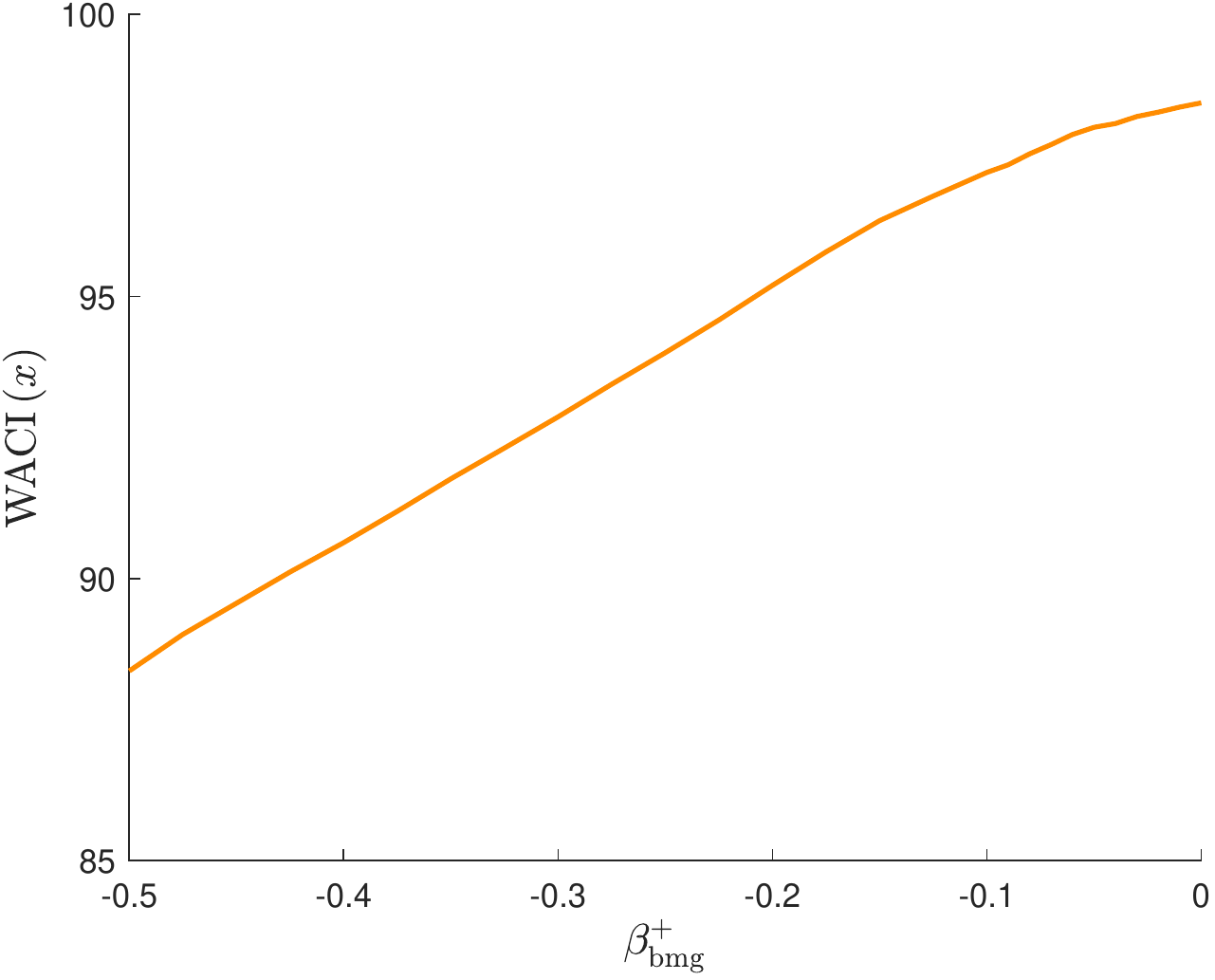}
\end{figure}

\begin{figure}[tbph]
\centering
\caption{Volatility of the constrained MV portfolios}
\label{fig:gmv13}
\figureskip
\includegraphics[width = \figurewidth, height = \figureheight]{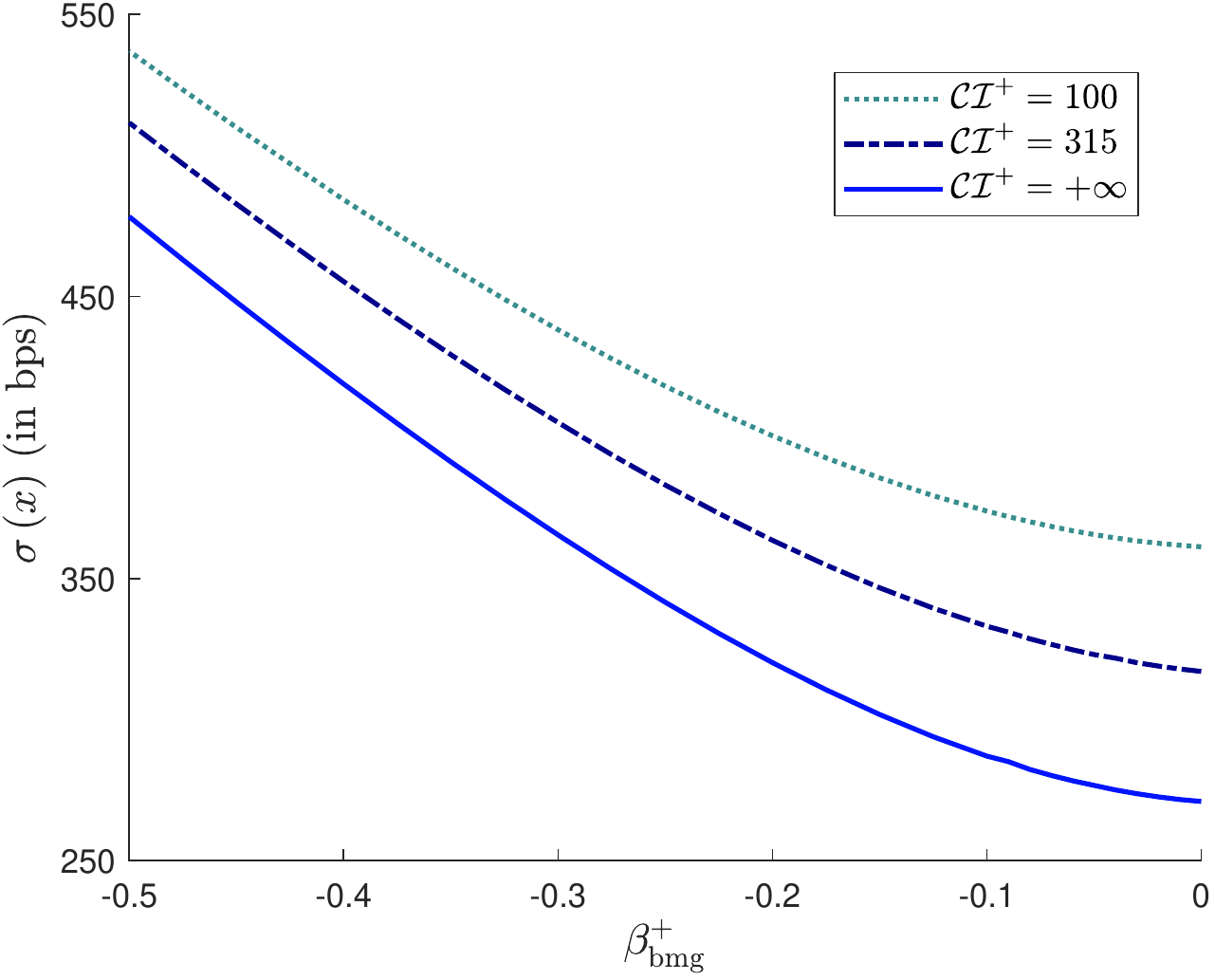}
\justify \small Based on the MKT+BMG model, the monthly volatility of the capitalisation-weighted MSCI World index is $13.25\%$.
\end{figure}

\begin{figure}[tbph]
\centering
\caption{Market and fundamental measures of carbon risk}
\label{fig:factor20}
\includegraphics[width = \figurewidth, height = \figureheight]{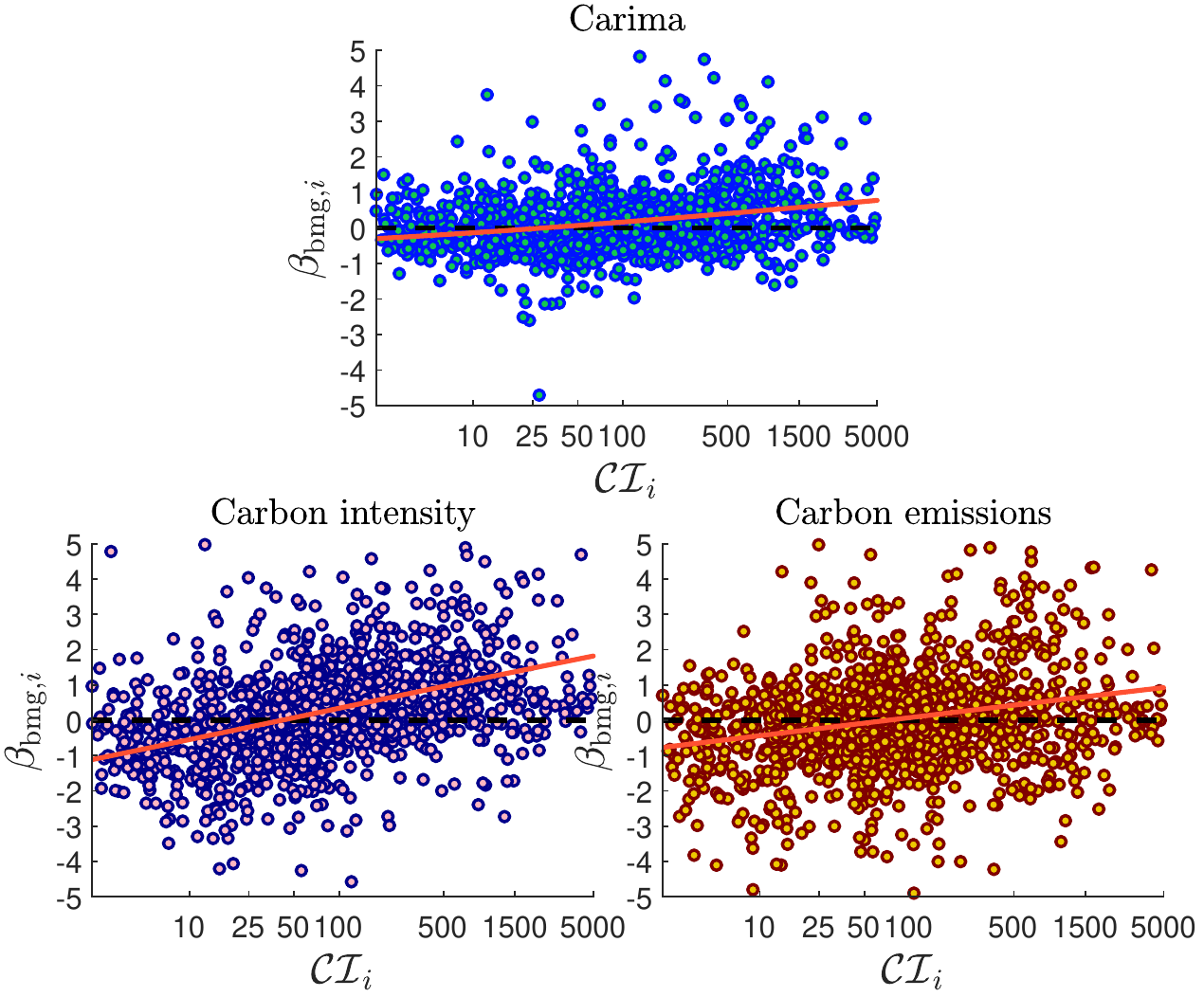}
\end{figure}

\begin{figure}[tbph]
\centering
\caption{Scatter plot of CW weights and BMG sensitivities}
\label{fig:te8}
\figureskip
\includegraphics[width = \figurewidth, height = \figureheight]{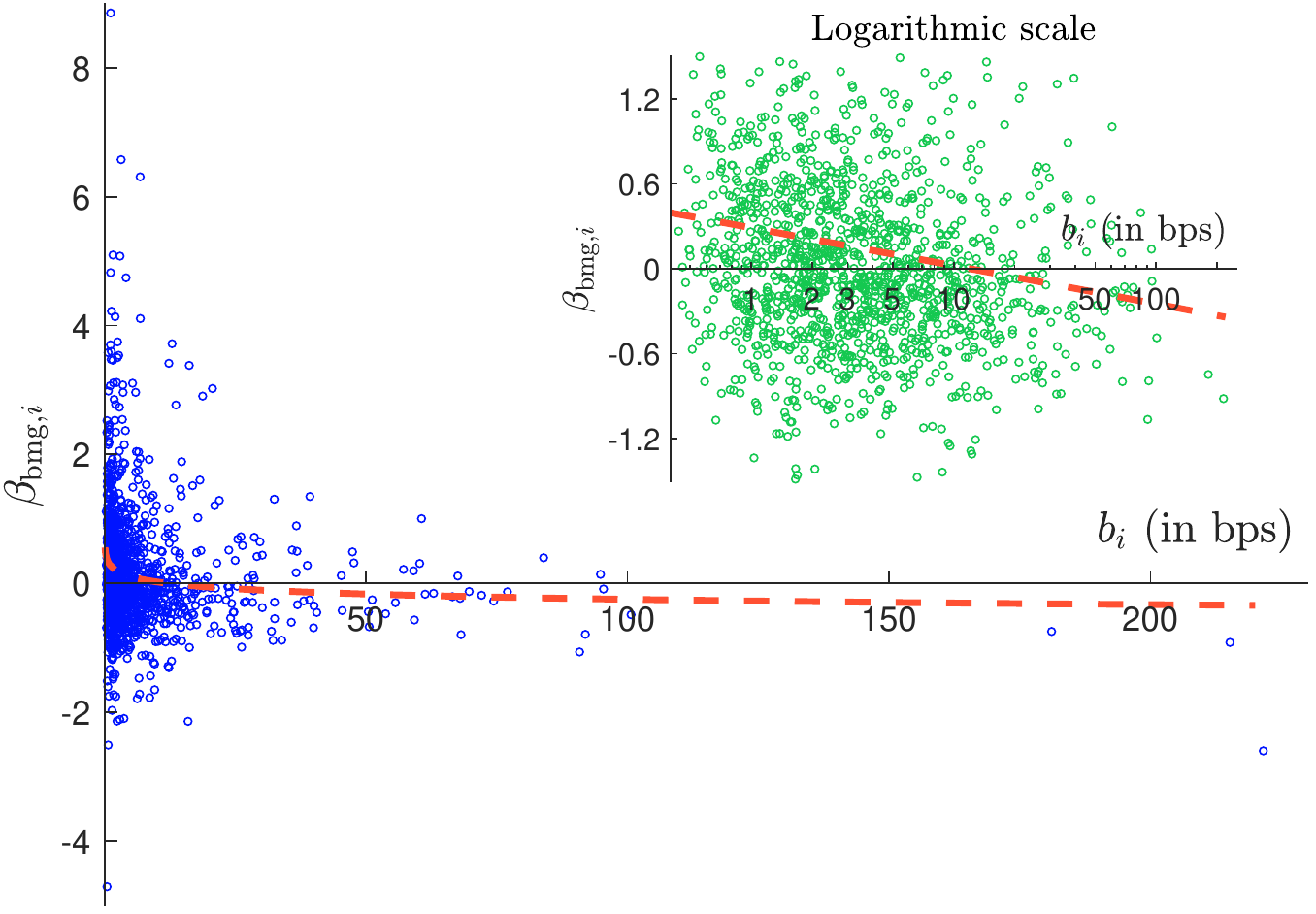}
\end{figure}

\begin{figure}[tbph]
\centering
\caption{Relationship between $\beta_{\bmg,i}$ and $\Delta_{i}=x_{i}^{\star }-b_{i}$ for the CW benchmark}
\label{fig:te6}
\includegraphics[width = \figurewidth, height = \figureheight]{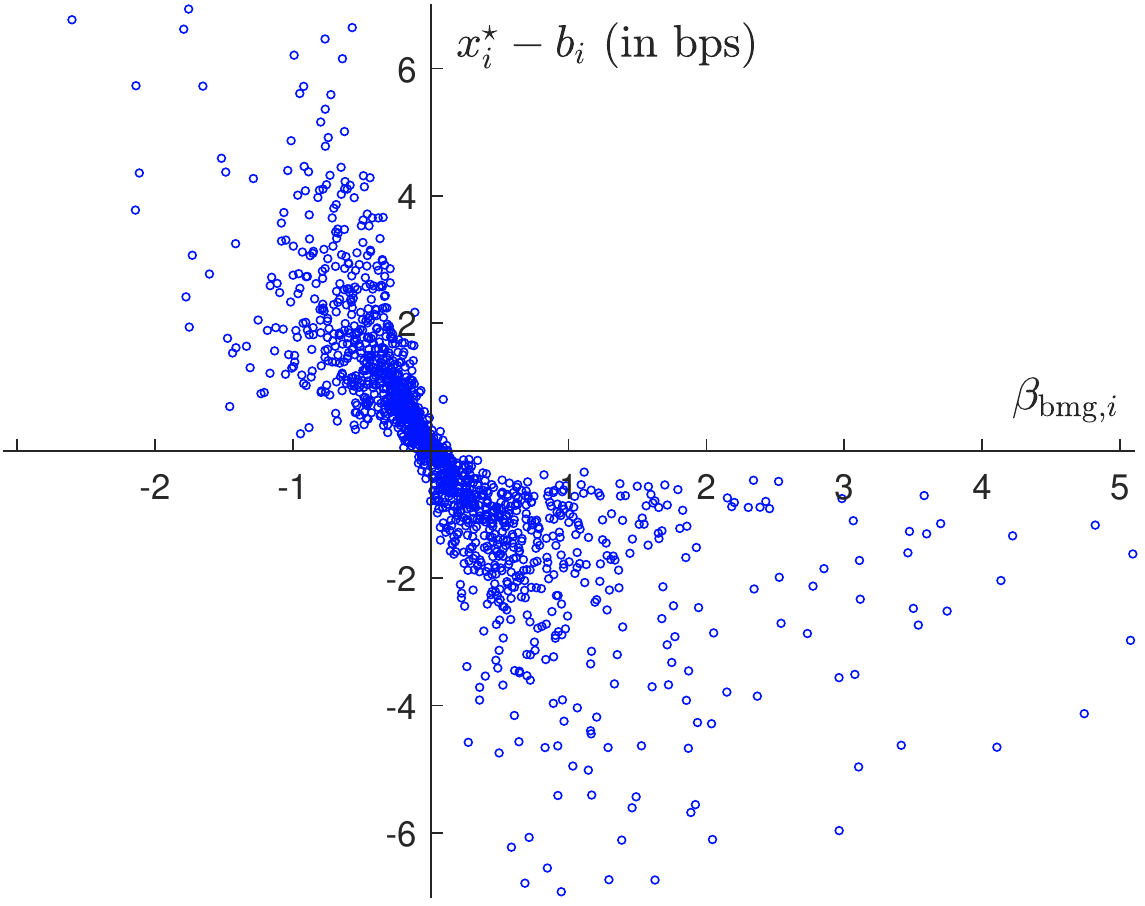}
\end{figure}

\begin{figure}[tbph]
\centering
\caption{Relationship between $\breve{\beta}_{\bmg,i}$ and $\Delta _{i}=x_{i}^{\star}-b_{i}$ for the CW benchmark}
\label{fig:te7}
\includegraphics[width = \figurewidth, height = \figureheight]{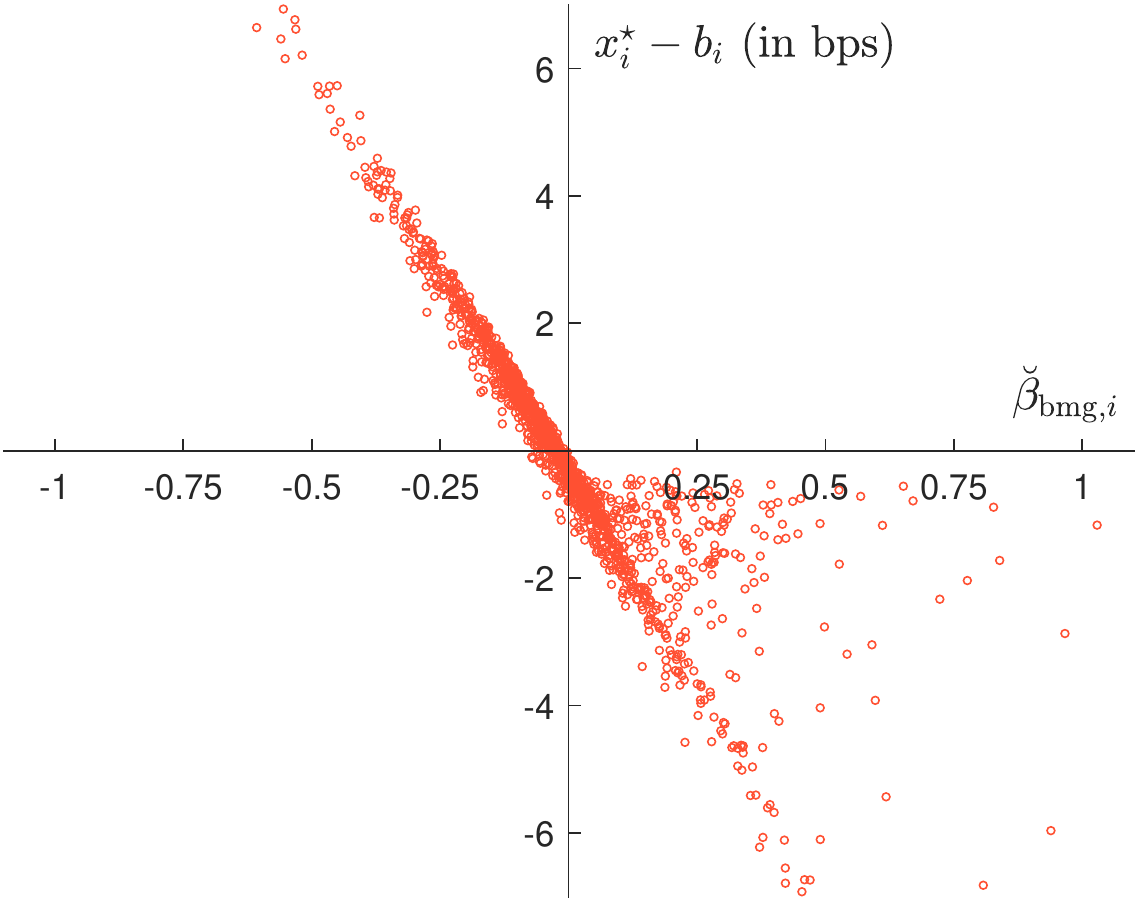}
\end{figure}

\begin{figure}[tbph]
\centering
\caption{Solution of the order-statistic optimization problem}
\label{fig:te11}
\includegraphics[width = \figurewidth, height = \figureheight]{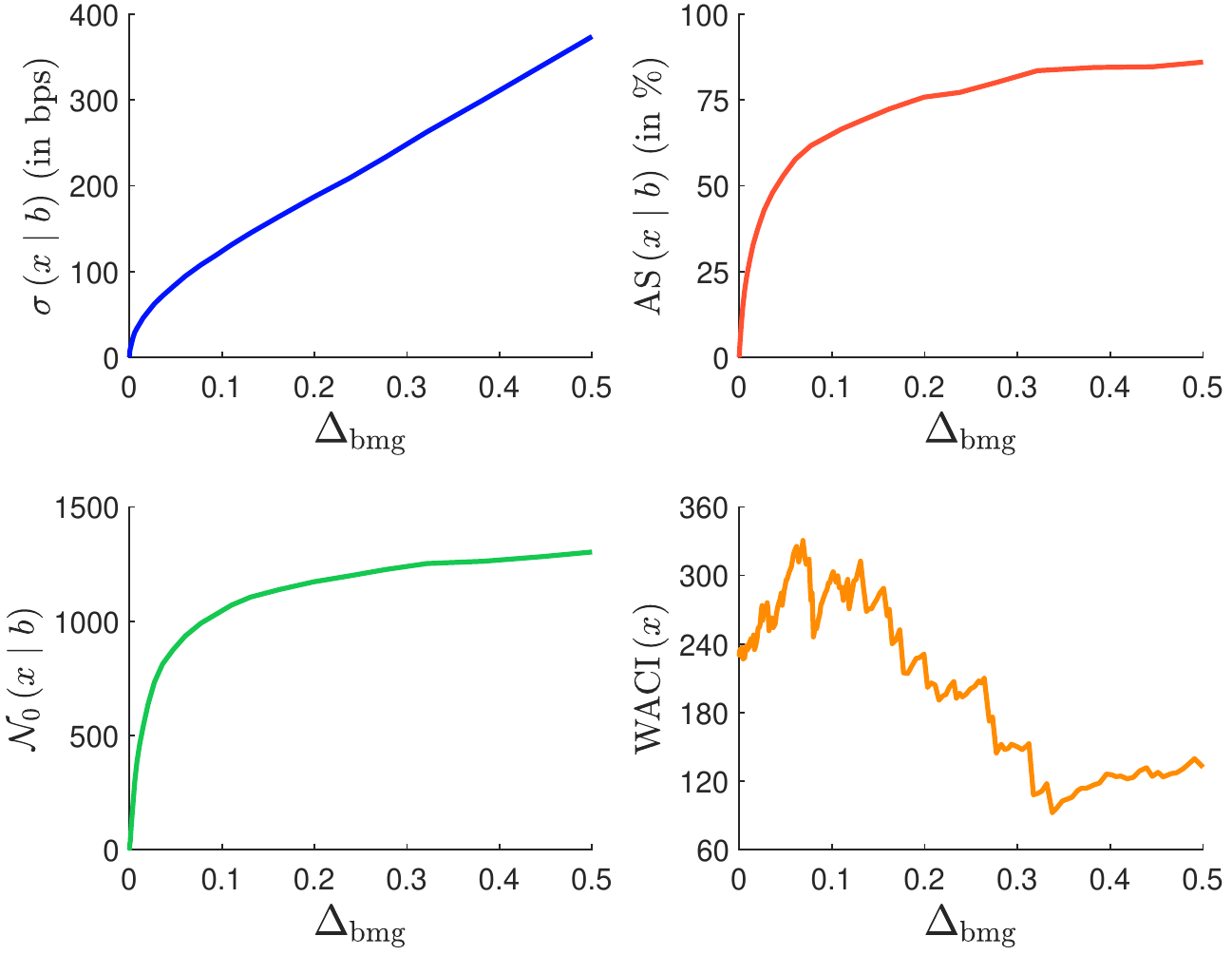}
\end{figure}

\begin{figure}[tbph]
\centering
\caption{Solution of the alternative order-statistic optimization problem}
\label{fig:te12}
\includegraphics[width = \figurewidth, height = \figureheight]{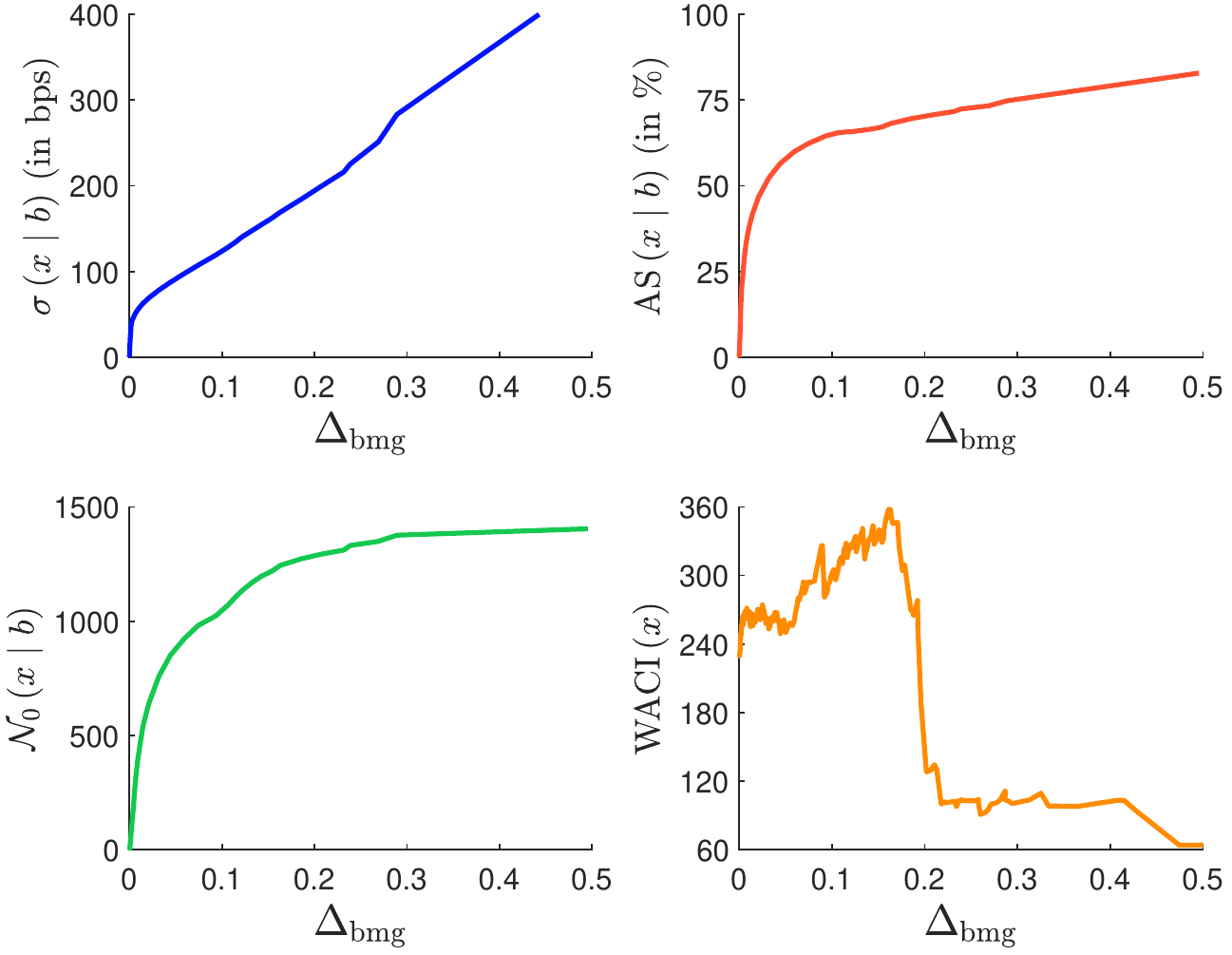}
\end{figure}


\end{document}